\documentclass[a4paper,fleqn,usenatbib]{mnras}

\usepackage{newtxtext,newtxmath}

\usepackage[T1]{fontenc}
\usepackage{ae,aecompl}

\usepackage{graphicx}	
\usepackage{amsmath}	
\usepackage{amssymb}	
\usepackage{subfig}
\usepackage[font=small,labelfont=bf]{caption} 

\title[Revisiting the \textit{Kepler} non-Blazhko sample]
{Revisiting the \textit{Kepler} non-Blazhko RR Lyrae sample:
Cycle-to-cyle variations and additional modes}

\author[Benk\H{o} et al.]{
J\'ozsef M. Benk\H{o}$^{1}$\thanks{E-mail: benko@konkoly.hu},
Johanna Jurcsik$^{1}$,
Aliz  Derekas$^{2,1}$
\\
$^{1}$Konkoly Observatory, MTA CSFK, 
Konkoly Thege M. u. 15-17., H-1121 Budapest, Hungary\\
$^{2}$ELTE E\"otv\"os Lor\'and University, Gothard Astrophysical Observatory, 
Szent Imre herceg u. 112., H-9704 Szombathely, Hungary\\
}

\date{Accepted 2019 March 18. Received 2019 March 9; in original form 2019 February 14}

\pubyear{2019}

\begin{document}
\label{firstpage}
\pagerange{\pageref{firstpage}--\pageref{lastpage}}
\maketitle

\begin{abstract}
We analyzed the long and short cadence light curves of the \textit{Kepler} non-Blazhko RRab stars.
We prepared the Fourier spectra, the Fourier amplitude and phase variation functions, time-frequency
representation, the O$-$C diagrams and their Fourier contents.
Our main findings are: (i) All stars which are brighter a certain magnitude limit show
significant cycle-to-cycle light curve variations. (ii) We found permanently excited additional modes
for at least one third of the sample and some other stars show temporarily excited additional modes.
(iii) The presence of the Blazhko effect was carefully checked and identified one
new Blazhko candidate but for at least 16 stars the effect can be excluded. This fact has important
consequences. Either the cycle-to-cycle variation phenomenon is independent from the Blazhko effect and
the Blazhko incidence ratio is still much lower (51\%-55\%) than the extremely large (>90\%)
ratio published recently.
The connection between the extra modes and the  cycle-to-cycle variations is marginal.
\end{abstract}

\begin{keywords}
stars: oscillations -- stars: variables: RR\,Lyrae
-- methods: data analysis -- space vehicles
\end{keywords}



\section{Introduction}

When the first pulsating variable stars were discovered at the end of the 19th century, 
seeing their accurately repetitive light curves, it was even suggested that they 
could be the basis of the time measurement as standard oscillators.
The discovery of incredible accuracy of the atomic vibration frequencies made all 
such suggestions of the past.
With the development of stellar pulsation and evolution theories it became evident 
that the periods of pulsating variables are changing during their evolution.
This type of variation was intensively searched in the first half of the past
century. The main tool of this  work was the O$-$C diagram (see \citealt{Sterken05} and
references therein). The decades or century-long diagrams of RR\,Lyrae stars, however, yielded
rather controversial results.
Only the smaller part of the investigated stars showed
evolution origin period change, the larger part showed irregular large
amplitude period variations (e.g. \citealt{Szeidl65, Szeidl73, Barlai89}).

A possible explanation of this finding was the sum up of the small random changes
of the pulsation cycles.
In other words: we see random walk in O$-$C diagrams \citep{Detre,Koen06}. Later,
several authors suggested possible irregular changes in the period of the classic
radially pulsating variables (Cepheids and RR Lyrae type) on various theoretical
bases \citep{Sweigart79, Deasy85, Cox98}. These ideas, however, have never been included 
into any standard pulsation codes.

The more recent and more extended period change studies of RR\,Lyrae stars
in the Galactic field and globular clusters \citep{LeBorgne07,Jurcsik01,Jurcsik12,Szeidl11}
came to the conclusion that most of the non-Blazhko stars show smooth evolution origin
period changes while Blazhko stars have large amplitude short time-scale irregular
period fluctuations. The possibility of cycle-to-cycle variation of the non-Blazhko stars 
has been removed from the agenda. 

The first direct detection of a random period jitter
of V1154\,Cyg, the only classical Cepheid of the original {\it Kepler} field \citep{Derekas12, Derekas17},
however, changed the situation. A similar phenomenon was suspected 
for CM\,Ori a mono-periodic (non-Blazhko) RR\,Lyrae star observed by the {\it CoRoT} space 
telescope \citep{Benko16}. In both cases the detected period variations
were about some thousandths or ten thousandths of the pulsation periods.
The earth-based observation typically neither precise nor well-covered enough 
to discover such a small random period fluctuations.   
They need to be not only  precise and uninterrupted but high 
cadence data as well. Might be, this is the reason why \citet{Nemec11} systematic
stability analysis on the non-Blazhko stars of the {\it Kepler} field resulted 
in a null result: the used long cadence (LC, $\sim$29~min sampled) {\it Kepler}
observations were too sparse to detect such tiny variations.
While the {\it CoRoT} and {\it K2} Cepheids' light curves are too short \citep{Poretti15}
 {\it Kepler} short cadence RR\,Lyrae data are promising for searching the effect.
This paper presents the investigations of RR Lyrae stars in the {\it Kepler} field based on the short cadence observations completed with some connecting
analysis using the long cadence data. 

\section{The sample and its data}

We used the non-Blazhko sample observed in the original {\it Kepler} field.
The latest detailed work on this sample was \citet{Nemec13} who listed 21 non-Blazhko RRab stars.
In the meanwhile the Blazhko behaviour of two stars (V350\,Lyr and KIC\,7021124) 
has been identified \citep{Benko15} so these two stars were omitted from the present sample. 
The investigated stars are listed in Table~\ref{tab:sample}.   

The {\it Kepler} mission was introduced in \citet{Borucki10} and all the technical details are discussed in the 
handbooks of \citet{KIH, DPH}, and \citet{KDCH}.
This work used  two light curves for each star: the total four-years-long 
normally $\sim29$~min sampled  so-called long cadence (LC) light curve, and 
the $\sim1$~min (over)sampled short cadence (SC) data of the same stars.
Tipically, a given star was observed in SC  mode in a few quarters
(see column 3 in Table~\ref{tab:sample}). 
In both cases the light curves have been produced by using our
own tailor-mode aperture photometry carried out on the publicly available 
original CCD frame parts 
(`pixel data')\footnote{{\it Kepler} pixel data can be downloaded from the web page of
MAST: \url{http://archive.stsci.edu/kepler/}, while the light curves used this
work from our web site: \url{http://www.konkoly.hu/KIK/}}. 
The data handling and the photometric process are described in \citet{Benko14}.
Here we mention only that for the sake of uniform handling the same parameters (apertures,
zero point shifts and scaling ratios) 
were used for both the SC and the LC data.  
\begin{table}
        \centering
        \caption{The used {\it Kepler} RR Lyrae sample.
The columns show the star's KIC ID; variable name, if exits;
the observed SC quarters; and the total observed SC time}
        \label{tab:sample}
        \begin{tabular}{rrcr} 
                \hline
KIC & Var. name & SC quarters &  $T$\\
&  &   &  (d)\\
                \hline
3733346& NR\,Lyr   &  Q11.1 & 31.1\\
3866709& V715\,Cyg &  Q7, Q9 & 186.8\\ 
5299596& V782\,Cyg &  Q7, Q9 & 186.8 \\ 
6070714& V784\,Cyg &  Q8, Q13-Q17 & 475.3\\ 
6100702&           &  Q8 & 67.0\\
6763132& NQ\,Lyr   &  Q10 & 93.4\\
6936115& FN\,Lyr   &  Q0, Q5, Q11.3 &  138.1\\ 
7030715&           &  Q9 & 97.4\\
7176080& V349\,Lyr &  Q9 & 97.4\\
7742534& V368\,Lyr &  Q10 & 93.4\\
7988343& V1510\,Cyg&  Q8 & 67.0\\
8344381& V346\,Lyr &  Q10 & 93.4\\
9591503& V894\,Cyg &  Q9 & 97.4\\
9658012&           &  Q11.1-Q11.2 & 62.0\\ 
9717032&           &  Q11 & 97.1\\
9947026& V2470\,Cyg&  Q7,Q9-Q10 & 281.2\\ 
10136240& V1107\,Cyg& Q9 & 97.4\\
10136603& V839\,Cyg & Q11.2 & 30.2\\
11802860& AW\,Dra   & Q0, Q5, Q11.3 & 138.2\\ 
                \hline
        \end{tabular}
\end{table}

\section{The SC light curves}

\begin{figure*}
\includegraphics[scale=.5]{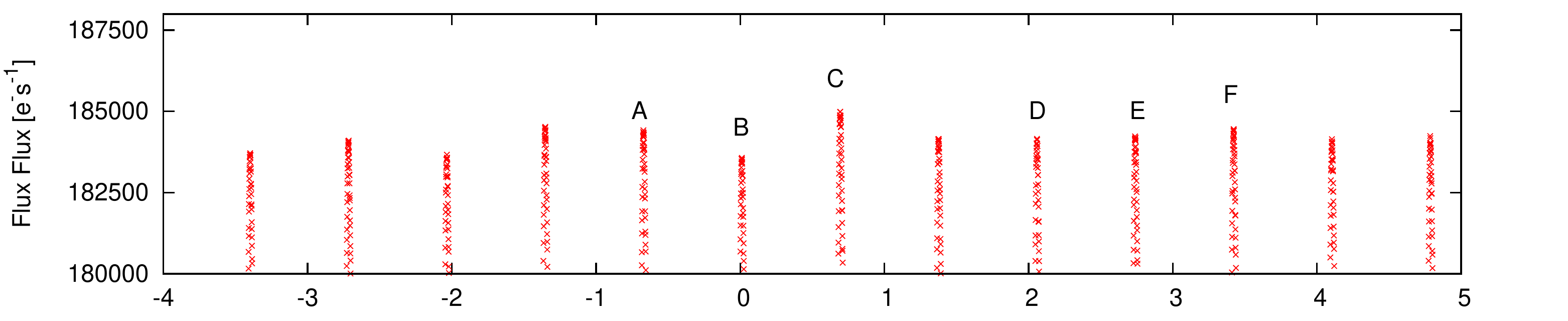}
\includegraphics[scale=.555, trim=0 0 0.8cm 0]{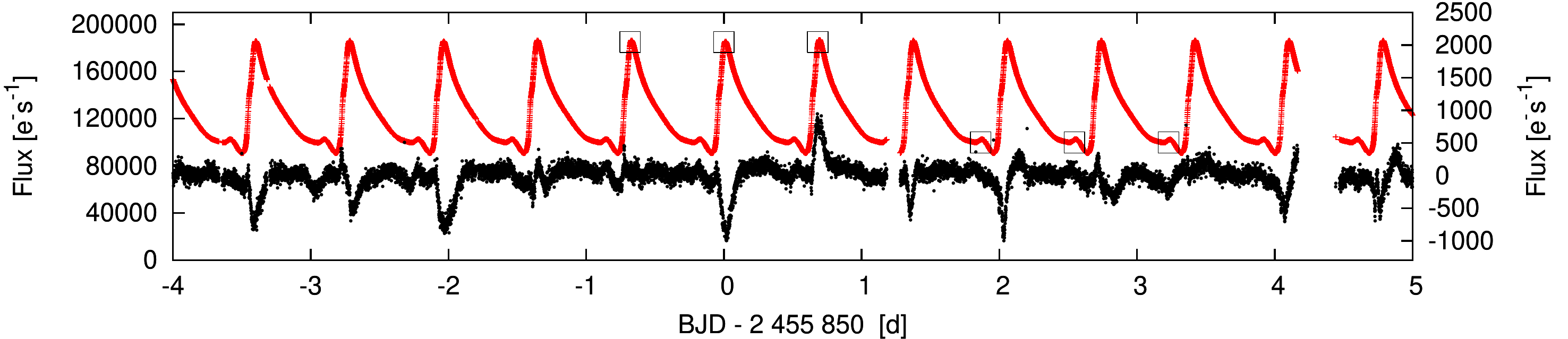}
\includegraphics[scale=0.39,trim=3cm 0 0 0]{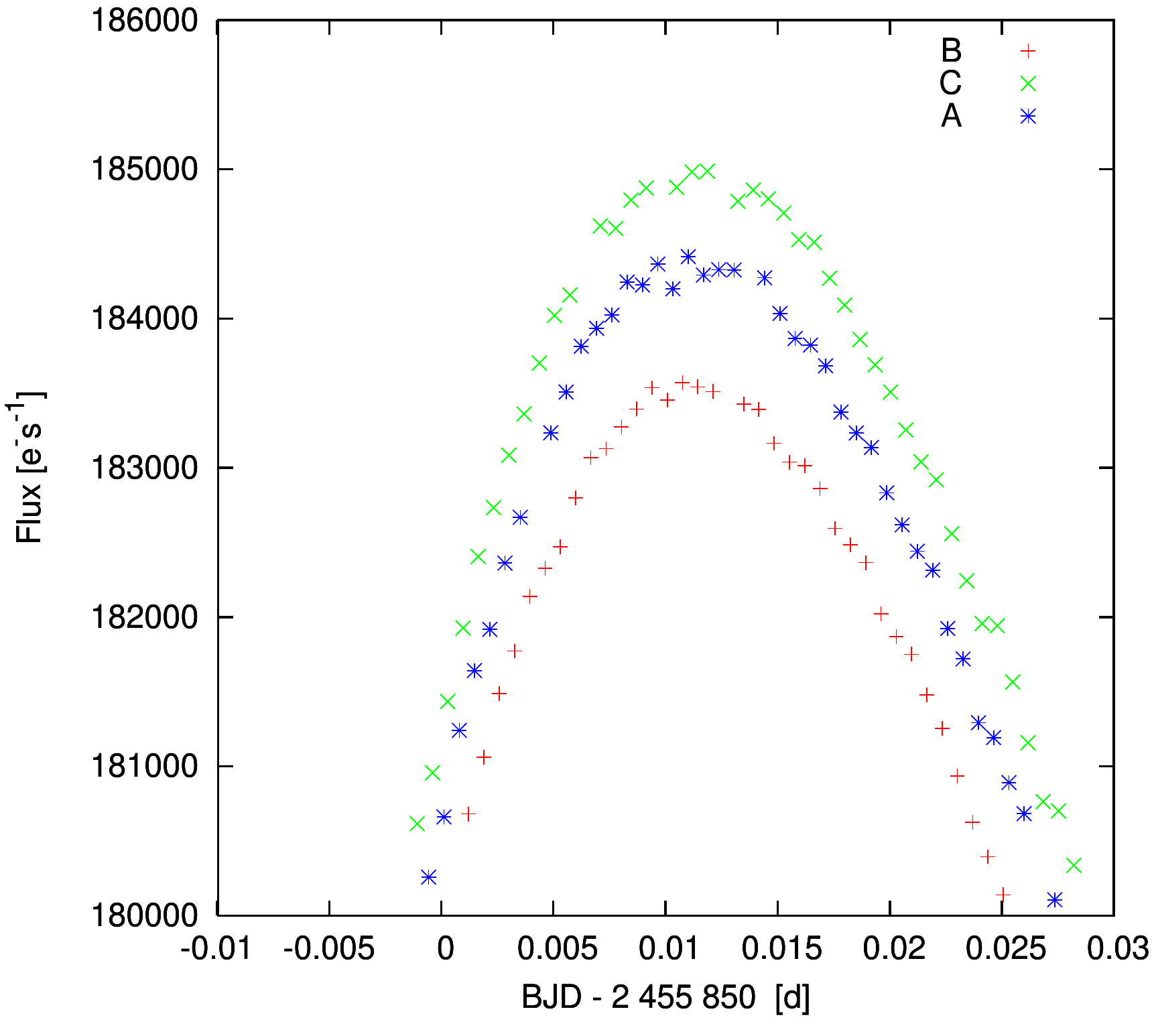}
\includegraphics[scale=0.4,trim=0 0 0 0]{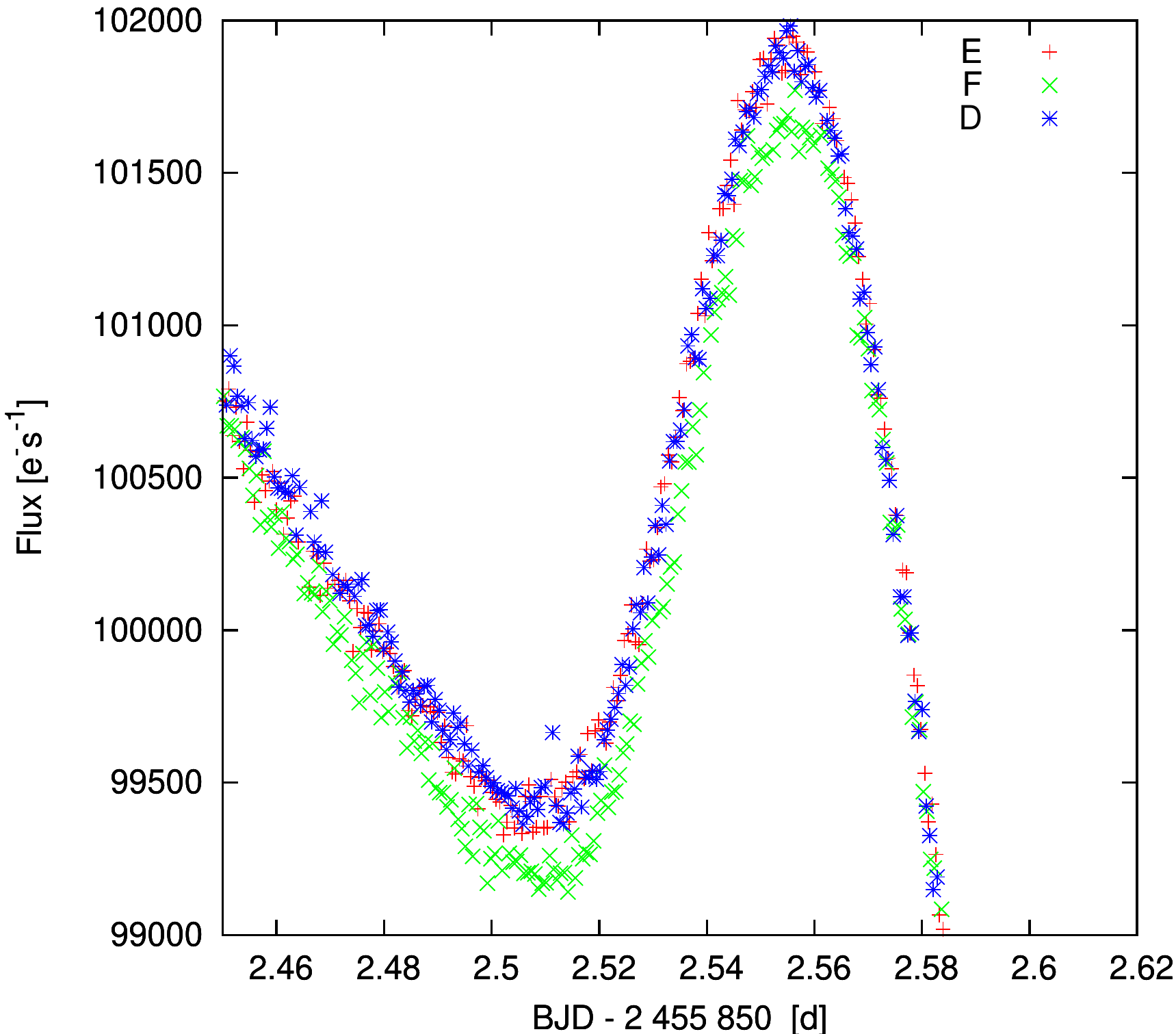}
\caption{Cycle-to-cyle variation of the SC flux curve of NR\,Lyr.
Top panel: a part of the flux curve around the pulsation maxima; 
middle panel: residual curve after we pre-whitened the data with 55 
significant harmonics of the main pulsation frequency 
(right scale, black points) and for comparison the original flux
curve (left scale, red points); 
bottom panels: three consecutive cycles from the flux curve
signed by different colours and
symbols folded to the cycle `B' and `E', respectively.
The small boxes in the middle panel indicate
the positions of these flux curve parts.
}\label{fig:sc_amp}
\end{figure*} 

\subsection{Cycle-to-cycle variation of the light curves}\label{sec:C2C}

First we examined the SC time series.
We used the raw flux data obtained from our tailor-made 
aperture photometry, which is practically a simple
pixel flux value summation  without any further processing.

While checking the flux curves we realised that the 
pulsation cycles are different to each other.
As an example we show a part of the SC light curve of NR\,Lyr in Fig~\ref{fig:sc_amp}.
The top panel shows the light curve around maxima of 13 consecutive pulsation cycles. The most striking feature is the different height of maxima. We marked three consecutive pulsation cycles with the letters `A', `B'
and `C'. In the bottom left panel, the same three cycles are plotted by shifting `A' and `C' cycles to the position
of `B' (red plus), i.e. `A' is shifted in the positive direction (blue asterisk) and `C' is shifted in the negative direction (green x). As we can see, all the maxima are well-covered by
observations and differ to each other significantly. The difference 
between maxima `B' and `C'  is $\sim1500$~e$^{-}$s$^{-1}$ 
(in magnitude scale is around 0.008~mag) which is huge compared
to the observational error of individual data points ($\sim2\times10^{-5}$~mag).

Although the most striking feature is the different maxima,
other parts of the light curves are also different. Looking at the 
three consecutive cycles marked with `D', `E' and `F' in the top panel of Fig~\ref{fig:sc_amp}.
The height of maxima of these cycles are almost the same, while cycle `F' is
a bit higher. Shifting the cycles `D' (blue asterisk) and `F' (green x) with plus or minus one 
pulsation cycle to the position of the cycle `E' (red plus) and crop around the bumps 
(pulsation phase $\phi\sim 0.6-0.8$), 
we get the bottom right panel of Fig~\ref{fig:sc_amp}. We see that 
the light curves of cycles `D' and `E' are overlapped but cycle `F' goes bellow these
two. The difference is abut 300~e$^{-}$s$^{-1}$ (0.003~mag). Since cycle `F' has the
largest maximum amongst these three cycles there is no vertical shift which could
eliminate both the maximum and the minimum differences simultaneously.  

The complex structure of the light curve changes can be studied in detail by preparing
the residual flux curve. 
A 55-element harmonic fit was removed from the data. 
The resulted curve is shown in the middle 
panel of Fig.~\ref{fig:sc_amp} (black dots) with
the original flux curve (red dots). The residual shows sharp
spikes at around the light curve maxima. These spikes are
positive or negative according to that the certain cycle flux curve is above or below
the fit, respectively. 
Spikes can also be found at different phases than maxima ($\phi=0$).
These phases are $\phi\sim0.92, 0.95, 0.7, 0.75$, and 0.1. The first two phases
are the beginning and the end of the light curve feature of the ascending branch
often called `hump' while $\phi\sim 0.75$ is the position of the `bump'.
The light curve of NR\,Lyr shows no evident features at the positions of
$\phi \sim 0.7$ and 0.1 but these phases are the same that were 
defined by \citet{Chadid14} as the positions of `rump' and `jump' recently.
Maxima and these phases are those parts of the light curves 
where the most prominent shock waves are generated \citep{Simon86,Fokin92,Chadid08,Chadid13}.

We found similar cycle-to-cycle (hereafter C2C) variations for all stars 
which are brighter than $K_{\mathrm{p}} \sim 15.4$~mag
(see 'yes' sign in the fourth column of Table~\ref{tab:features}).
The KIC $K_{\mathrm p}$ magnitudes given in Column 2 of Table~\ref{tab:features} 
were determined by ground-based photometry by \citet{KIC}, who observed each star 
in three different epochs. This  observing strategy is well-suited to constant stars
but it could result in inaccurate average magnitudes for large amplitude 
variable stars as RR\,Lyrae. The brightness of our stars are, therefore, better
characterized by the measured average flux (Col. 3 in Table~\ref{tab:features}) 
than the KIC magnitudes.

\begin{figure}
\includegraphics[scale=.32,angle=270,trim=4.5cm 0 4cm 0]{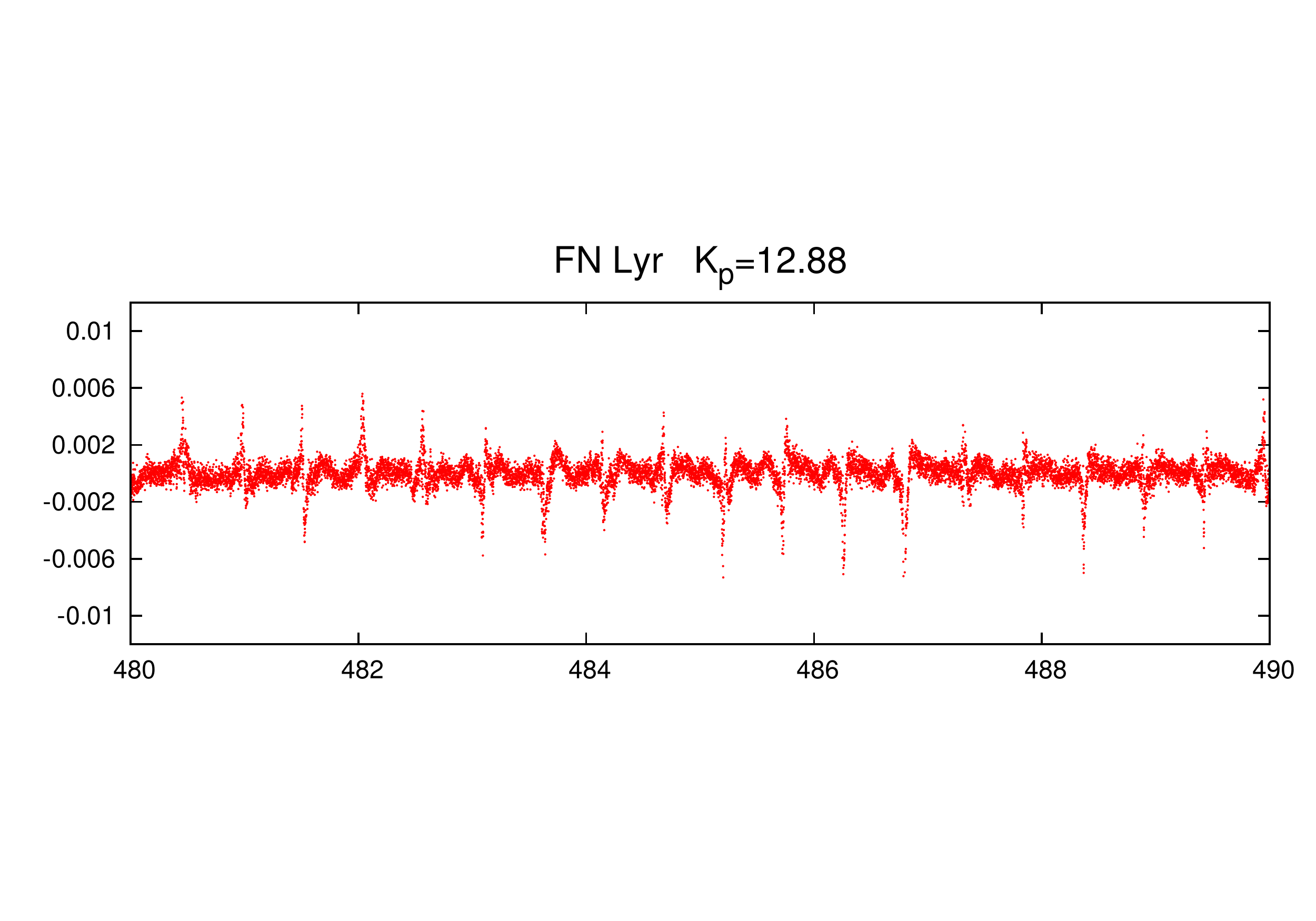}
\includegraphics[scale=.32,angle=270,trim=4.5cm 0 4cm 0]{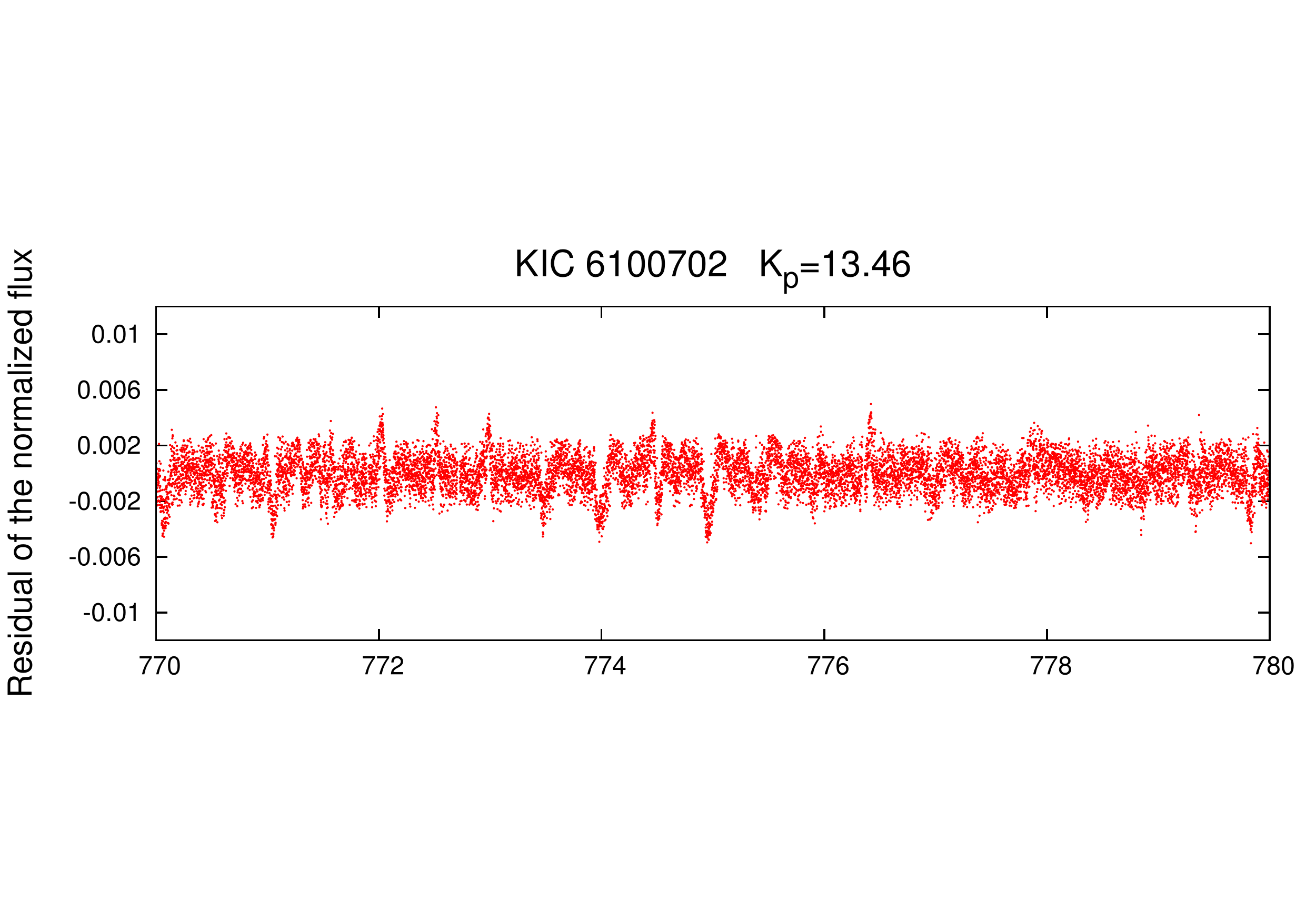}
\includegraphics[scale=.32,angle=270,trim=4.5cm 0 4cm 0]{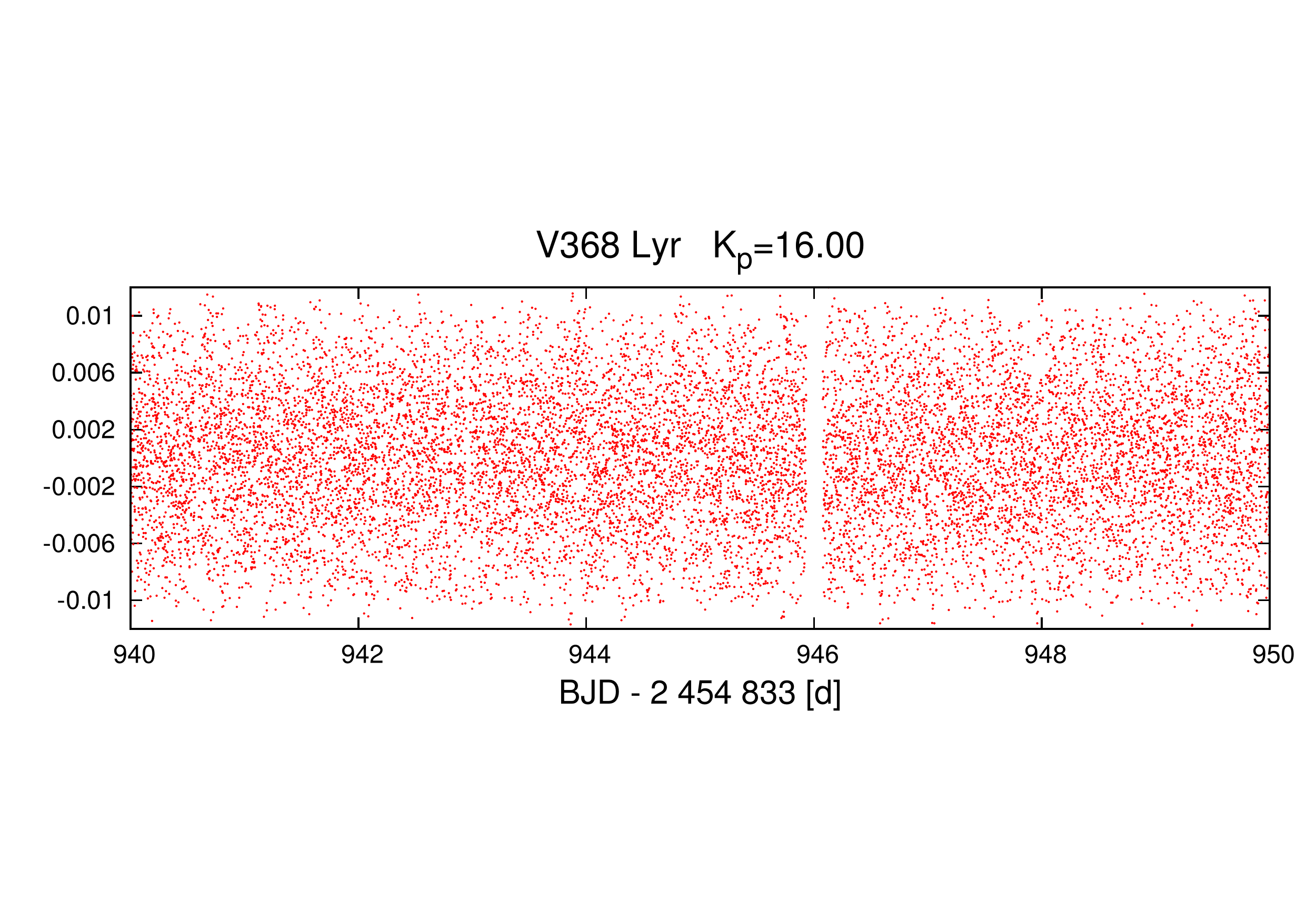}
\caption{Ten-days-long parts of residual flux curves.
The three different brightness stars are shown in the same relative scale.
The apparent brightness of the stars are decreasing from top to bottom.
The detectability of the C2C variation features are highly depends on
the brightness. The fainter of the star the hardest to detect the effect.
}\label{fig:sc_res}
\end{figure}

The maximal brightness deviations is similar for all stars:
the difference between the highest and lowest maxima is about 0.006-0.008~mag.
This general value might be responsible for the lack of C2C variation 
of fainter stars: the higher observation noise make the effect to be undetectable.
The situation is illustrated with Fig.~\ref{fig:sc_res} where we plotted 
the residual of the normalized flux ($F/\langle F\rangle$, where $F$ means
the flux in e$^{-}$s$^{-1}$ and $\langle F\rangle$ is the average flux) 
curves of three stars with different brightness in the same scale.
While the C2C variations of FN\,Lyr 
($K_{\mathrm p}=12.88$~mag) in top panel of Fig.~\ref{fig:sc_res} is very similar to NR\,Lyr,
the spikes are less detectable for the fainter KIC\,6100702 
($K_{\mathrm p}=13.46$~mag, middle panel). Finally, no structure can
be recognized within the higher noise level of the faintest star V368\,Lyr 
($K_{\mathrm p}=16.00$~mag, bottom panel).

The C2C behaviour of NR\,Lyr showed in Fig.~\ref{fig:sc_amp} is typical
not just in its amplitude but in its other characteristics as well.
The difference in maxima are generally higher than the minima (or other parts
of the light curves). The maxima (and minima) value variation seems to be random.
Sometimes increasing or decreasing amplitude cycles follow 
each other but in many other cases a small amplitude cycle follows a large amplitude one 
or vice versa (see also top panels of Fig~\ref{fig:sc_amp} and Fig~\ref{fig:sc_res}). 

\begin{table}
        \centering
        \caption{Detection of the C2C variation. Name;
\textit{Kepler} $K_{\mathrm p}$ brightness from the KIC catalogue \citep{KIC}; 
C2C variation detection by eye; Detection index $D$ (see the text for the details).
}
        \label{tab:features}
        \begin{tabular}{rcrcr} 
                \hline
Name & $K_{\mathrm p}$ & $\langle F \rangle$& C2C & $D_5$  \\
          & (mag)        & e$^{-}$s$^{-1}$ &  &     \\
                \hline
NR\,Lyr   &12.684& 128717 & yes    & 69.8   \\ 
V715\,Cyg & 16.265& 4731 &         & 14.5   \\ 
V782\,Cyg & 15.392& 12892 & yes    & 35.7   \\ 
V784\,Cyg & 15.370& 10129 & ?      & 76.2   \\ 
KIC\,6100702& 13.458&48145 & yes   & 92.9    \\
NQ\,Lyr     & 13.075&63394 & yes   & 98.7    \\
FN\,Lyr     & 12.876&115746 & yes  & 96.6    \\ 
KIC\,7030715 & 13.452&76707 & yes  & 112.8   \\
V349\,Lyr    & 17.433&1638  &      & 21.0    \\ 
V368\,Lyr   & 16.002&3772 &        & 21.6    \\ 
V1510\,Cyg  & 14.494&19762 & yes   & 28.2    \\
V346\,Lyr   & 16.421&2404 & ?      & 28.2    \\ 
V894\,Cyg   & 13.293&91854 & yes   & 109.8   \\ 
KIC\,9658012& 16.001&6692 & yes    & 26.8    \\ 
KIC\,9717032& 17.194&2521 &        & 11.6    \\ 
V2470\,Cyg& 13.300&64935  & yes    & 114.3   \\ 
V1107\,Cyg & 15.648&6293 & yes     & 34.0    \\
V839\,Cyg  & 14.066&25339 & yes    & 32.0    \\
AW\,Dra    & 13.053&108385 & yes   & 95.4    \\ 
                \hline
        \end{tabular}
\end{table}

\subsection{Origin of the C2C variations}\label{sec:real}

Although \citet{Chadid00} and \citet{Chadid13} reported spectroscopic 
C2C variations of RR\,Lyrae stars, on ground-based photometric basis only marginal
signs of such an effect were published  (e.g. \citealt{Barcza02, Jurcsik08}). 
On space photometric basis
no similar C2C variation of non-Blazhko RRab stars have been
reported ever before, so we checked our finding carefully.

(i) It is known that disruptions such as safe modes, 
the regular monthly downloads of data or quarterly rolls could cause 
abrupt changes in the row \textit{Kepler} fluxes \citep{DPH}.
We indeed detected small flux curve changes for many stars after such events but
the C2C variations are appeared continuously over the
entire data sets and are not concentrated around the discontinuity events. 
This rules out that the C2C variations would result from this technical problem. 

(ii) To avoid possible data handling problems which may cause such
an effect we used the raw tailor-made aperture photometric fluxes.
The local instrumental trends were handled in three different ways. 
(1) For 13 stars the SC data show no serious instrumental trends so we
used these data without any further processing.
(2) The raw data of six stars, however, show noticeable trends
which were removed by subtraction of fitted polinomials.
(3) As an independent check we applied a method to all SC data sets
in which we adjust each pulsation cycle to a common zero point.
For a given star a Fourier sum was fitted to each cycle separately,
the determined zero points were connected with a smooth continuous curve
which was then subtracted from the data.   	
This algorithm works well and removes the tiniest instrumental trends
but it has an assumption that zero point variations  
can only be caused by instrumental effects.
Although no systematic amplitude changes connected to this small zero point corrections were
detected in any of the studied stars,
it is known that, for example, the Blazhko effect also causes zero point 
variations \citep{Jurcsik05,Jurcsik06,Jurcsik08}.
In this respect, we know nothing about the C2C variations, 
so we did not use these zero point corrected data except for this test.

We compared the C2C variations of the raw (1) or the globally
corrected (2) data to the zero point corrected (3) data. 
These comparisons resulted in qualitatively similar C2C variations though the actual value 
of the quantitative properties 
(e.g. amplitude difference between consecutive cycles) was slightly different.
This test showed that the C2C variation is not caused by our data handling.      
\begin{figure}
\includegraphics[scale=0.37]{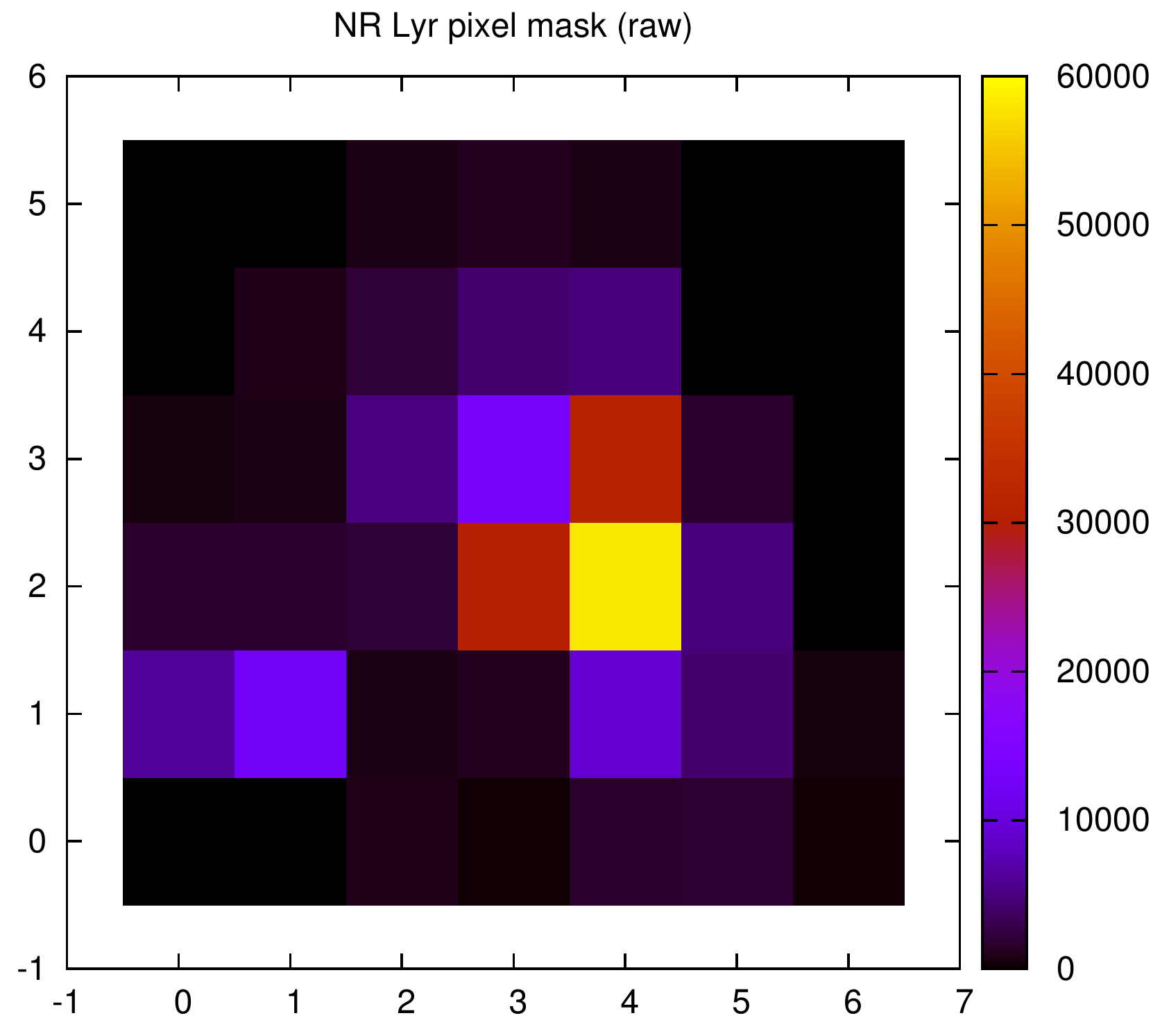}
\includegraphics[scale=0.37]{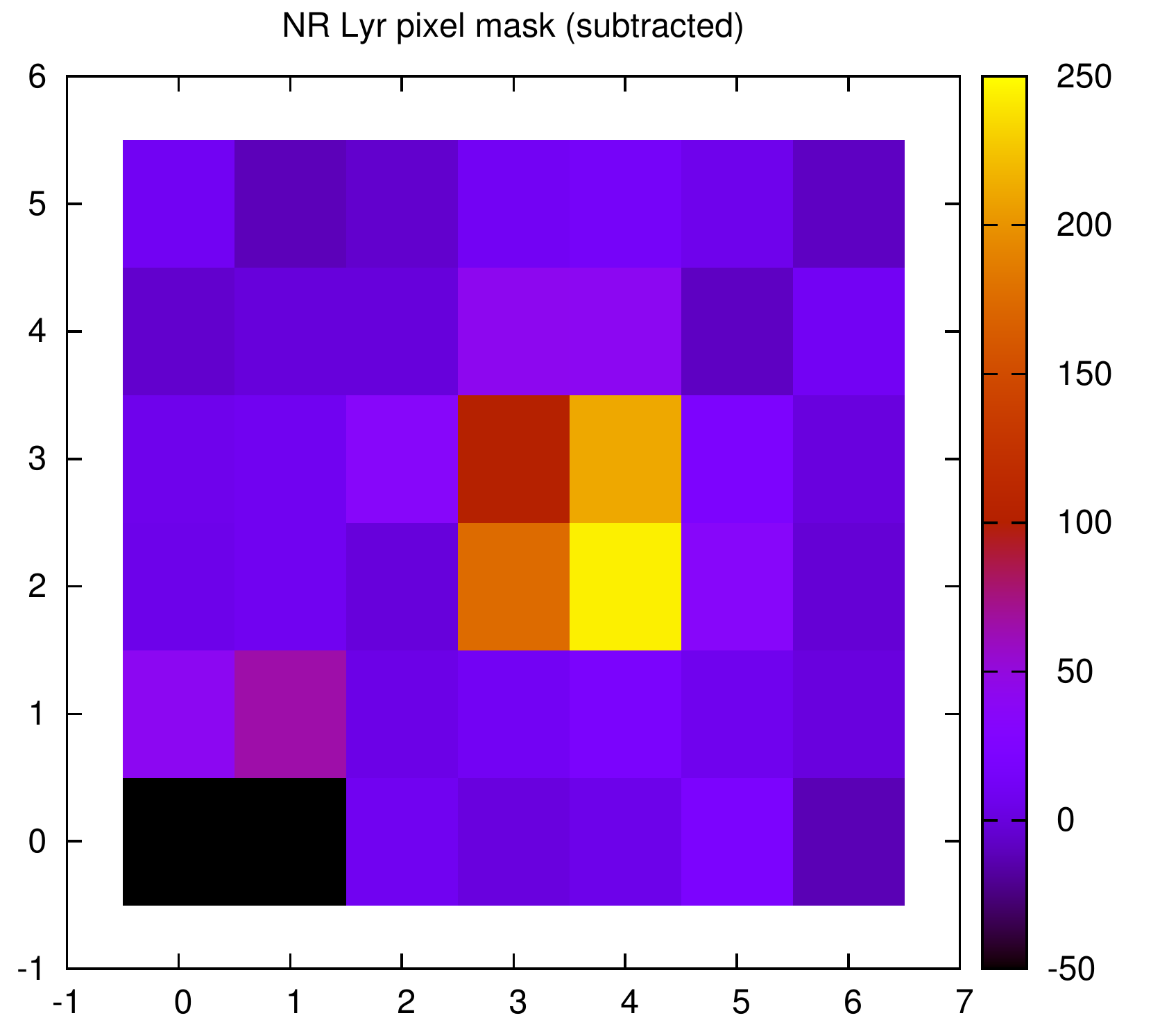}
\caption{The pixel mask of NR\,Lyr during the SC (Q11.1) observation.
The upper panel shows the flux in `high' maximum signed by `C' in Fig.~\ref{fig:sc_amp}, 
while the lower panel contains the flux difference of the `high'maximum `C' and `low' maximum `B'.
}\label{fig:pixel}
\end{figure} 

(iii) The next potential cause can come from the photometry, such as background sources, 
drift of the stars in the CCD frame, etc.
We have chosen high and low maxima pairs from the light curves 
and plotted the flux in the pixel maps at the high maximum phase and also
the flux differences between the high and low maxima phases.  
This is plotted for NR\,Lyr in Fig.~\ref{fig:pixel}.
The figure shows that (1) the amplitude difference is connected to the
star and there is not any other sources of light and (2) 
the position of the star is fixed within the pixel mask.
These image properties minimize the chance of C2C variations
being caused by serious photometric problems.

The investigation of the pixel masks resulted in a by-product. 
We found a faint variable source in the frame of V784\,Cyg. 
The source was identified with the star KIS\,J195622.44+412013.9
($g=20.19$, $r=19.32$ and $i=18.67$~mag) 
observed by the \textit{Kepler}-INT survey \citep{Greiss12}
 (see also last paragraph in Sec~\ref{sec:add}).
Other variable sources have not been found in any other frames. 
 
(iv) For testing unknown instrumental effects as an explanation of C2C variation, we investigated 
similar observations with a different instrument.  
Currently the only independent instrument which observed high precision 
time series for non-Blazhko RR\,Lyrae stars is the {\it CoRoT} space telescope \citep{Baglin06}.  
To our knowledge three non-Blazhko stars were observed with the oversampled
mode which mean 32~sec sampling.
(The time coverage of the normal 8~min sampling of {\it CoRoT} is too sparse for 
our porpose.) These are: CoRoT\,103800818 ({\it r}$_{\mathrm{CoRoT}}=14.39$~mag, \citealt{Szabo14}), 
CM\,Ori (CoRoT\,617282043, {\it r}$_{\mathrm{CoRoT}}=12.64$~mag, \citealt{Benko16}) 
and the BT\,Ser (CoRoT\,105173544, {\it r}$_{\mathrm{CoRoT}}=12.99$~mag) 
which was overlooked by previous {\it CoRoT} RR\,Lyr studies. 
\begin{figure}
\includegraphics[scale=0.37]{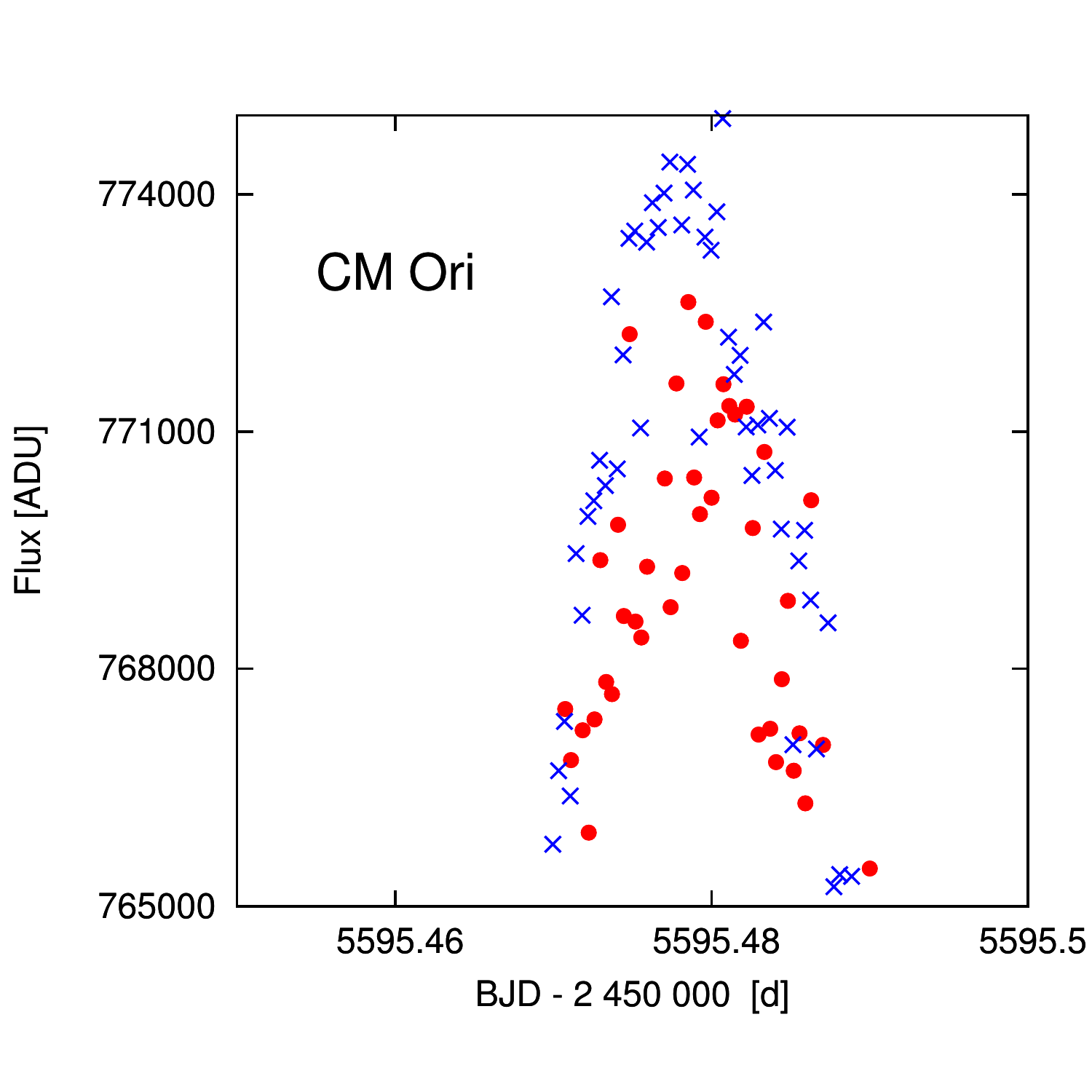}
\includegraphics[scale=0.37, trim=0 0 0 2cm]{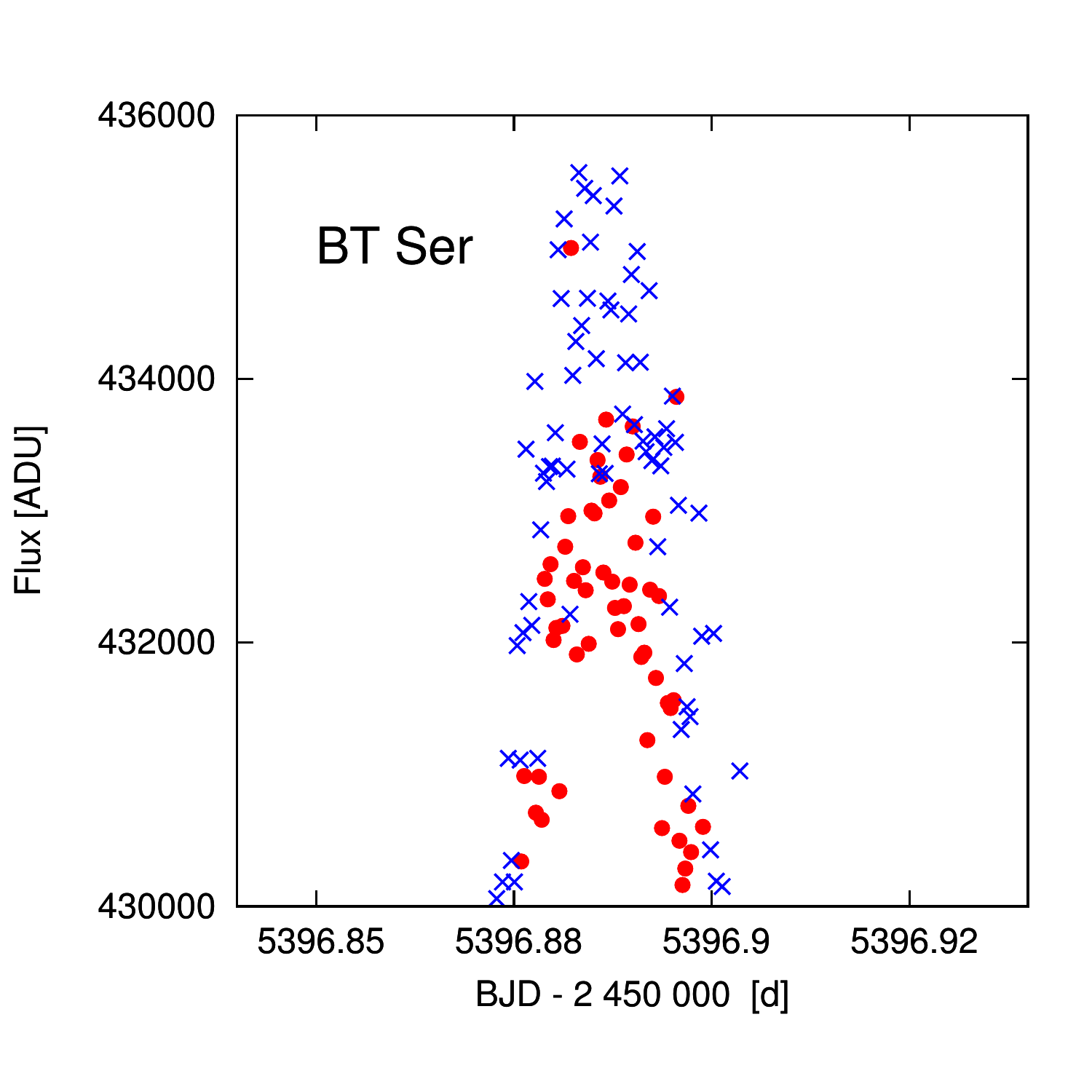}
\caption{The amplitude difference of consecutive cycles in \textit{CoRoT}
non-Blazhko stars. The red dots mean the original light curve points while the 
blue `x' symbols show the light curve points shifted one pulsation cycle.
}\label{fig:corot}
\end{figure} 

We used the oversampled flux time 
series\footnote{The data can be downloaded from the IAS CoRoT Public Archive
\url{http://idoc-corot.ias.u-psud.fr/sitools/client-user/COROT_N2_PUBLIC_DATA/project-index.html}}
 of CoRoT\,103800818 
from LRc04 run (74.6~d long oversampled part,  176\,871 data points), 
CM\,Ori LRa05 (90.5~d, 200\,999 observations),
and BT\,Ser which was observed in two subsequent {\it CoRoT} runs
LRc05 and LRc06, meaning 168.4~d-long almost continuous observations with 
391\,455 individual data points. 
These amount of data are comparable with the SC data of present {\it Kepler}
sample. CM\,Ori and BT\,Ser are relatively bright:  
despite the smaller aperture of {\it CoRoT}  we have similarly
accurate light curves for these stars as for the fainter {\it Kepler} stars.

We performed similar investigation of the {\it CoRoT} light curves as we did for {\it Kepler}
stars and we found C2C variation for the two brighter stars CM\,Ori and BT\,Ser.
Fig.~\ref{fig:corot} shows their amplitude variation in the same way that was plotted in the bottom left panel of 
Fig.~\ref{fig:sc_amp} for NR\,Lyr.
Even though the scatter is evidently higher, the
amplitude difference is obvious. The largest difference between high and
low amplitude maxima is about 0.005-0.006~mag. This value is similar to our estimation
obtained from {\it Kepler} stars.  The $\sim$2~mag fainter third star CoRoT\,103800818 
show no C2C variation as we expected on the basis of {\it Kepler} sample where also seems to be
exist a detection limit at about 15.4~mag.

These tests suggest that the detected C2C variations 
are predominantly belong to the stars. Of course, serious
time- and flux-dependent non-linearity of the detectors
might cause similar effects, however, no such problems have been reported
neither for \textit{CoRoT} nor for \textit{Kepler}. 
A promising independent check opportunity will 
be the analysis of \textit{TESS} \citep{Tess} oversampled (2-min) data. 

(v) There is an additional argument that the C2C light curve variations belong to the stars:
the shape of the residual light curves. The spikes described in Sec~\ref{sec:C2C}
are not randomly distributed in the pulsation phase but
appeared exactly at the phase of the hydrodynamic shocks.
This findings argees well with the results of 
radial velocity studies \citep{Chadid00,Chadid13} where the C2C radial velocity
curve variations were explained cycle to cycle variation 
of the hydrodynamic phenomena induced the shock waves in RR Lyrae atmosphere.

\subsection{Characterising the C2C variations}\label{sec:charC2C}

\begin{figure}
\includegraphics[scale=0.33,angle=270,trim=0 0 3.9cm 0]{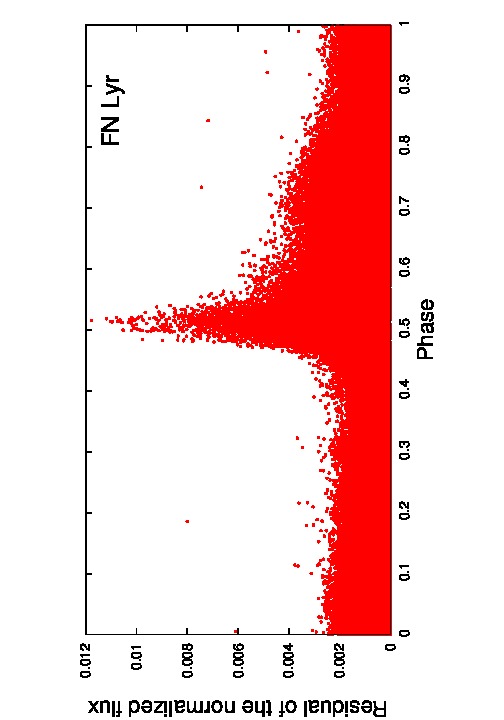}
\includegraphics[scale=0.33,angle=270,trim=0 0 2.5cm 0]{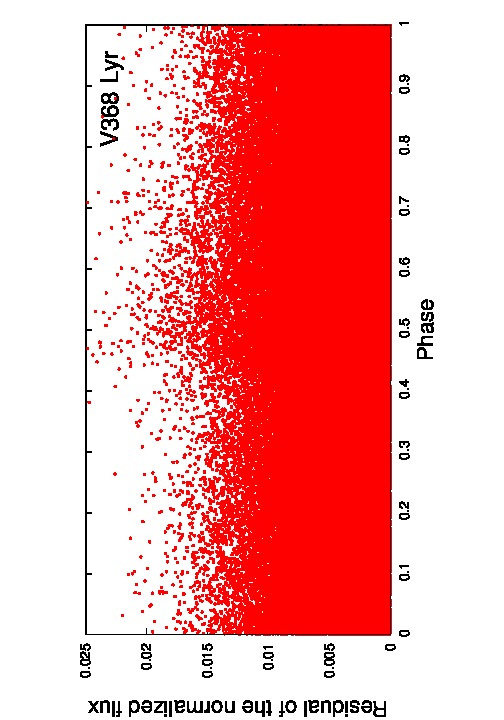}
\caption{The absolute values of the residual of the normalized flux vs. phase diagram 
of FN\,Lyr (up) and V368\,Lyr (down). 
}
\label{fig:folded_res}
\end{figure}

Beyond the visual inspection done in Sec.~\ref{sec:C2C},
we defined a quantity which numerically measures the detectability of the C2C variations.
As we have seen the C2C variations focus around the pulsation maxima therefore
the residual flux curves show spikes around these positions (Fig.~\ref{fig:sc_res}). 
The phase diagrams of these residual flux curves show a broadening around the phase of the
pulsation maxima ($\phi=0.5$ see Fig~\ref{fig:folded_res}). By comparing the amplitudes of these broadenings to the
amplitude of non-broadened phases, we can define a numerical value wich typify the 
detectability of C2C variation. 

A simple statistical approach was implemented. 
We folded the SC residual flux curves $r(t)$
with their periods then the obtained $r(\phi)$ phase
diagrams were splitted into few bins: $r_1 ,r_2 , \dots r_n$ ($n$ is integer). 
In each bin the average of the absolute values of the
residual fluxes $\langle \vert r_i\vert\rangle$
and its standard error $s_i$ was determined. 
The difference between the maximal and minimal bin values is
\begin{equation}
\Delta_n:= \left[ \langle \vert r_j\vert\rangle^{\mathrm{(max)}}- 
     {\langle\vert r_k\vert \rangle^{\mathrm{(min)}}} \right], \ \ \ j,k \in {1,2,\dots, n}.
\end{equation}
We can define the detectation parameter as
\begin{equation}\label{eq:D}
D_n:=\frac{\Delta_n}{s_n}, \ \ \ {\mathrm{where}} \ \ \ s_n=\max(s_j, s_k).
\end{equation}
The $D_n$ is a significance-like parameter. It  
measures how much larger the average flux of the central bin
which contains the spike than a bin which definitely not contains it.
The difference is expressed in the ratio of the standard error.
In column 4 of Table~\ref{tab:features} the $D_5$ values are given.
If a star has more than one distinct SC quarter observations, we determined
$D$ for each quarter separately, and here we show their averages.

The $D$ seem to be good C2C variations detection indicator: if $D$ value is high ($D > 30$)
we can detect evident C2C light curve variations by eye and if this value is low ($D < 25$) 
we cannot see anything. The faint variable in the frame of V784\,Cyg disturbs the visual inspection,
but the $D$ parameter clearly show the existence of the C2C variation. 
There is a trend between the parameter $D$  and the average flux $\langle F \rangle$
(Column 3 in Table~\ref{tab:features}): 
the brighter the star, the higher the associated $D$ parameter.
It suggests that the phenomenon is similar in strength for all stars and the differences of the detection 
are mainly  because of the brightness differences. 

The C2C variations seem to be random. To investigate this,
we prepared the Fourier amplitude and phase variation functions.
The SC light curves were divided into period-long bins and so
each bin contained abut 600-800 points depending on the cycle size. 
This handling minimize numerous possible technical problems such as 
 zero point fluctuations or short time-scale trends.
The amplitude and phase variation functions $A_n(t)$, $\phi_n(t)$ 
were calculated for each star by applying ten-element harmonic fits.
This calculation has been done with the {\sc{LCfit}} \citep{lcfit}  non-linear Fourier fitting package.

\citet{Plachy13} investigated RR\,Lyrae models  corresponding to resonance states 
and chaotic pulsation. Their synthetic chaotic luminosity curves show similar changes
than we presented here: the random-like changes are concentrated around the maxima
and the amplitudes are also in similar magnitude range. 
These raise the possibility that by the C2C variations
we observed might be the sign of chaotic pulsation. Detailed testing of such a 
possibility is far beyond the goal of this paper but we investigated
the Poincare return maps of $A_1$ and $\phi_1$ values as a fast and easy check.
We prepared four maps for each star: $(A_1^{(j)}, A_1^{(j+1)})$, $(A_1^{(j)}, A_1^{(j+2)})$,
$(\phi_1^{(j)}, \phi_1^{(j+1)})$ and $(\phi_1^{(j)}, \phi_1^{(j+2)})$.
Here $j$ integers mean the cycle numbers of the pulsation.  
Most of these maps have an oval shape showing the amplitude and phase variations
but no other evident structures can be detected. That is, the observed 
C2C variations might be chaotic but we cannot verify this at least with the
return maps, certainly.

\section{Fourier spectra}

\begin{figure*}
   \centering
\setlength{\tabcolsep}{0.0cm}
\begin{tabular}{ccc}
\includegraphics[width=3.5cm,angle=270,trim=30 0  20 10]{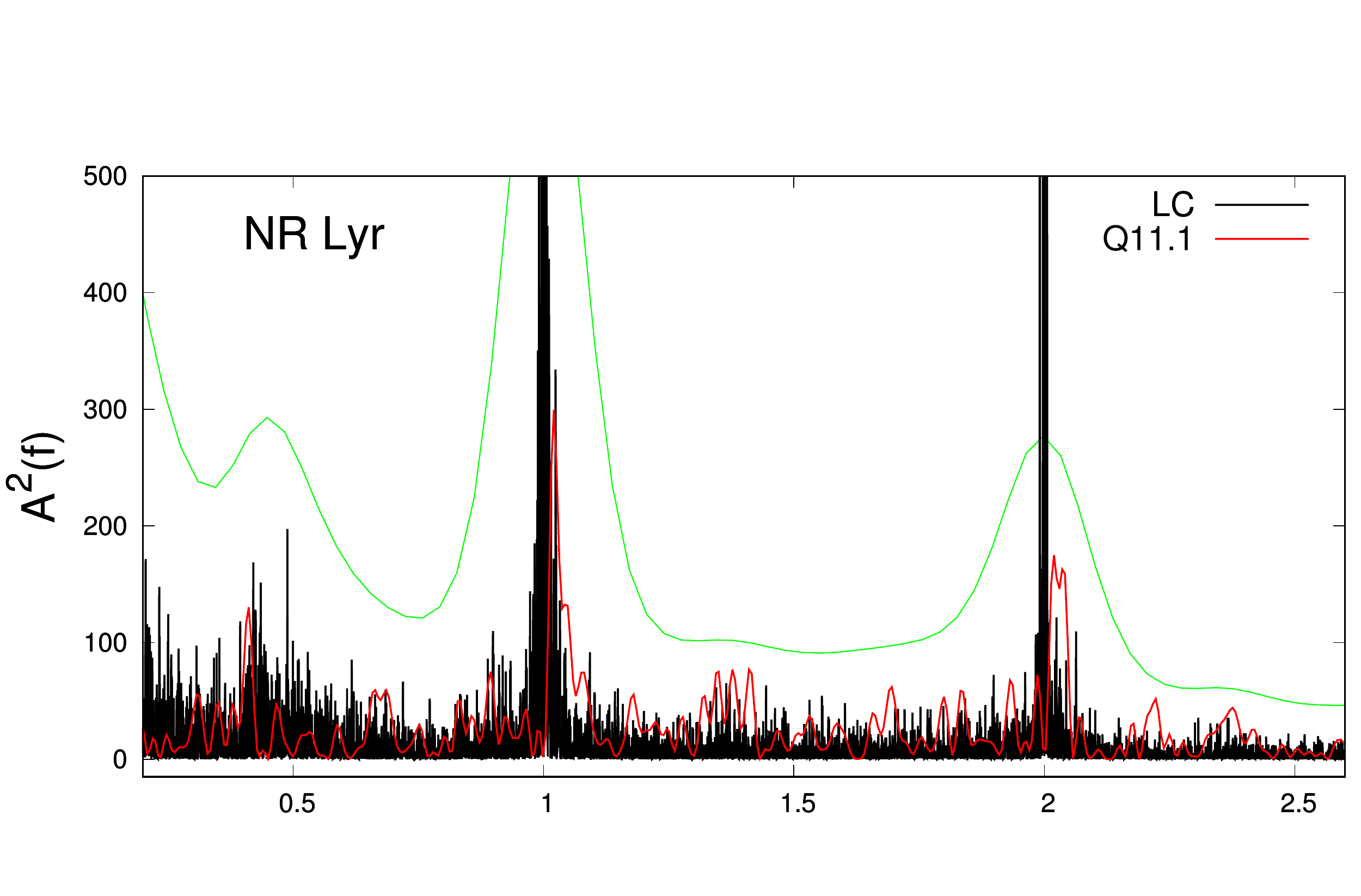}&
\includegraphics[width=3.5cm,angle=270,trim=30 20 20 10]{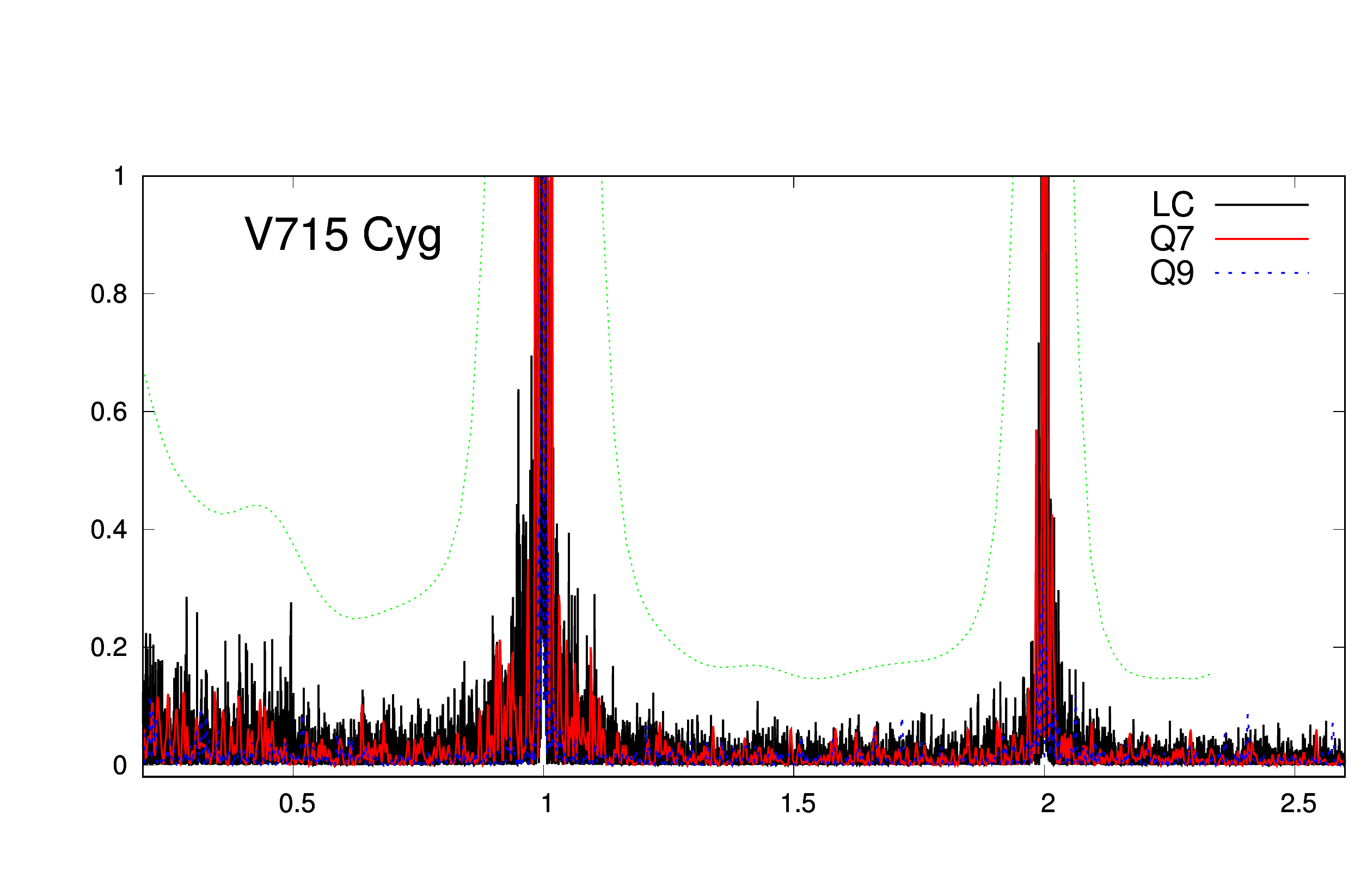}&
\includegraphics[width=3.5cm,angle=270,trim=30 20 20 10]{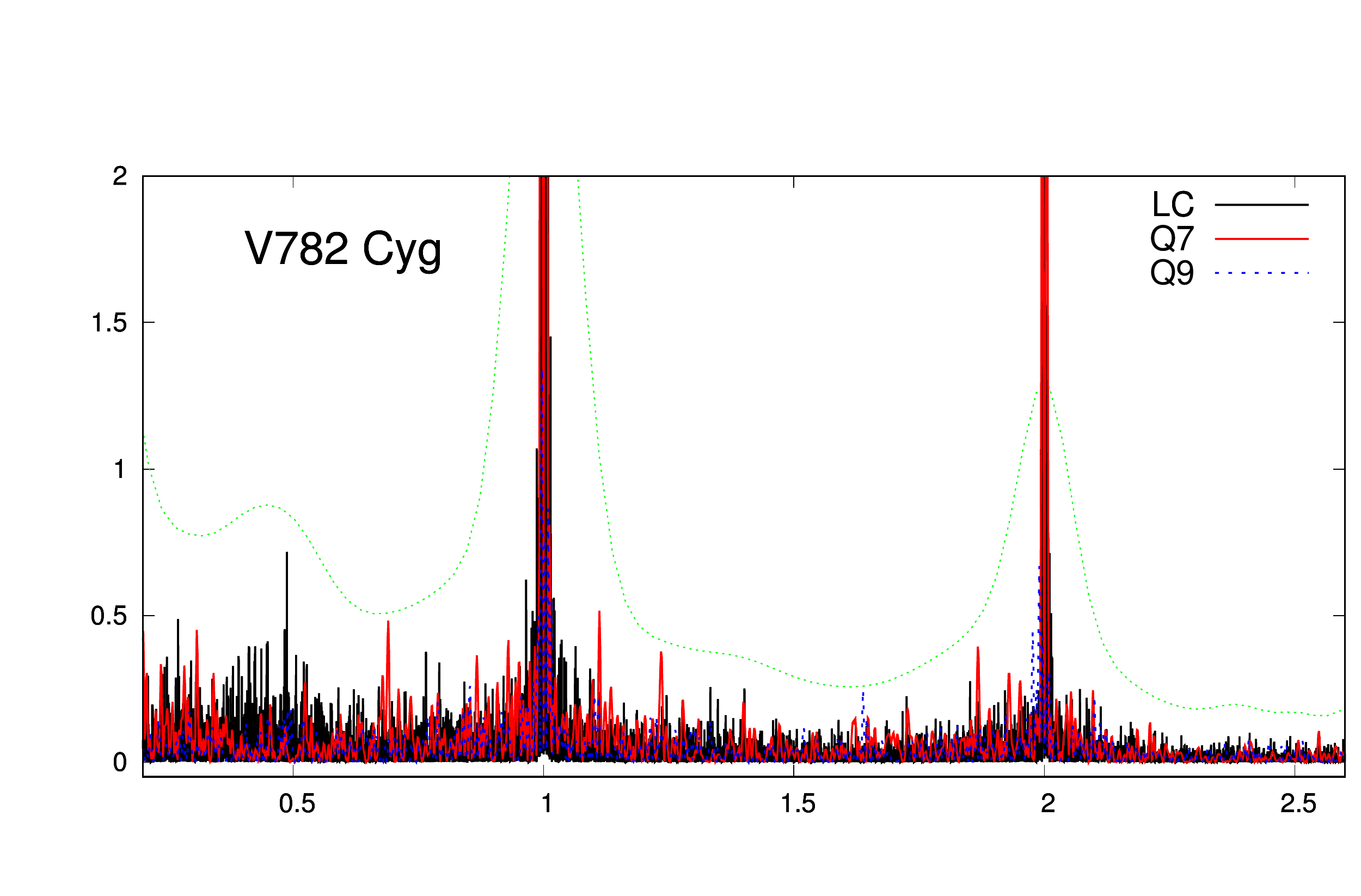}\\
\includegraphics[width=3.5cm,angle=270,trim=30 0  20 10]{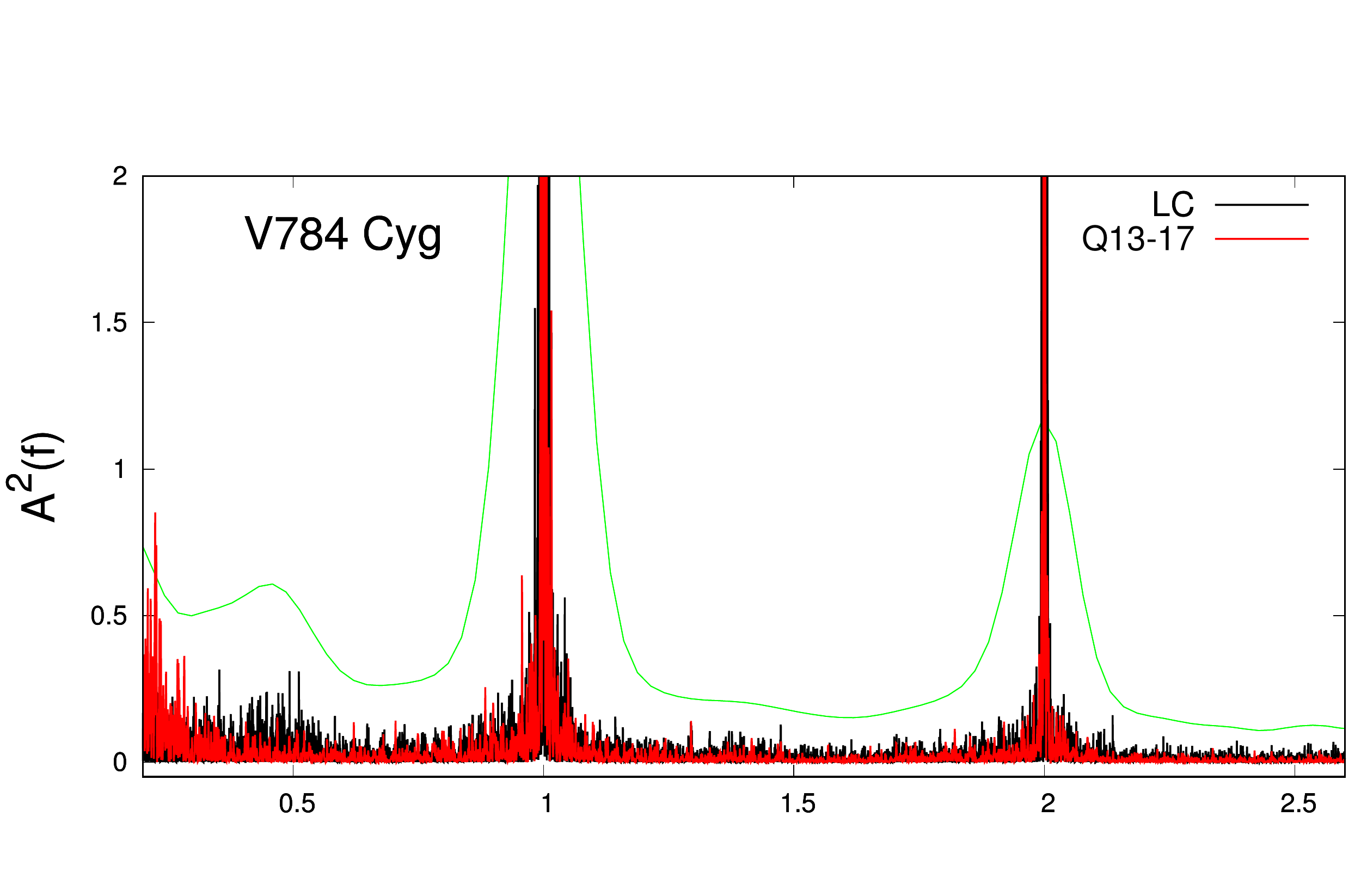}&
\includegraphics[width=3.5cm,angle=270,trim=30 20 20 10]{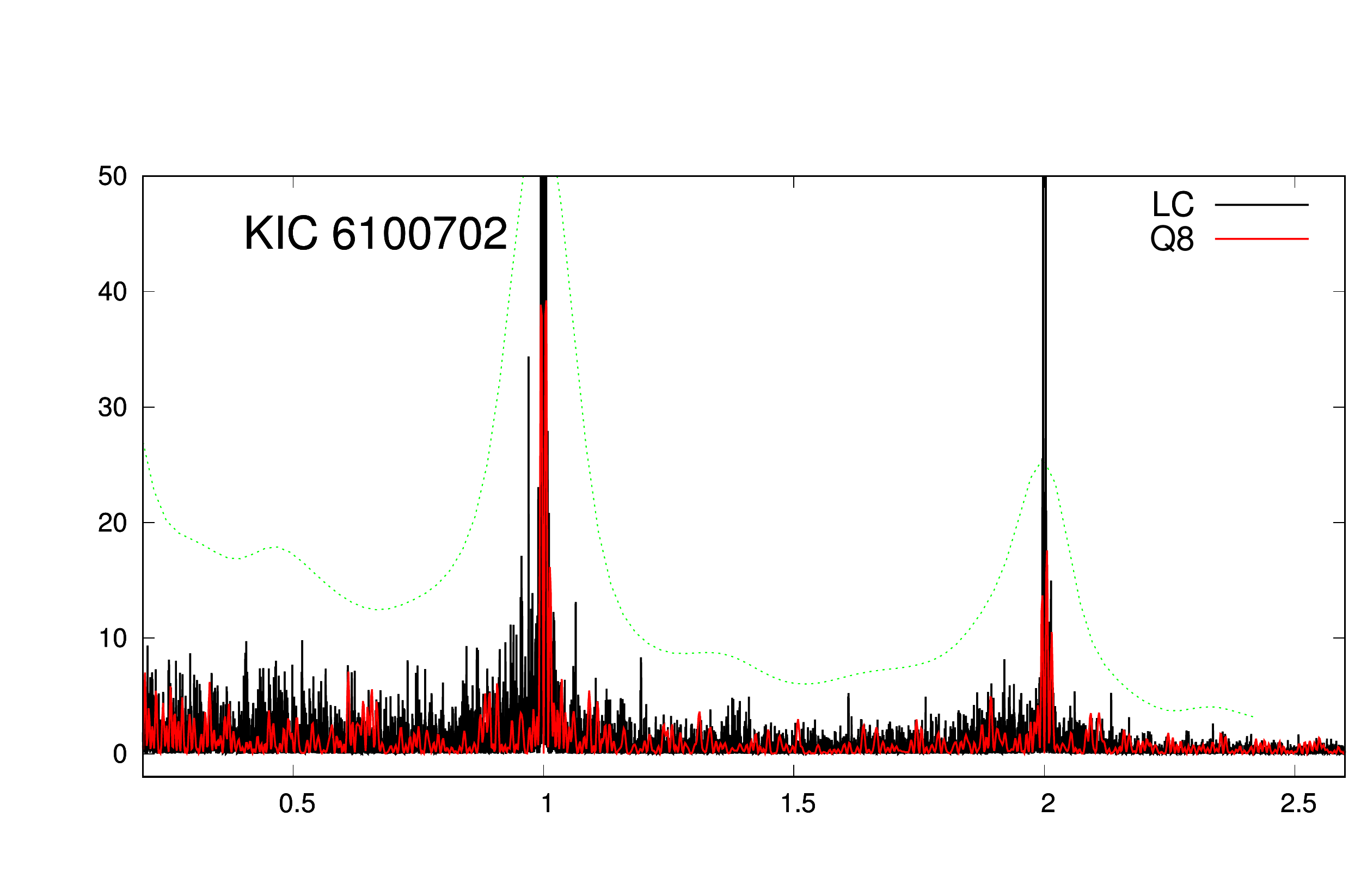}&
\includegraphics[width=3.5cm,angle=270,trim=30 20 20 10]{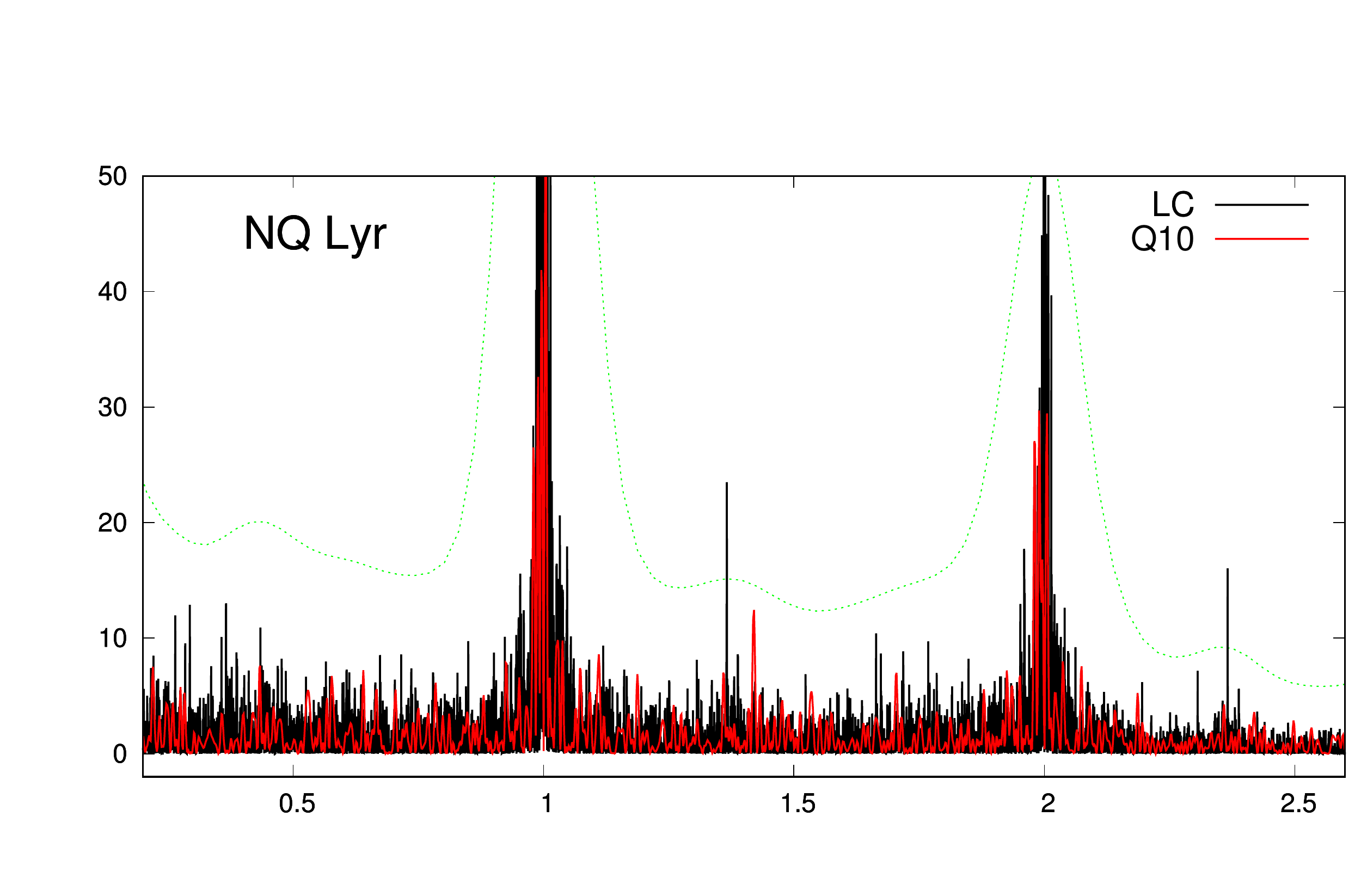}\\
\includegraphics[width=3.5cm,angle=270,trim=30 0  20 10]{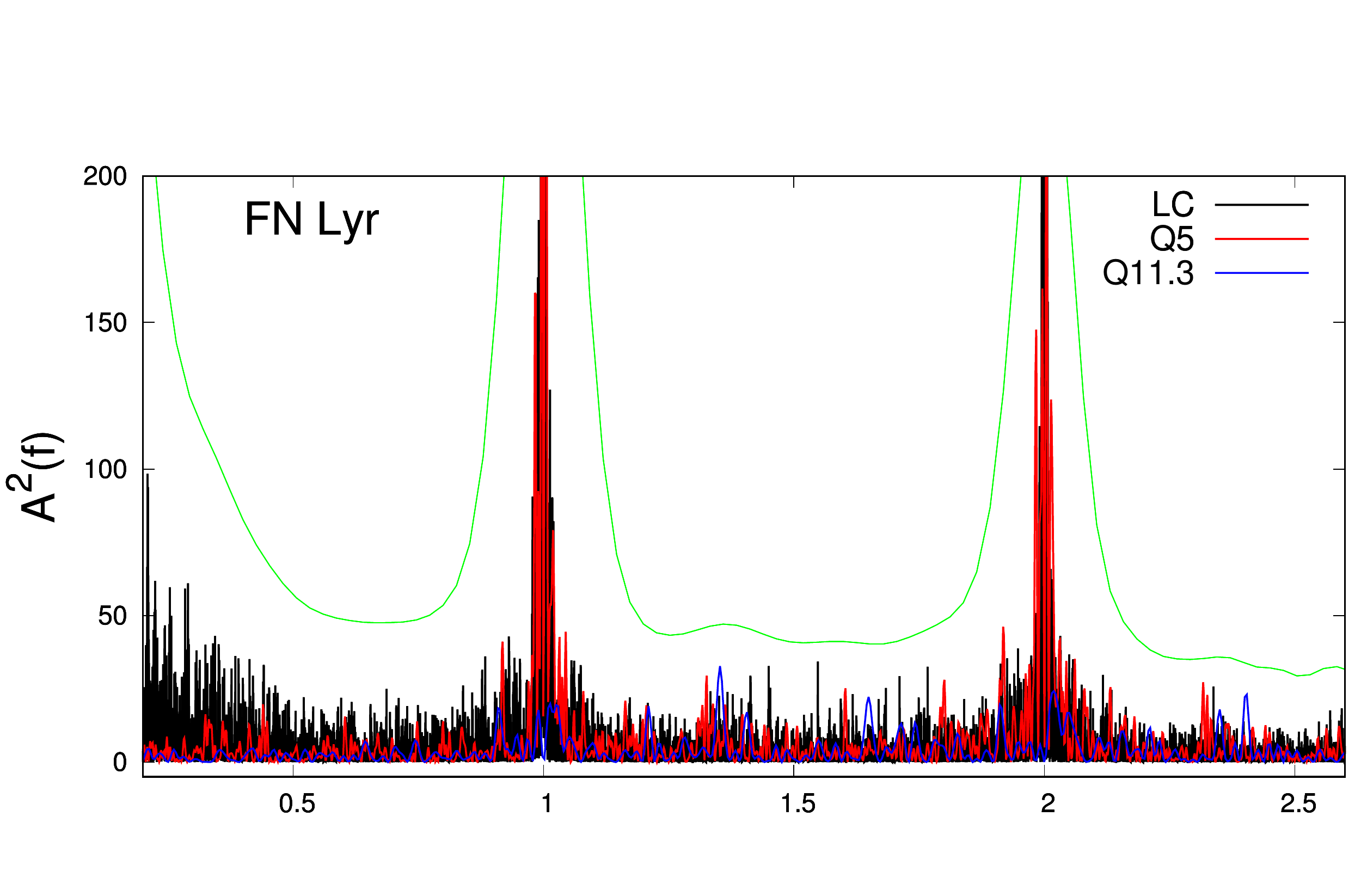}&
\includegraphics[width=3.5cm,angle=270,trim=30 20 20 10]{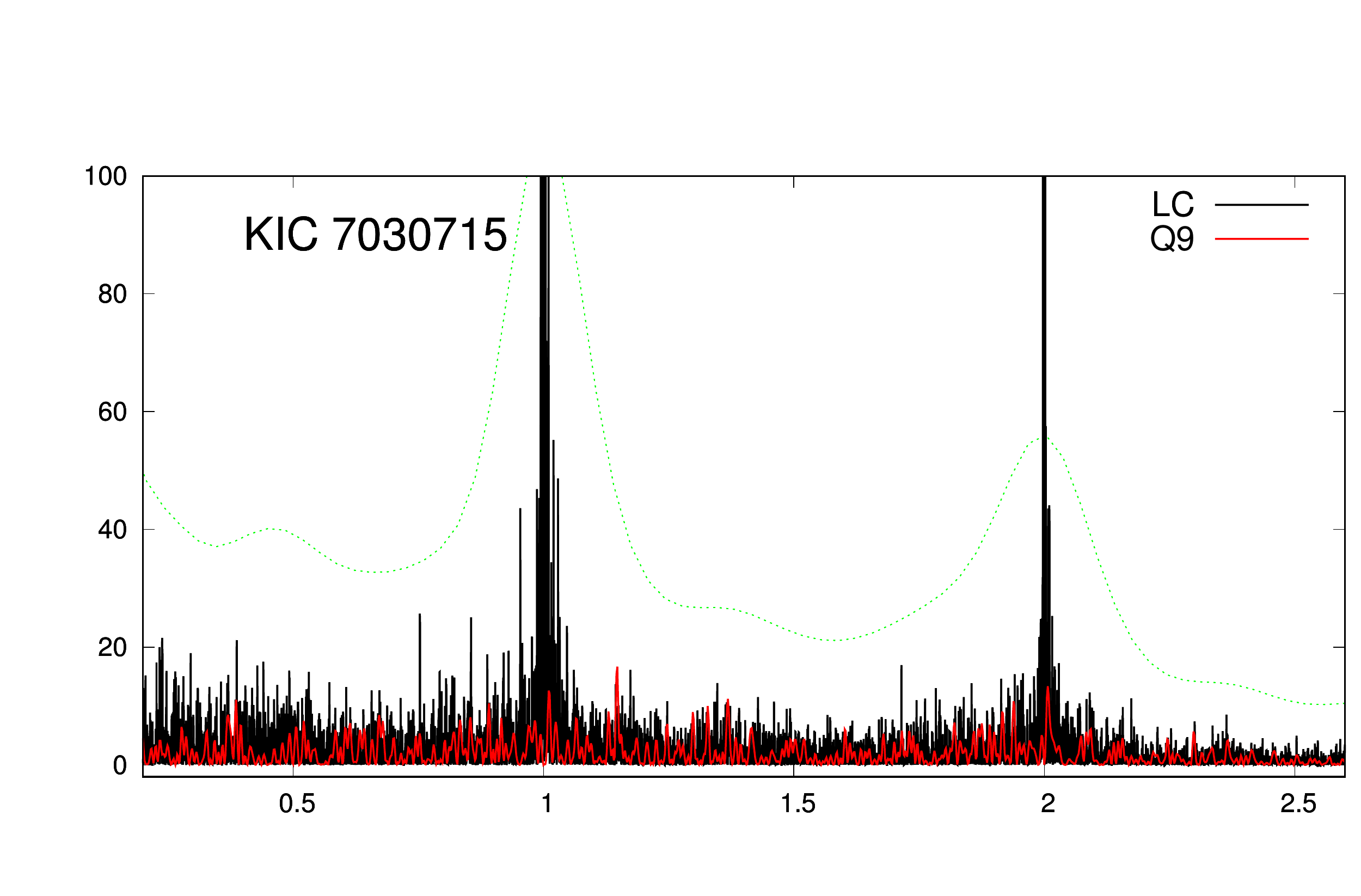}&
\includegraphics[width=3.5cm,angle=270,trim=30 20 20 10]{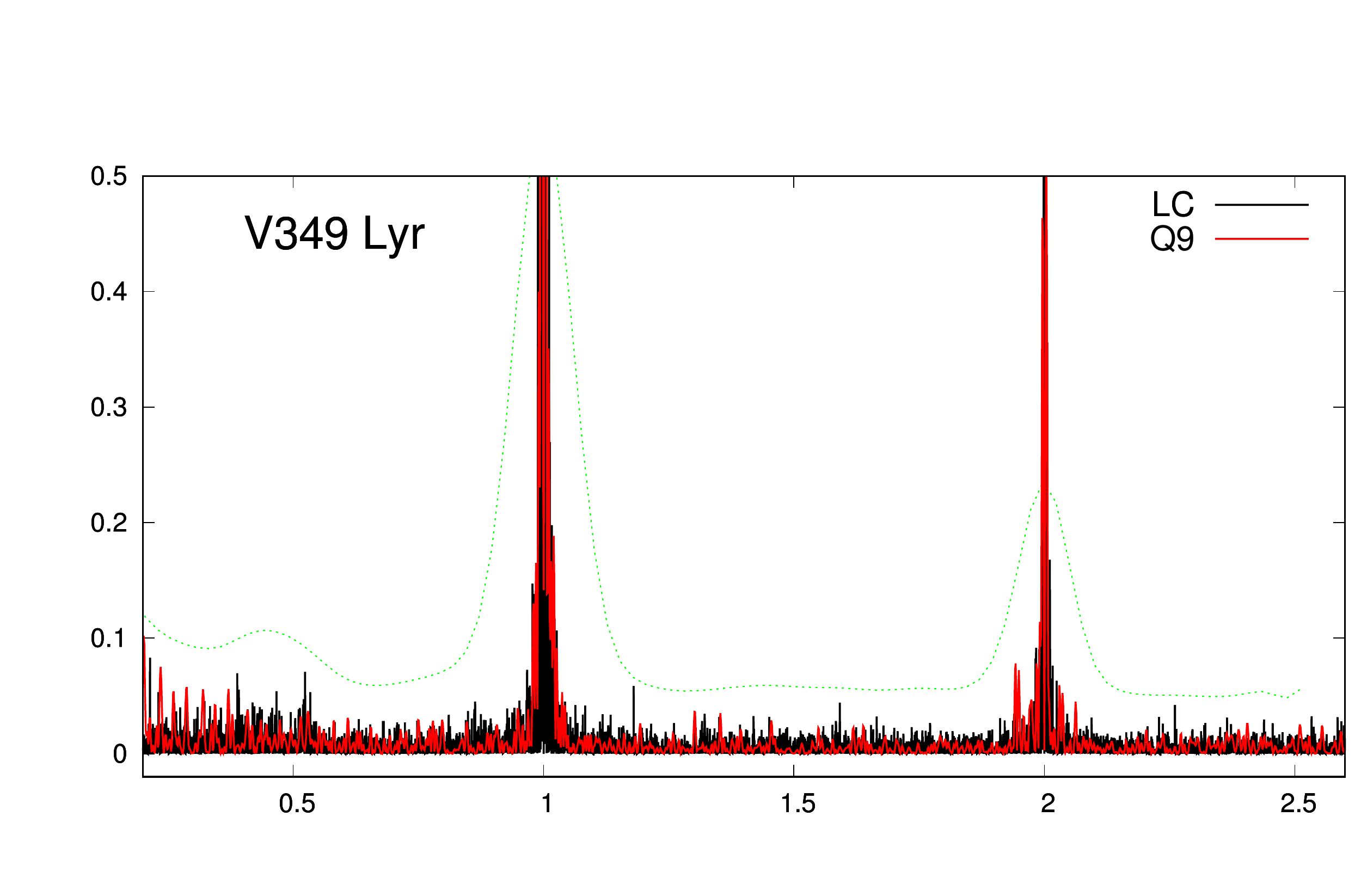}\\
\includegraphics[width=3.5cm,angle=270,trim=30 0  20 10]{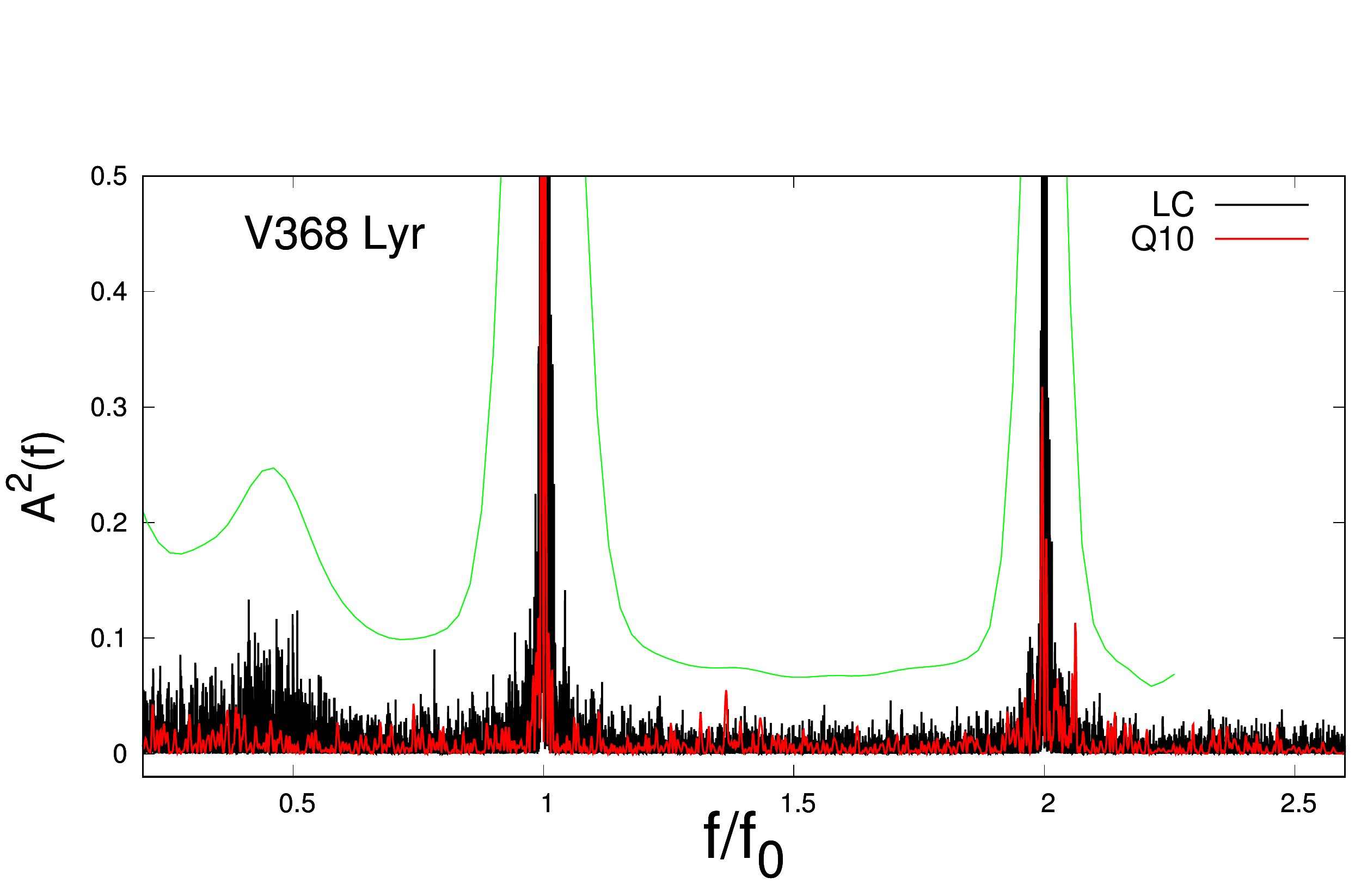}&
\includegraphics[width=3.5cm,angle=270,trim=30 20 20 10]{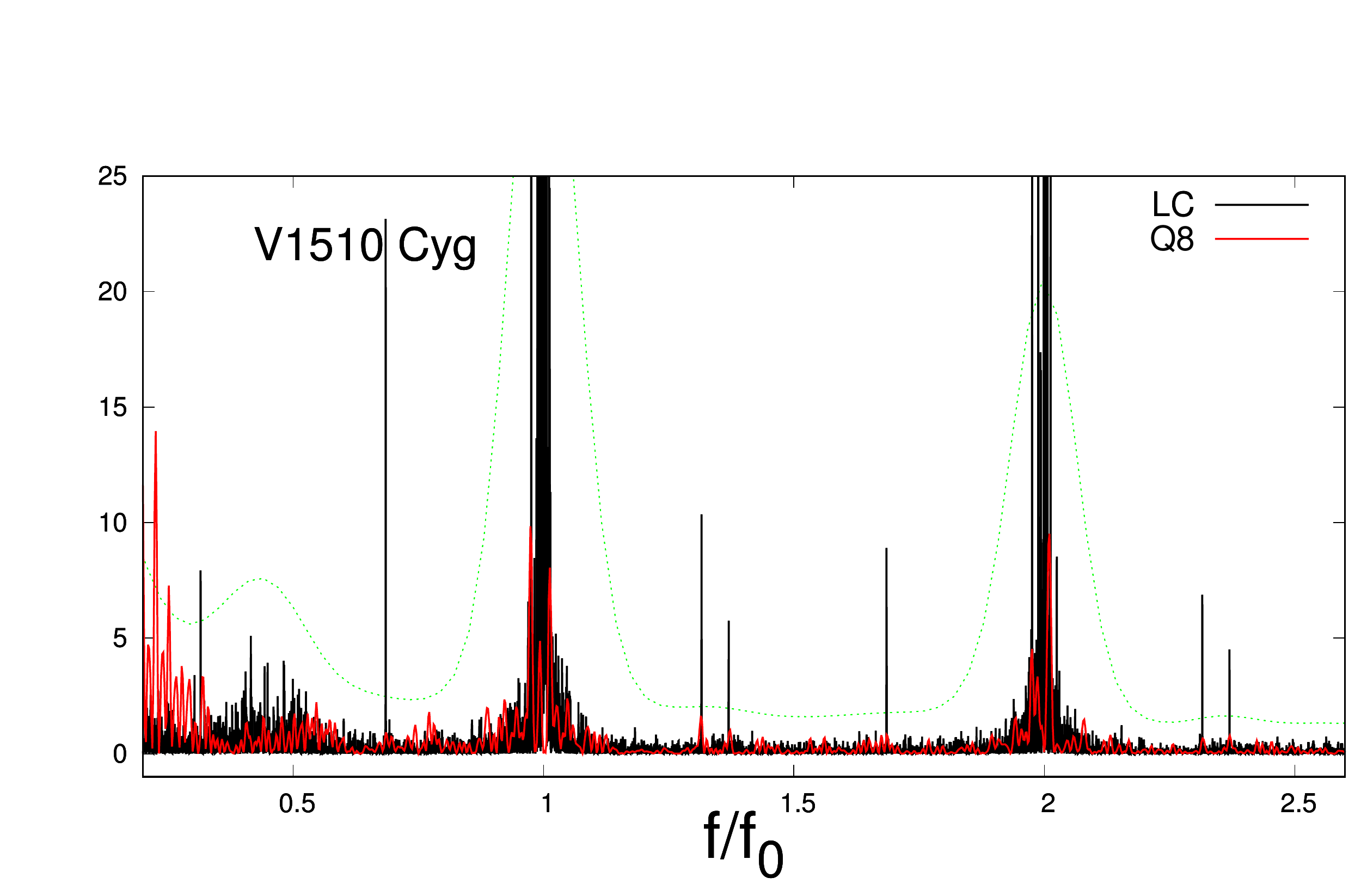}&
\includegraphics[width=3.5cm,angle=270,trim=30 20 20 10]{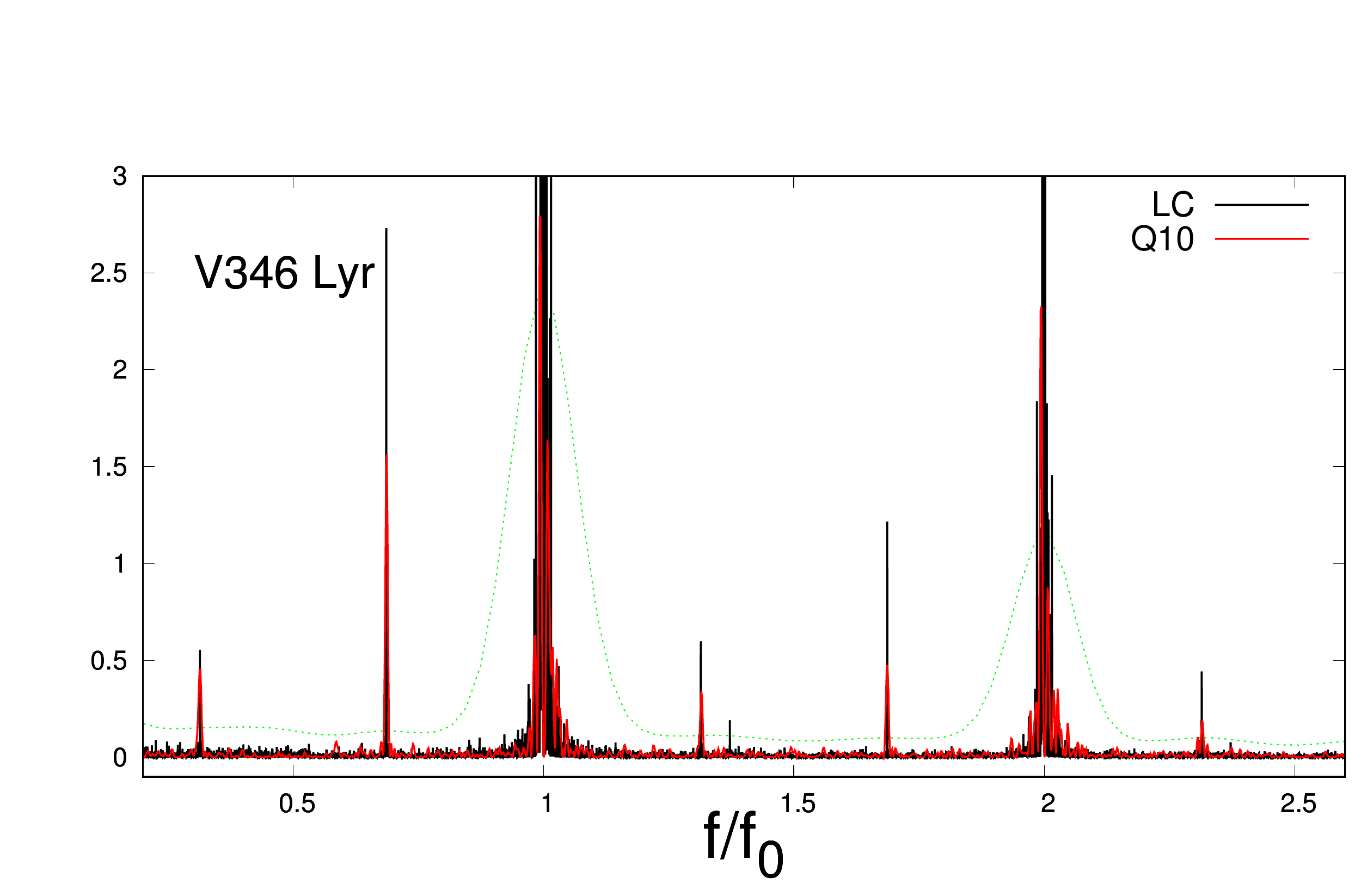}\\
\end{tabular}
    \caption{Power spectra of the residuals around the main pulsation frequency 
$f_0$ and its first harmonic 2$f_0$. Black curves show the LC power spectra. 
Spectra computed from the SC data are signed red and blue curves 
(for the actual quarter numbers see the labels in the panels). 
Providing comparable spectra we scaled the horizontal axes  
with the value of $f/f_0$. The vertical scales of SC power spectra
are also normalized to the signal to noise ratio of the LC spectra.
Dotted green curves show the estimated $S/N=4$ values of the LC spectra.}
    \label{fig:sc_spectra1} 
\end{figure*}
\begin{figure*}
   \centering
\setlength{\tabcolsep}{0.0cm}
\begin{tabular}{ccc}
\includegraphics[width=3.5cm,angle=270,trim=30 0  20 10]{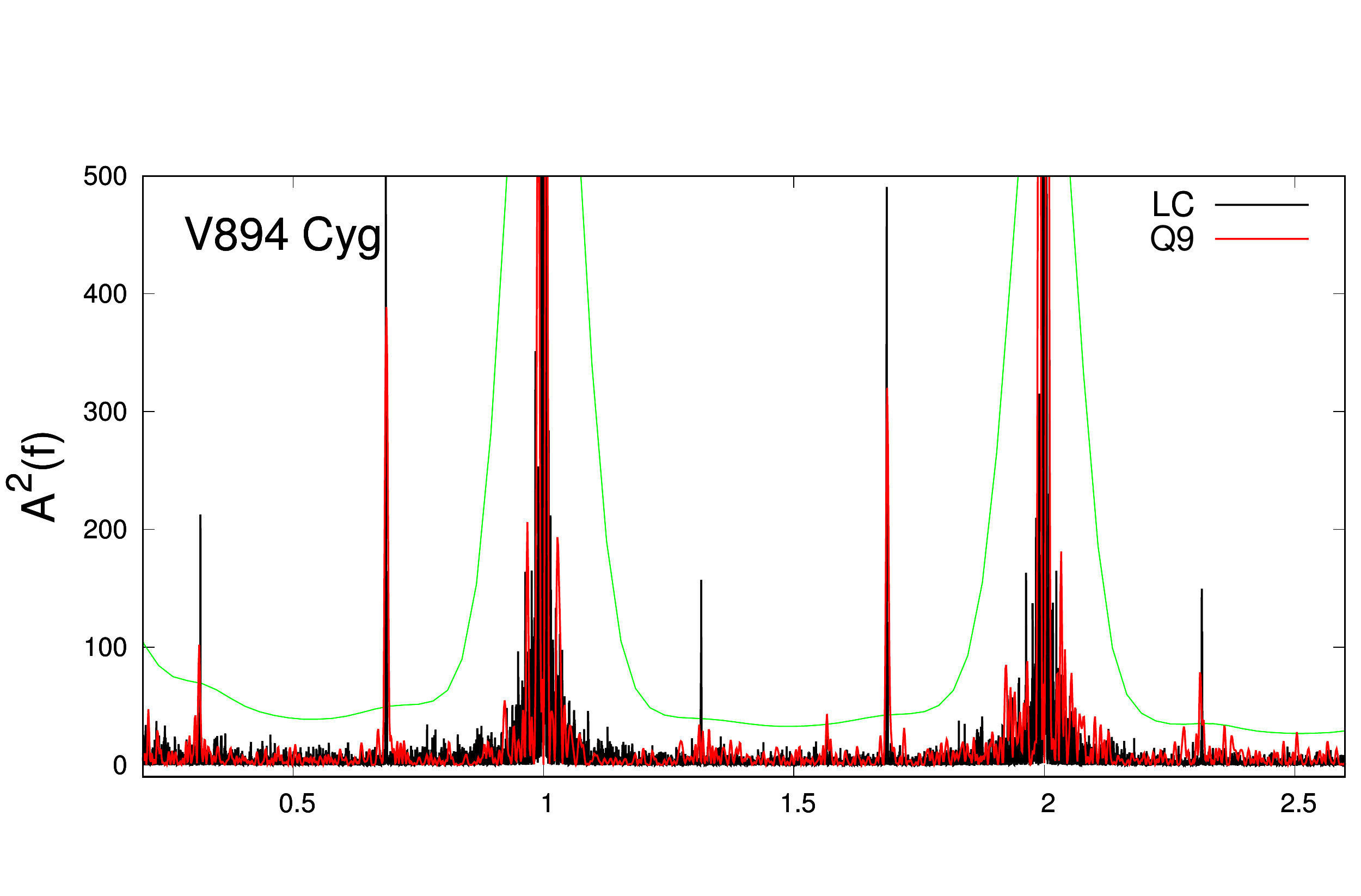}&
\includegraphics[width=3.5cm,angle=270,trim=30 20 20 10]{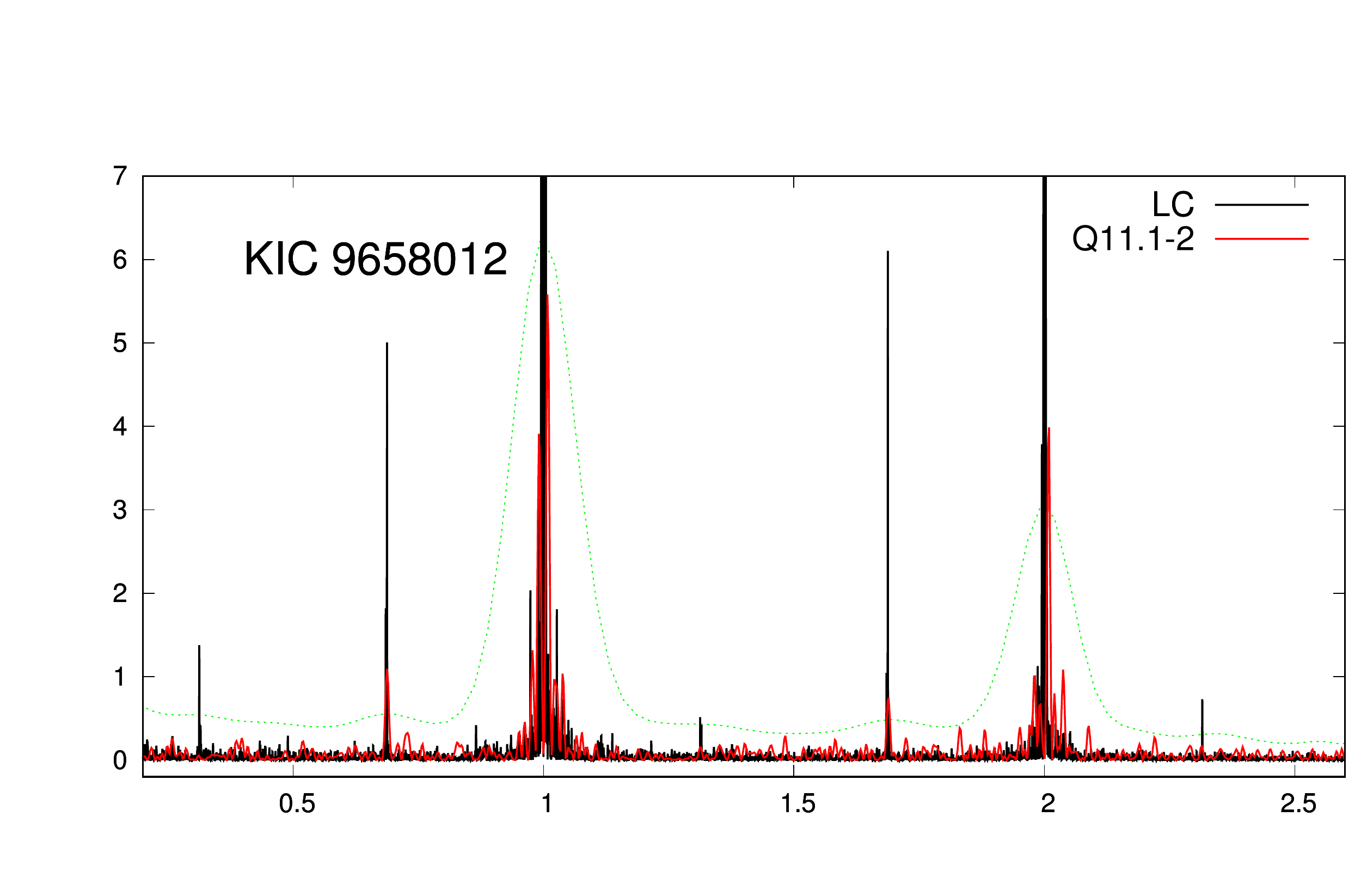}&
\includegraphics[width=3.5cm,angle=270,trim=30 20 20 10]{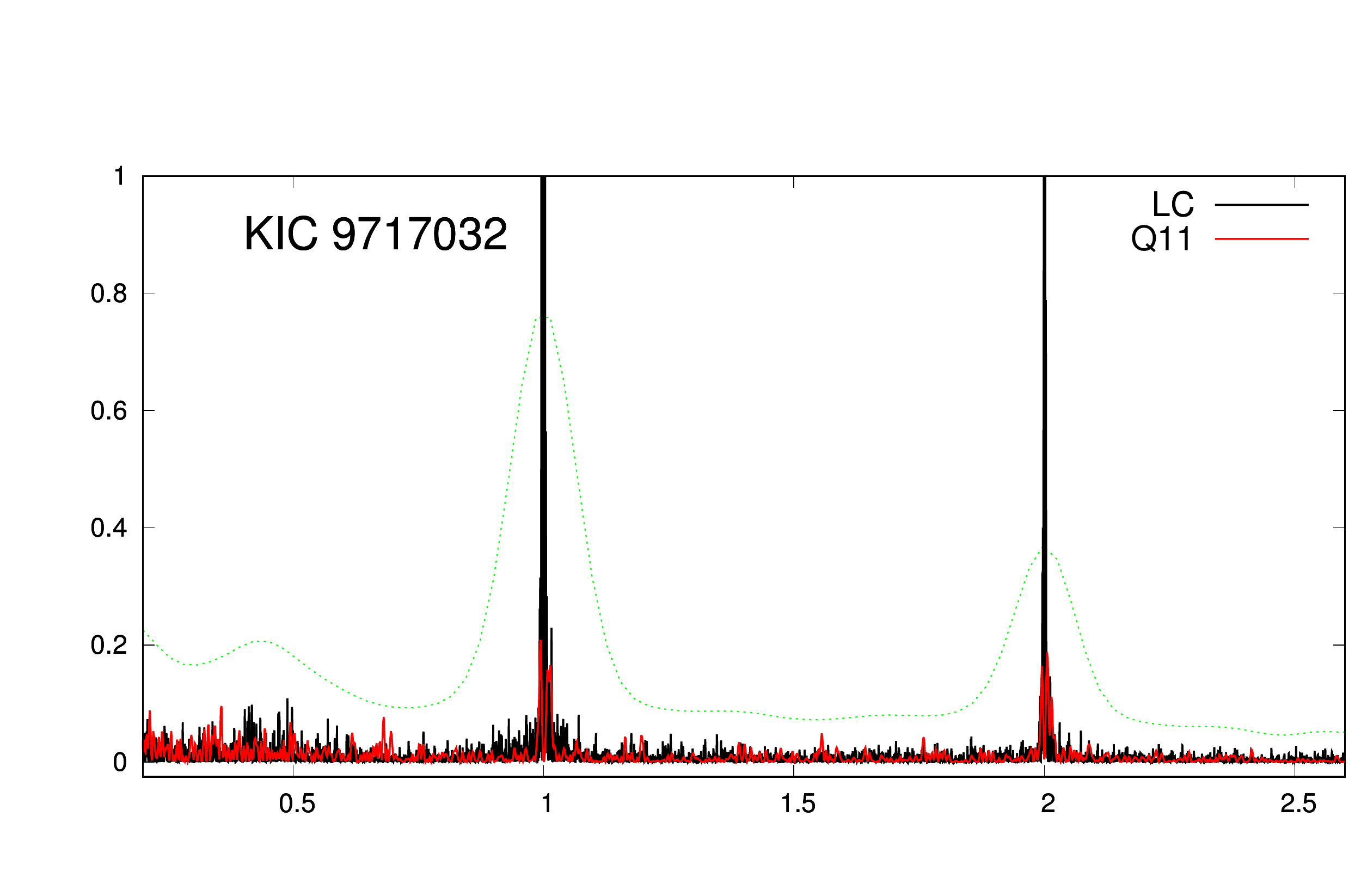}\\
\includegraphics[width=3.5cm,angle=270,trim=30 0  20 10]{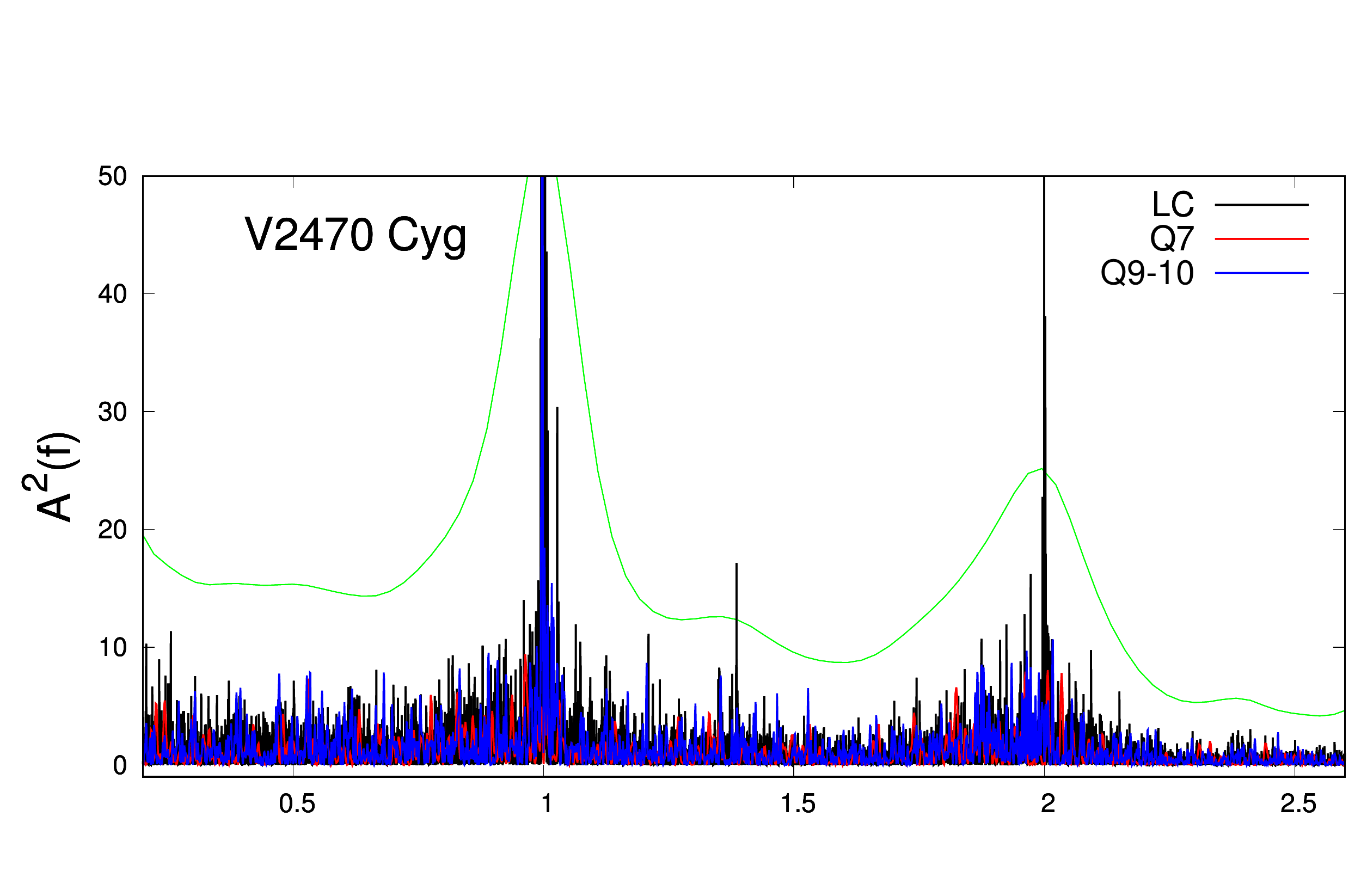}&
\includegraphics[width=3.5cm,angle=270,trim=30 20 20 10]{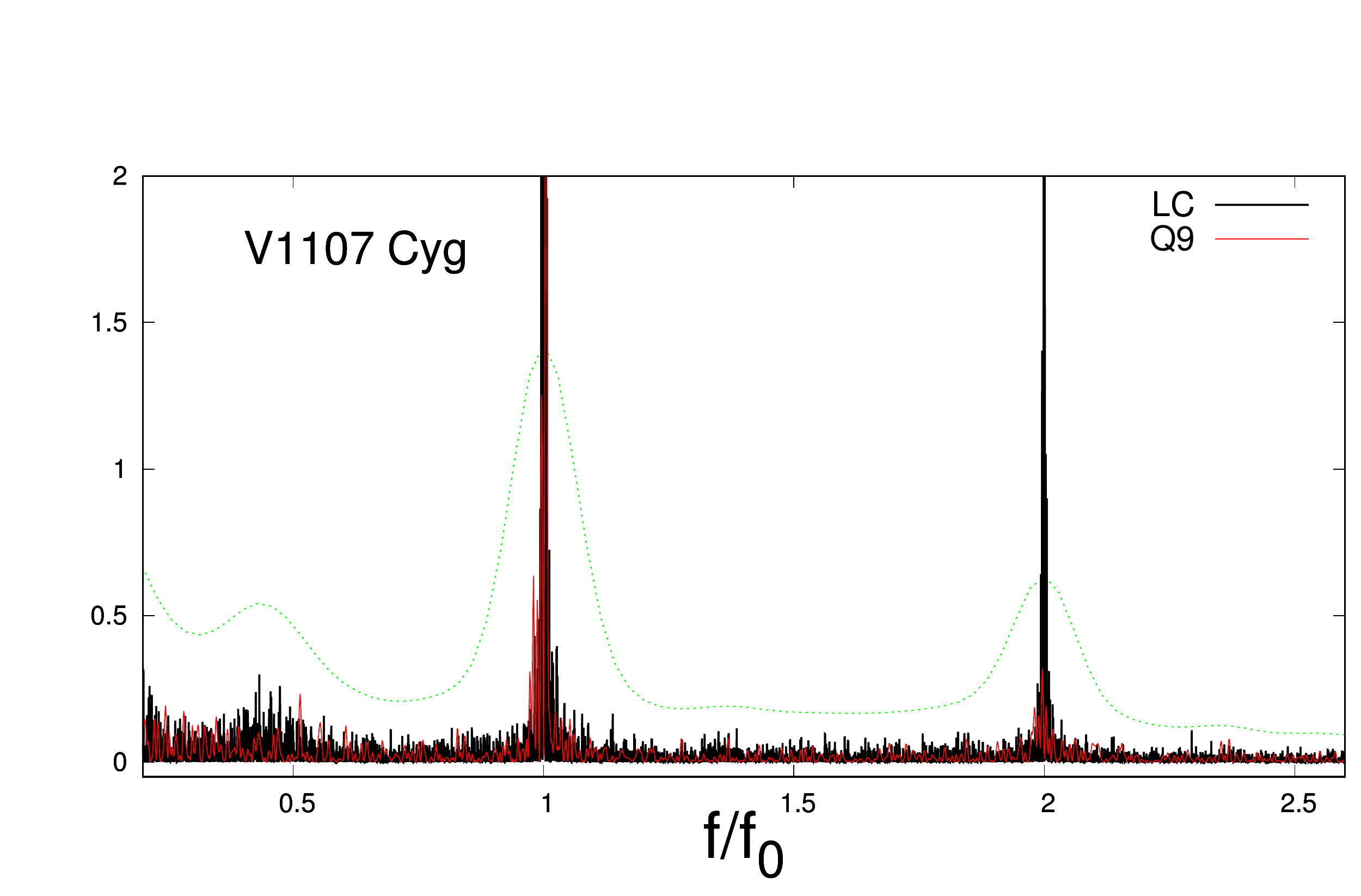}&
\includegraphics[width=3.5cm,angle=270,trim=30 20 20 10]{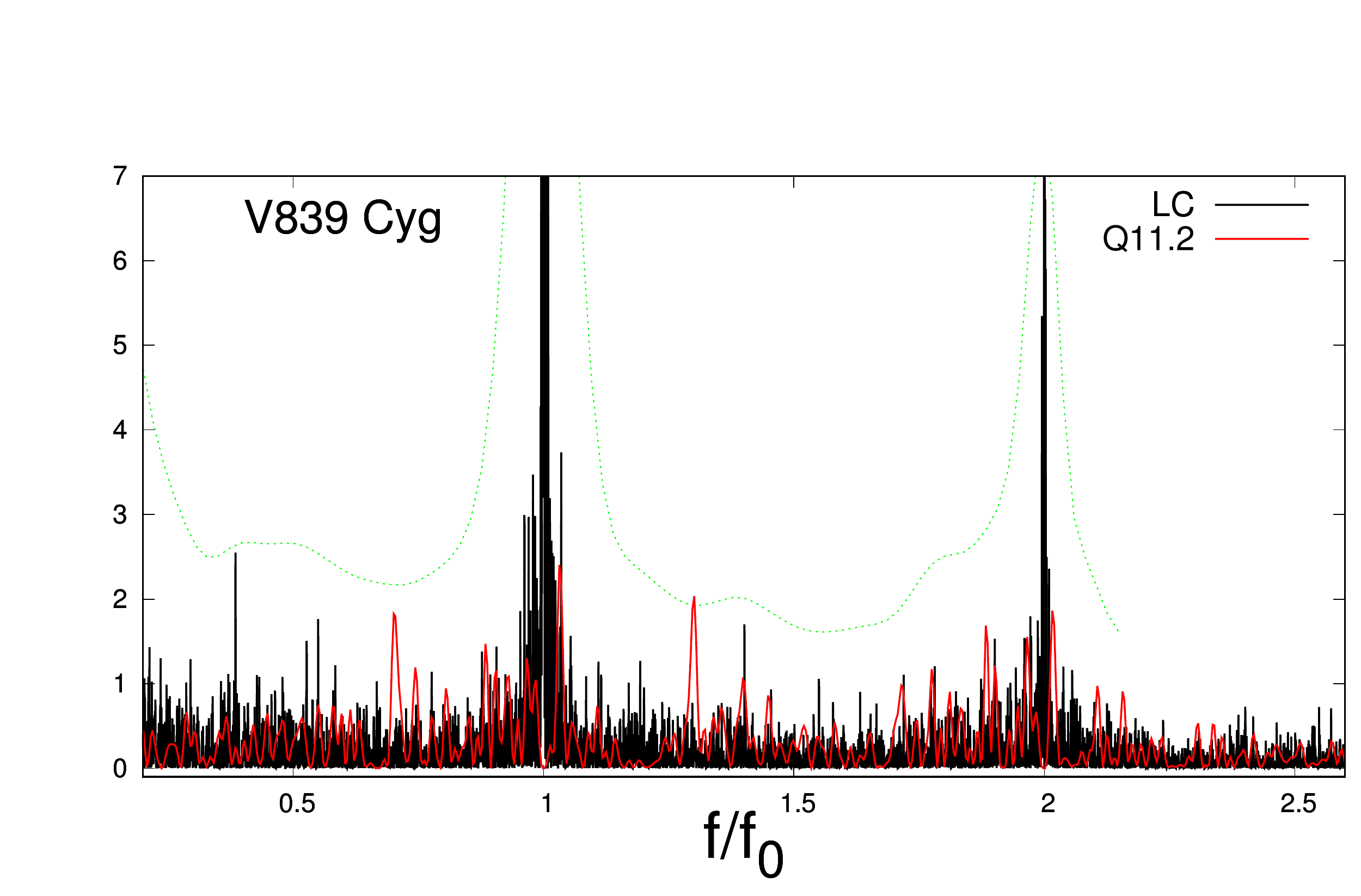}\\
\includegraphics[width=3.5cm,angle=270,trim=30 0  20 10]{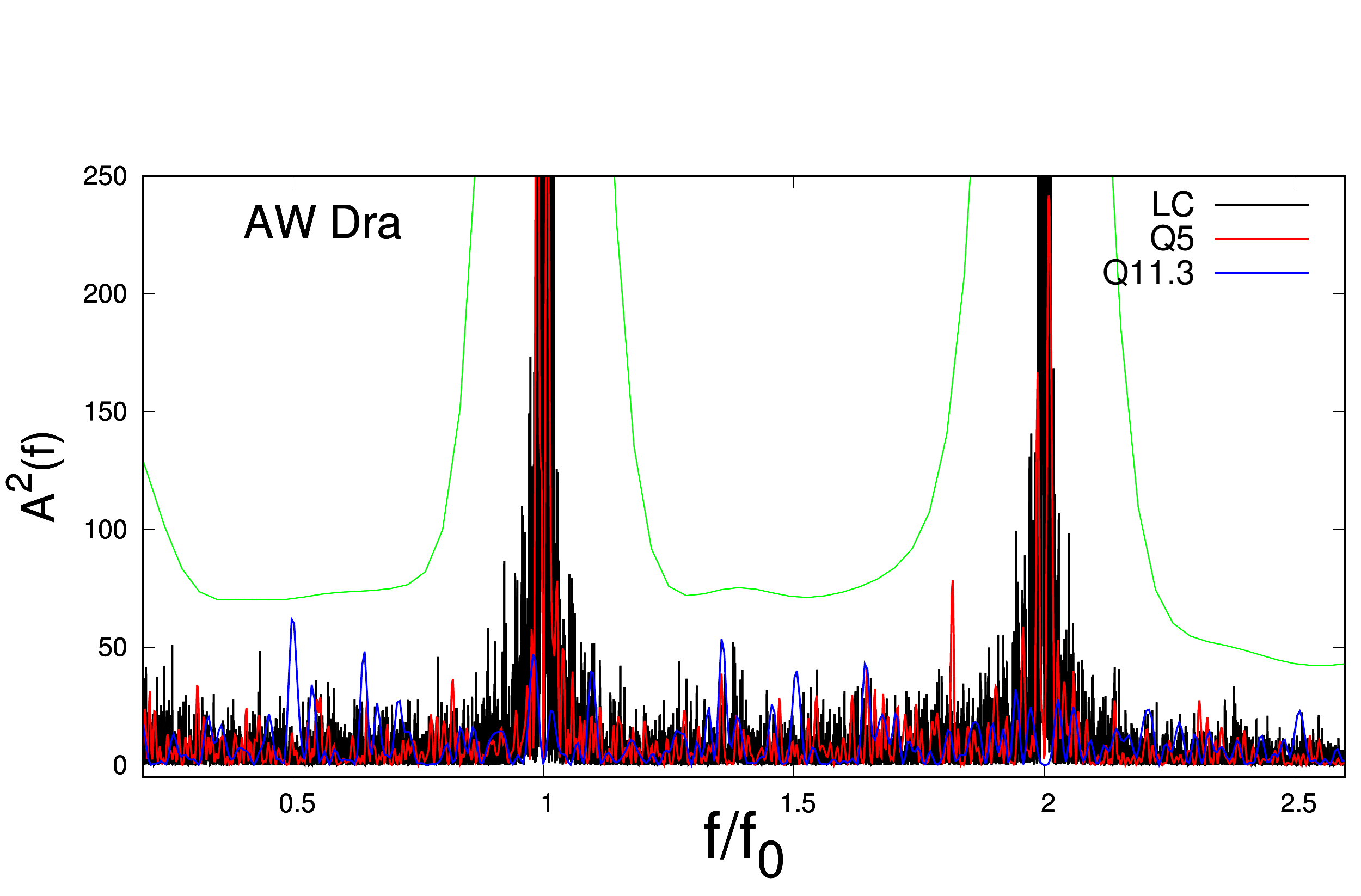}\\
\end{tabular}
    \caption{Continuation of Fig.~\ref{fig:sc_spectra1}}
    \label{fig:sc_spectra2}
\end{figure*}
We prepared the Fourier spectra of both the LC and the SC light curves using the  
discrete Fourier transform tool of the program package {\sc MuFrAn} \citep{Kollath90}.
The spectra are dominated by the main pulsation frequencies $f_0$ and
their harmonics $nf_0$ ($n$ is a positive integer).
After pre-whitening the data for a significant number (35-55) of harmonics 
we obtainded the residual spectra. 
In Fig.~\ref{fig:sc_spectra1}-\ref{fig:sc_spectra2} parts of 
the residual power spectra are shown around the main pulsation frequencies 
($f_0$) and their first harmonics ($2f_0$). The black lines indicate the LC spectra, 
while the spectra of the SC light curves are shown with thin red lines. 
For those stars where two distinct SC observations are available 
(e.g. Q7 and Q9 for V715\,Cyg, etc., see Table~\ref{tab:sample}) the second
SC spectra are plotted by dotted blue lines (see the labels in the panels).  
 
The power spectra are vertically normalized with (in practice divided to) 
the signal-to-noise ratio S/N \citep{Breger93} of the LC spectra. 
The S/N=4 ratio functions of the LC data are plotted in green dotted lines
in  Fig.~\ref{fig:sc_spectra1}-\ref{fig:sc_spectra2}.
Strictly speaking, the shape of the SC and the LC S/N ratio vs. frequency functions are
different, so we cannot transform them to each other by a simple normalization  
but such a normalization can give an approximate agreement in a shorter
frequency interval. That is, the S/N ratio of the LC spectra is approximately 
valid for all spectra within the $f_0$ and $2f_0$ intervals plotted in the panels.
The noise level around the harmonics are overestimated because of
the instrumental origin side peaks appearing in the LC spectra (see fig.~4. in \citealt{Benko19}).
Instead of the frequency, the horizontal axes show the $f/f_0$ values
because this way the spectra can be compared directly.   

\subsection{Signs of the C2C variations}

How does the Fourier spectrum of a randomly C2C varying light curve residual 
look like? In order to check this, we prepared synthetic light curves, for which we used the formulae
of simultaneous amplitude and frequency modulation summarised in \citet{Benko11, Benko18}.
The carrier wave coefficients (frequency, harmonic
amplitudes and phases) defined a simplified RR\,Lyrae-like light curve
with nine harmonics, and  C2C
randomly changing amplitude modulation functions for both amplitude and frequency modulation parts
were assumed. The random values were set for each amplitude and phase separately.
The synthetic light curves were 
sampled in the same points as the observed Q7 SC data.   
The spectra of the synthetic light curves after we removed the nine harmonics from the data show
significant peaks at the first few (4-5) harmonics. The surroundings of the peaks have
a red noise profile as we expected from a random process.

Comparing these synthetic spectra with the spectra of the observed SC data,
we found them fairly similar.
The observed SC residual spectra are also dominated by  frequencies at around $f_0$ and its harmonics. 
This is true for the entire sample not just for the bright stars which show evident C2C variations 
and high $D$ values but for the faintest stars as well. 
The high-$D$ stars show many ($>10$) significant
harmonic peaks while low-$D$ stars have typically few (3-4) significant harmonics.
This can be explained with that the fine structure of the spikes at the higher 
harmonics are veiled by the higher noise of low-$D$ stars. 
For most cases we detect more than one single peak around the harmonic 
positions $kf_{0}, k=1,2,\dots$ which is again a similarity to the synthetic data spectra.
These side peaks due to the C2C variation might explain the distinct group with extremely 
small Blazhko amplitude found by \citet{Kovacs18}.
Double or multiple peaks, however, could not be just because of the C2C variations. 
This can also be the consequences of long time-scale 
(longer than the observed time span) light curve variations caused by instrumental problems 
or very long period Blazhko effect. 
Since we found some evidence for such effects (see later in Sec.~\ref{sec:o-c}), 
we cannot declare undoubtedly the detection of the C2C variations on all stars.
This also means that the Fourier spectra alone are not discriminative enough to find  
C2C variations.

\subsection{Additional modes}\label{sec:add}

Six stars' LC spectra show significant ($S/N>4$) additional peaks around the main pulsation
frequency and its first harmonics. These are: NQ\,Lyr, V1510\,Cyg, 
V346\,Lyr, V894\,Cyg, KIC\,9658012 and V2470\,Cyg.  
SC spectra of three of these stars (V346\,Lyr, V894\,Cyg and KIC\,9658012) 
contain significant additional frequencies. 
Several other stars 
show visible but strictly not significant (2$<$S/N$<$4) peaks in their LC or SC spectra.
The frequency of the highest additional peaks with their S/N ratio are given in 
Table~\ref{tab:per}.
\begin{figure*}
\includegraphics[width=6cm,angle=270,trim=50 30 50 30]{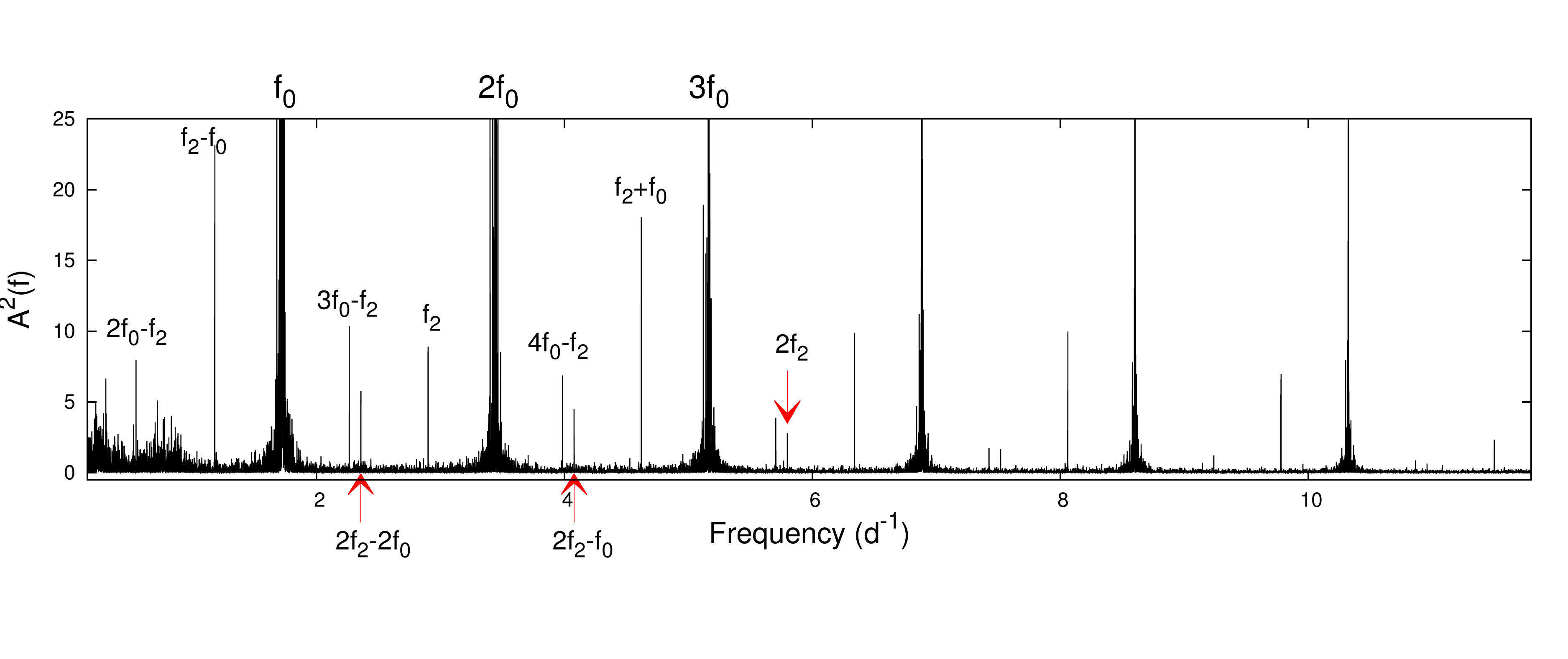}
    \caption{A possible identification of the additional mode frequencies in 
the pre-whitened spectrum of the LC light curve of V1510\,Cyg.
The positions of the main pulsation frequency and its first two harmonics are
also marked.}
    \label{fig:spec_id} 
\end{figure*}

In the past years low amplitude additional frequencies 
were found for many RR\,Lyrae stars (for a resent review see \citealt{Molnar17}). 
If we focus only on the fundamental mode pulsators (RRab stars) then the half-integer
frequencies ($f_0/2, 3f_0/2,\dots$) of the
period doubling (PD) effect \citep{Kolenberg10, Szabo10}
appearing in many Blazhko RRab stars was the first theoretically
modelled case. Other type of extra frequencies which
were  discovered in numerous Blazho RRab stars are the 
low order radial overtone frequencies ($f_1, f_2$ \citealt{Chadid10, Poretti10, Benko10})
and their linear combinations with the fundamental mode frequency.
Although the simultaneous appearing of the PD and the first overtone
frequencies was reproduced by radial numerical hydrodynamic codes
as triple resonance states \citep{Molnar12}, it is not evident 
that all of such frequencies can be explained on this purely radial basis. 
Especially thought-provoking that the amplitude of the 
`linear combination' frequencies are 
many times higher than their suspected basis frequencies.
Such a behaviour is detected for nonradial modes empirically for rhoAp stars
by \citet{Balona13} and explained by theoretically \citet{Kurtz15}
which suggests that these frequencies could belong to non-radial modes
excited at or near the radial mode positions.

The extra frequencies of V1510\,Cyg, V346\,Lyr, V894\,Cyg and KIC\,9658012 
could be identified as the second radial overtone mode $f_2$ and their 
linear combinations. This identification are
shown in Fig.~\ref{fig:spec_id} for V1510\,Cyg which has the richest
extra frequency pattern. 
As we see, a few linear combination frequencies (e.g. $f_2-f_0$, $f_2+f_0$, or
$3f_0-f_2$) have an amplitude higher than the amplitude of $f_2$.
The situation is the same for V346\,Lyr and V894\,Cyg: the highest amplitude
extra frequency is $f_2-f_0$, while for  KIC\,9658012 it is the $f_2$.
Stars where the highest amplitude additional frequency is lower than 
the fundamental one are rather rare. We found V838\,Cyg \citep{Benko14} the
only  published case. Additional mode frequencies at the position of $f_2-f_0$ are
known for few CoRoT and \textit{Kepler} Blazhko RRab stars (see \citealt{BenkoSzabo14} 
and references therein) but in all those cases the frequency
$2f_2-2f_0$ has higher amplitude.

On the basis of their additional frequency content NQ\,Lyr and V2470\,Cyg form a separate
subgroup in our sample. Their highest additional
peaks are around the position of the first radial overtone frequency $f_1$, but
their frequency ratios ($f_0/f_1=0.732$ for NQ\,Lyr and 0.722 for V2470\,Cyg)
are lying  bellow the values of the canonical Petersen diagram. Such ratios have 
been detected for the first time for two RRd stars in the globular cluster 
M3 by \citet{Clementini04}. Later similar ratio has been found for
the \textit{Kepler} Blazhko RRab stars V445\,Lyr \citep{Guggenberger12} 
and RR\,Lyr itself \citep{Molnar12}. In the OGLE survey data of the Galactic Bulge 
has been found numerous RRd stars showing similarly small frequency ratios 
\citep{Soszynski11,Soszynski14}. Studying the RRd stars in 
the globular cluster M3 \citet{Jurcsik14,Jurcsik15} found that all four Blazhko RRd
stars have anomalous frequency ratio and three of them have smaller
then the normal one as we found for the present stars. Significant amount
of such RRd stars were identified in the Large Magellanic Cloud by the OGLE survey
\citep{Soszynski16} and also in \textit{K2} data \citep{Molnar17}.

\citet{Soszynski16} defined these stars as
`anomalous double-mode RR\,Lyrae stars'. This group is characterized by not just 
its anomalous period (or frequency) ratio but the dominant pulsation 
mode is the fundamental one here while for the `normal' RRds
it is the first overtone. Additionally, most of these anomalous
RRd stars show the Blazhko effect \citep{Smolec15}. 
Since we analyzed RR\,Lyrae stars classified formerly as RRab type
it is evident that NQ\,Lyr and V2470\,Cyg is dominated by  
the fundamental mode. The amplitude ratios are $A(f_1)/A(f_0)=0.00025$ and 0.00032
for NQ\,Lyr and  V2470\,Cyg, respectively.
These ratios are two-three magnitudes smaller than the similar parameters 
of the anomalous RRd stars discovered from the ground \citep{Jurcsik14,Soszynski16}.
The anomalous RRd stars almost always show the Blazhko effect. However,
we did not detect any modulation for our stars (see the details later).
Of course, very small amplitude and very long period (longer than four years) 
modulation can not be ruled out.
\begin{figure*}
   \centering
\setlength{\tabcolsep}{0.0cm}
\begin{tabular}{cc}
\includegraphics[width=3.5cm,angle=270,trim=30 10 50 0]{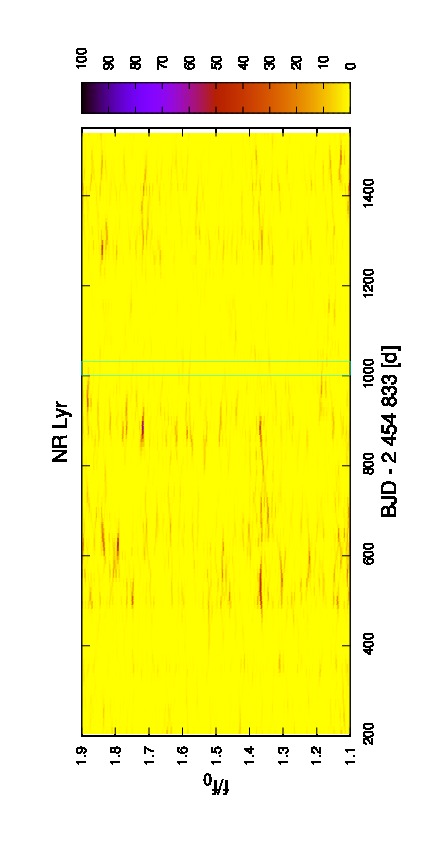}&
\includegraphics[width=3.5cm,angle=270,trim=30 10 50 0]{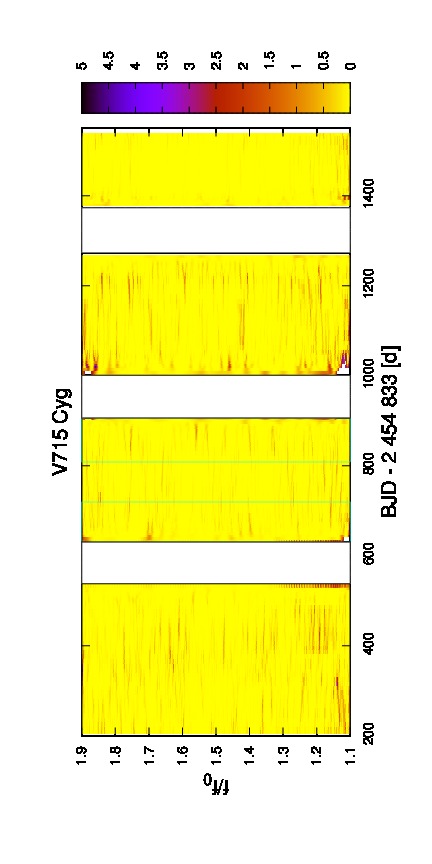}\\
\includegraphics[width=3.5cm,angle=270,trim=30 10 50 0]{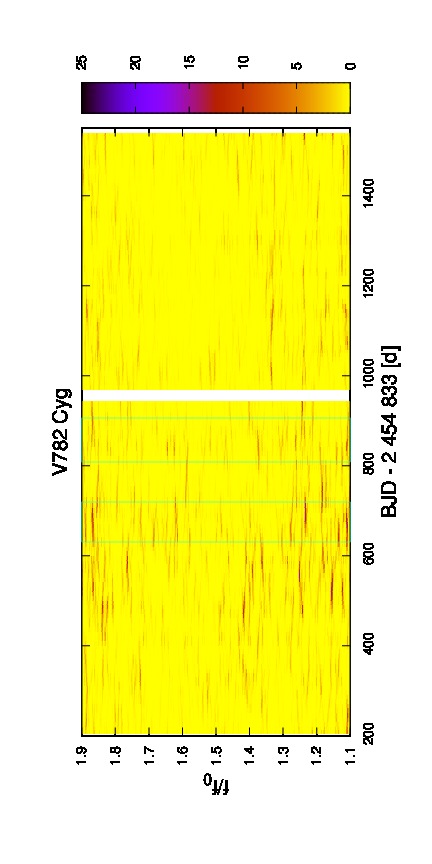}&
\includegraphics[width=3.5cm,angle=270,trim=30 10 50 0]{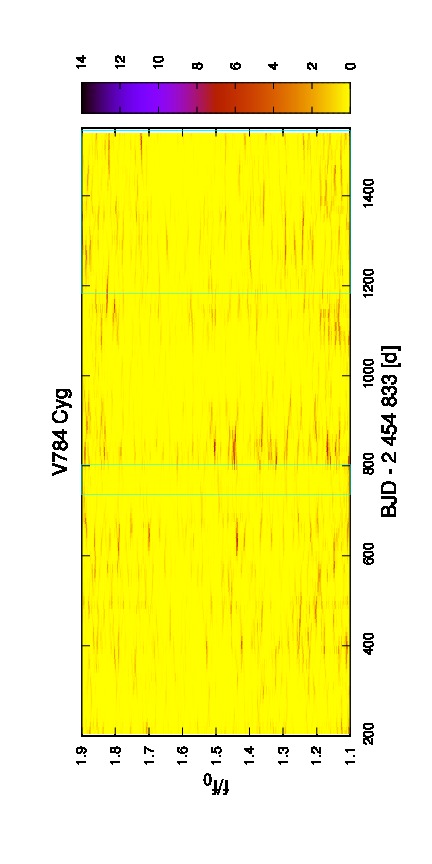}\\
\includegraphics[width=3.5cm,angle=270,trim=30 10 50 0]{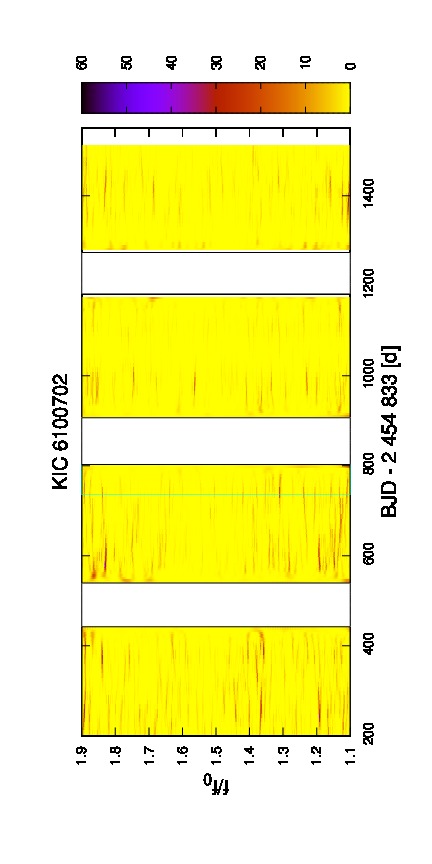}&
\includegraphics[width=3.5cm,angle=270,trim=30 10 50 0]{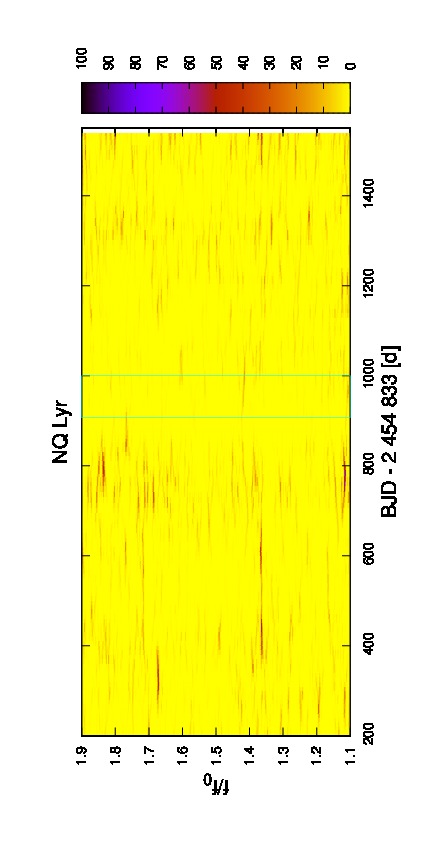}\\
\includegraphics[width=3.5cm,angle=270,trim=30 10 50 0]{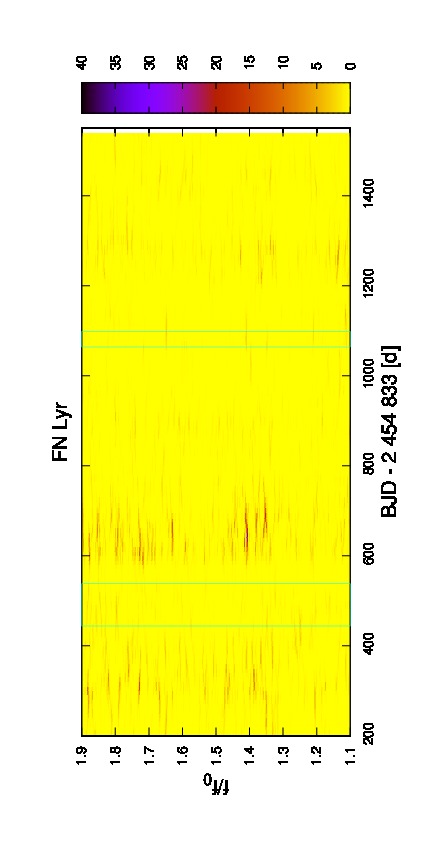}&
\includegraphics[width=3.5cm,angle=270,trim=30 10 50 0]{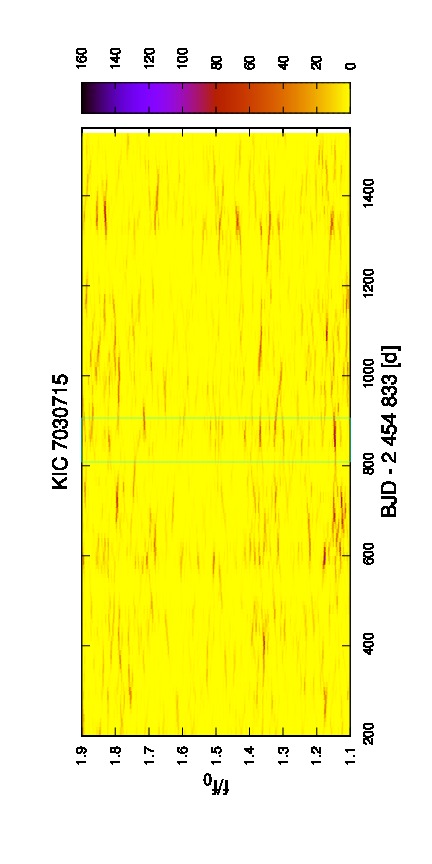}\\
\includegraphics[width=3.5cm,angle=270,trim=30 10 50 0]{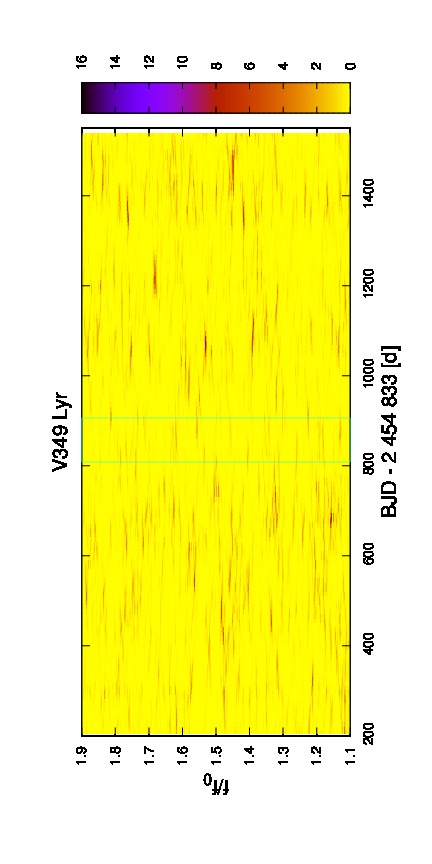}&
\includegraphics[width=3.5cm,angle=270,trim=30 10 50 0]{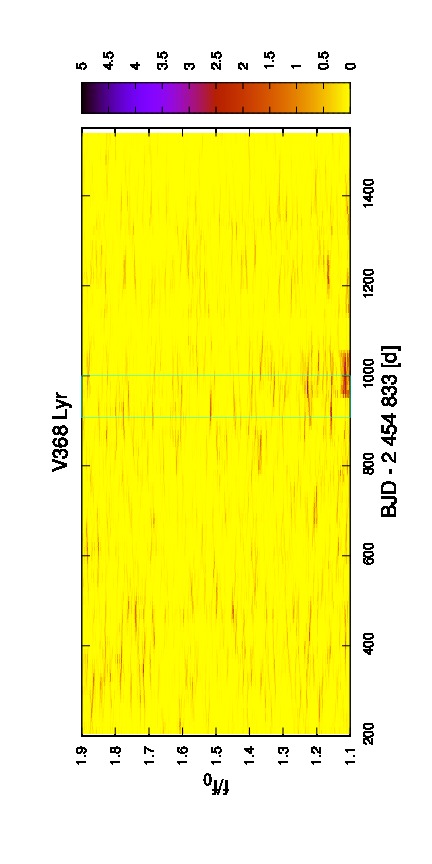}\\
\includegraphics[width=3.5cm,angle=270,trim=30 10 50 0]{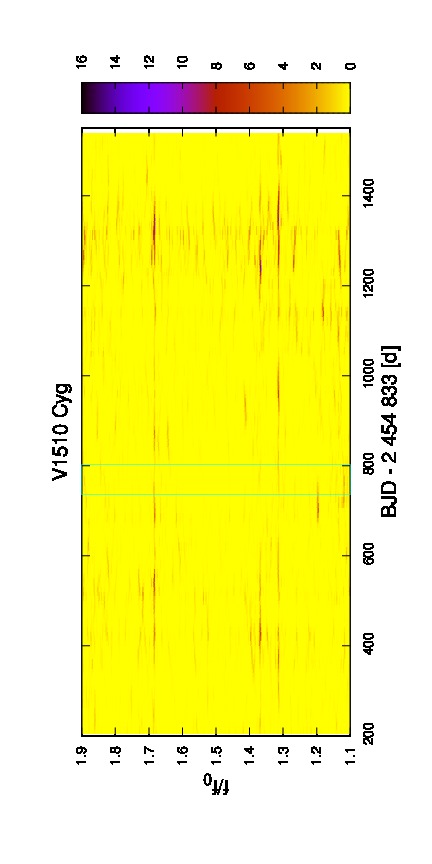}&
\includegraphics[width=3.5cm,angle=270,trim=30 10 50 0]{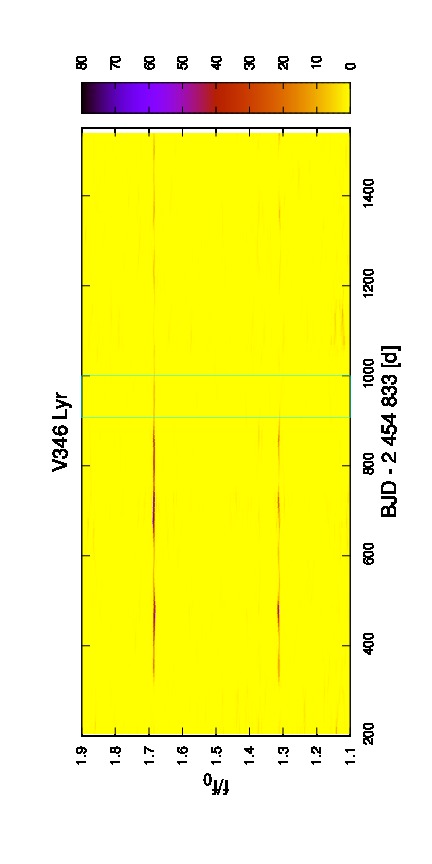}\\
\end{tabular}
    \caption{
Time-frequency variation of the frequencies around the main pulsation frequency 
$f_0$ and its first harmonic 2$f_0$. 
Showing comparable spectra we indicated the normalized frequency $f/f_0$ in the horizontal axes.}
    \label{fig:tf1} 
\end{figure*}
\begin{figure*}
   \centering
\setlength{\tabcolsep}{0.0cm}
\begin{tabular}{cc}
\includegraphics[width=3.5cm,angle=270,trim=30 10 50 0]{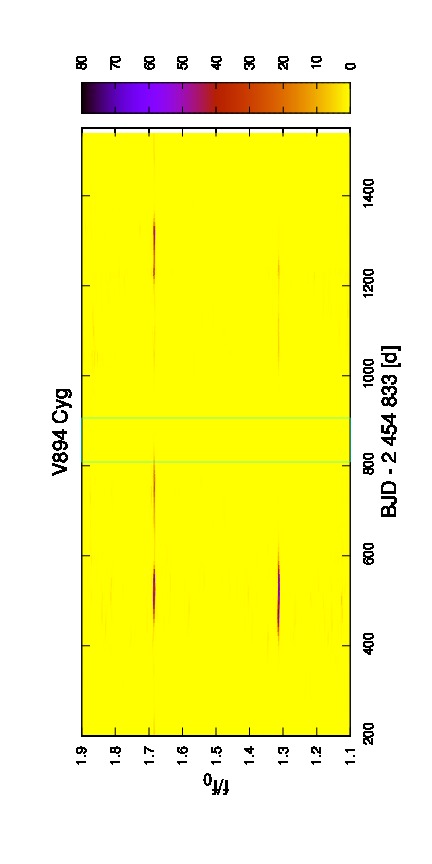}&
\includegraphics[width=3.5cm,angle=270,trim=30 10 50 0]{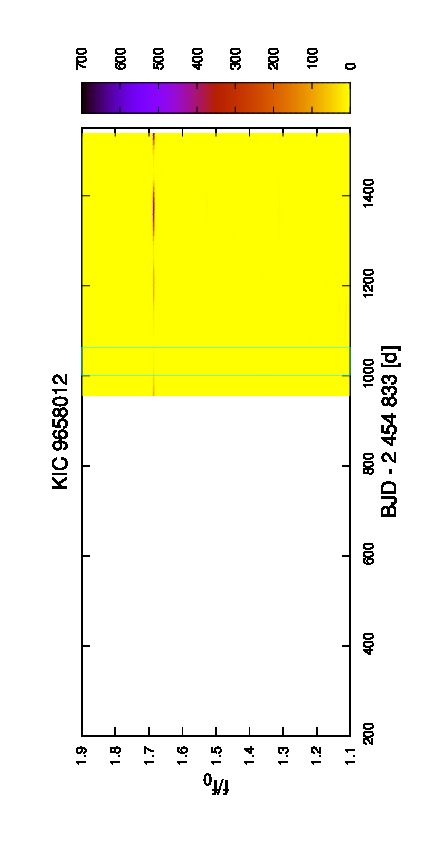}\\
\includegraphics[width=3.5cm,angle=270,trim=30 10 50 0]{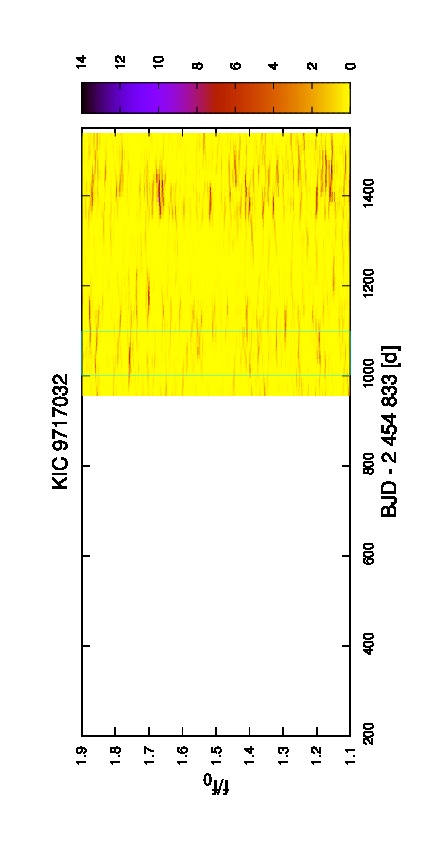}&
\includegraphics[width=3.5cm,angle=270,trim=30 10 50 0]{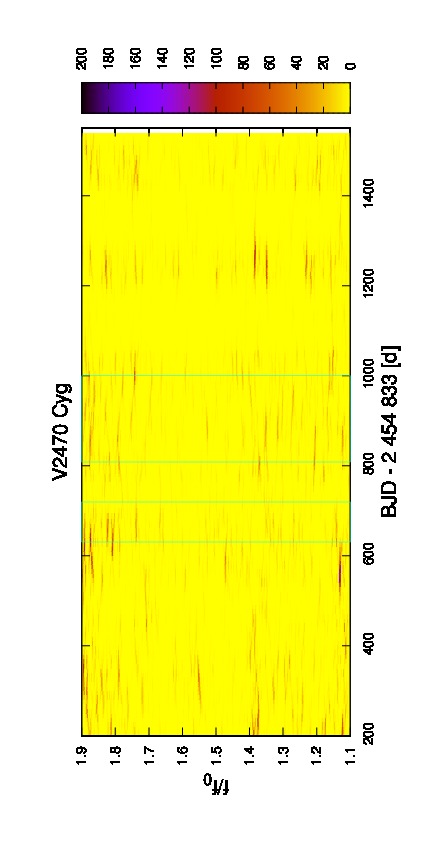}\\
\includegraphics[width=3.5cm,angle=270,trim=30 10 50 0]{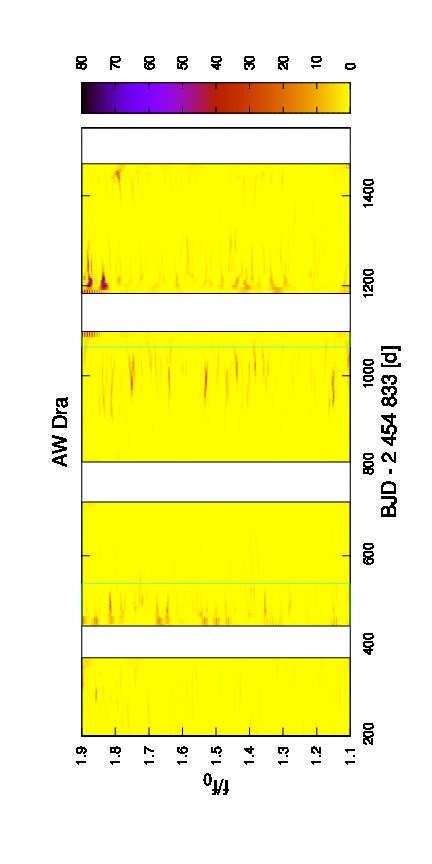}&
\includegraphics[width=3.5cm,angle=270,trim=30 10 50 0]{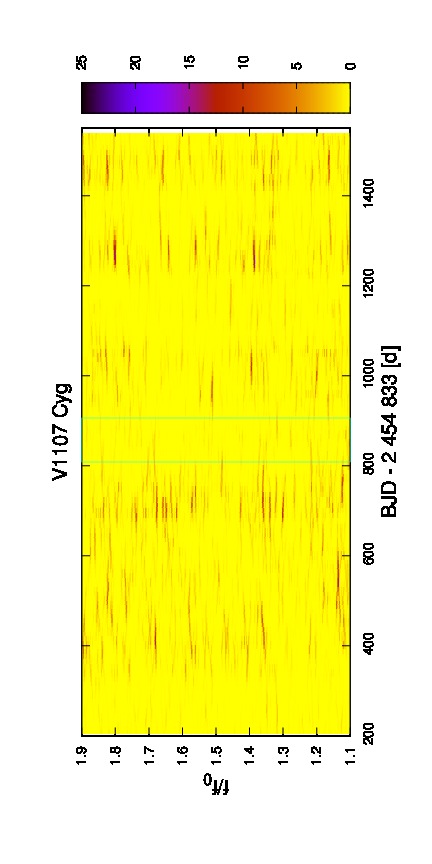}\\
\includegraphics[width=3.5cm,angle=270,trim=30 10 50 0]{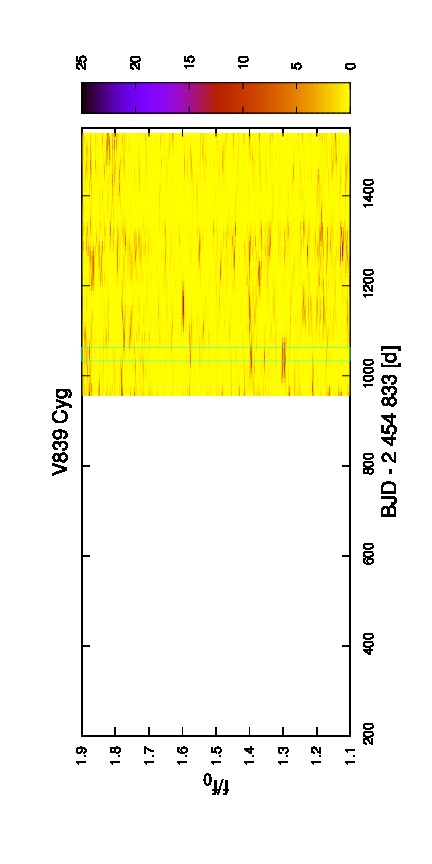}\\
\end{tabular}
    \caption{
Continuation of Fig.~\ref{fig:tf1}.
}
    \label{fig:tf2} 
\end{figure*}

As \citet{Clementini04} and \citet{Soszynski11,Soszynski16} pointed out the low frequency
ratio of anomalous RRd stars could only be obtained from the evolutionary models assuming  
either higher metallicity  ([Fe/H]$ > -0.5$) or smaller mass ($M < 0.55$~M$_{\sun}$)
than the usual parameters of RR\,Lyrae stars.  
For the \textit{Kepler} sample metallicities from high resolution spectroscopy
were published by \citet{Nemec13}. They found the metallicity of NQ\,Lyr and V2470\,Cyg
to be [Fe/H]$=-1.89\pm 0.10$~dex and [Fe/H]$=-0.59\pm 0.13$~dex, respectively. 
Comparing these values with the period ratios we can conclude that
the standard evolutionary theory cannot explain neither of these stars' 
present position in the instability strip (see fig.~8 in \citealt{Chadid10}).  
It needs an alternate evolutionary channel as it was suggested by \citet{Soszynski16}.
Since the mass seems to be lower than the normal RR\,Lyrae regime
we raise the possibility that this altenate tracks could belong to binaries similarly
to the case of OGLE-BLG-RRLYR-02792 \citep{Smolec13}.
This idea can be justified or refuted by a future spectroscopic work.
Alternatively \citet{Plachy13} found higher order resonant 
solutions in their hydrodynamic codes 
(e.g. with 8:11, or 14:19 ratios between $f_0/f_1$) 
which frequency ratios are outside 
the traditional RRd range but very similar to the ratio of the observed anomalous RRd stars.
In this case the mass and metallicity are not necessary anomalous.

We detected  in
many SC and/or LC  spectra an increase around the half of the main pulsation frequency ($\sim f_0/2$).
In some cases distinct visible but not really significant peaks ($S/N<4$) are
also appear (see e.g. NR\,Lyr, V782\,Cyg) around $0.48-0.49f_0$
but in most cases only the noise level 
increases around this position. It is shown well by the S/N ratio curves 
(green dotted lines in Fig.~\ref{fig:sc_spectra1}-\ref{fig:sc_spectra2}).
The reason of this feature is not clear.  Some possible explanations:
(i)
There is a rough trend within the signs of the residual light curve peaks:
a positive peak is followed by a negative and vice versa.
If this effect would be more regular we would seen a kind of `period doubling' and
its Fourier representation  would be the subharmonic $kf_0/2$ frequencies. 
However, we can see in Fig.~\ref{fig:sc_amp} and \ref{fig:sc_res} that this 
feature are far from the regularity and for all observed PD effect the 
highest amplitude frequency is the $3f_0/2$ and not $f_0/2$ as we see here.
(ii) The frequencies of the increase around $\sim f_0/2$ could be linear 
combination frequencies as $f'-f_0$. In this scenario $f'$ frequencies would 
be located around the first overtone $f_1$ with anomalous frequency ratio ($\sim 0.72-0.73$).
This way, almost all RRab stars would show anomalous RRd behaviour.
(iii) The formerly cited work of \citet{Plachy13} investigated the 
Fourier spectra of synthetic luminosity curves  
belongs to e.g. 6:8 resonance solutions. These models
show similar subharmonic structures (see their fig.~8) what we presented here
 but our light curves do not show any other signs of this resonance.
Higher order hardly detectable resonances might also explain the phenomenon but 
this scenario is rather speculative. (iv) No less than the assumption 
in which non-radial g modes are assumed as an explanation. 
The frequency range of the detected increases are 
bellow the Brunt-V\"ais\"al\"a frequency but calculations for
non-radial modes of RR\,Lyrae stars did not obtain considerable amplitudes around 
these regions \citep{VanHoolst98, Nowakowski03, Dziembowski16}. 
Therefore, this spectral feature requires further observational
and theoretical investigations.

Finally we note, that we found a significant peak of V784\,Cyg spectrum 
at 10.1393454~d$^{-1}$ ($S/N=7.1$). The frequency must belong to the background 
star KIS\,J195622.44+412013.9 which was mentioned in Sec.~\ref{sec:real}
because such a frequency would be very unusual for an RR\,Lyrae star and
it has no linear combination with the pulsation frequency of V784\,Cyg.
This frequency is typical of a $\delta$\,Scuti star, suggesting the variability type of
KIS\,J195622.44+412013.9.

\subsection{Time frequency variations}\label{sec:tfourier}

All the detected additional frequencies show noticeable time dependency. 
The relative amplitude of the peaks are different for different time spans
(LC, SCs) even so some peaks are undetectable in a given time series.
Some frequency changes can also be suspected. 

Time frequency analysis tools such as wavelet or Gabor transformations 
generally need strickly equidistant time series. So the observed data must be 
interpolated somehow. Avoiding this, we chose the simple time dependent Fourier tool
of the {\sc{SigSpec}} \citep{Reegen07,Reegen11} package. 
As inputs we used the LC light curve residuals which were obtained after 
removing 55-harmonic Fourier fits from the original light curves.
We set in {\sc{SigSpec}} one hundred days-long time bins for each star 
and used ten-day steps.
This resulted in $\sim 150$ Fourier spectra for each target.

Since additional peaks appear between $f_0$ and $2f_0$, we show this area of the spectra
in Figs.~\ref{fig:tf1} and \ref{fig:tf2} as contour plots.
For easier comparison, instead of the frequencies in the vertical axes, 
similar to the Figs.~\ref{fig:sc_spectra1} and \ref{fig:sc_spectra2}, 
the quantity $f/f_0$ is indicated.
The colour scales show the power values. 
The white area in panels indicate the missing data quarters when 
the given stars were located in any of the corrupted chips. 
The green boxes symbolise the time spans of the SC observations.  

Figs.~\ref{fig:tf1} and  \ref{fig:tf2} illustrate 
how the amplitudes of the additional frequencies dynamically change. 
Similar amplitude changes were revealed for the additional modes 
of Blazhko RRab and RRc stars \citep{Benko10,Szabo10,Szabo14,Moskalik15}.
The SC spectra in Figs.~\ref{fig:sc_spectra1}-\ref{fig:sc_spectra2}
represent snapshots of these variations. This explains the sometimes
different frequency content of the SC and LC spectra.
It is well traceable e.g.
how  the amplitude of $3f_0-f_2$ of V894\,Cyg decreased from a significant level
to below the detection limit
from the beginning of the observations to the time of the SC quarter.

The figures allow us to find such additional frequencies which are
significant only in a short time interval not observed any of the SC
quarters and averaged out from the spectra of the four-years LC data.  
The detected frequencies with their approximate visibility dates 
in the brackets are the followings: 
NR\,Lyr: $f_1$ ($t\sim 450-850$~d), and $f_2$ ($t\sim 900$~d); 
KIC\,6100702: $f_1$ ($t\sim 200-400$~d); 
NQ\,Lyr: $f_2$ ($t\sim 300-400$~d); 
FN\,Lyr: $f_1$ ($t\sim 650$~d). 
Three of these stars (NR\,Lyr, KIC\,6100702 and FN\,Lyr) do not show
significant additional modes in their SC and LC spectra.

\subsection{Connection between the C2C variations and the additional modes}

\begin{figure}
\includegraphics[scale=.32,angle=270,trim=4.5cm 0 7cm 0]{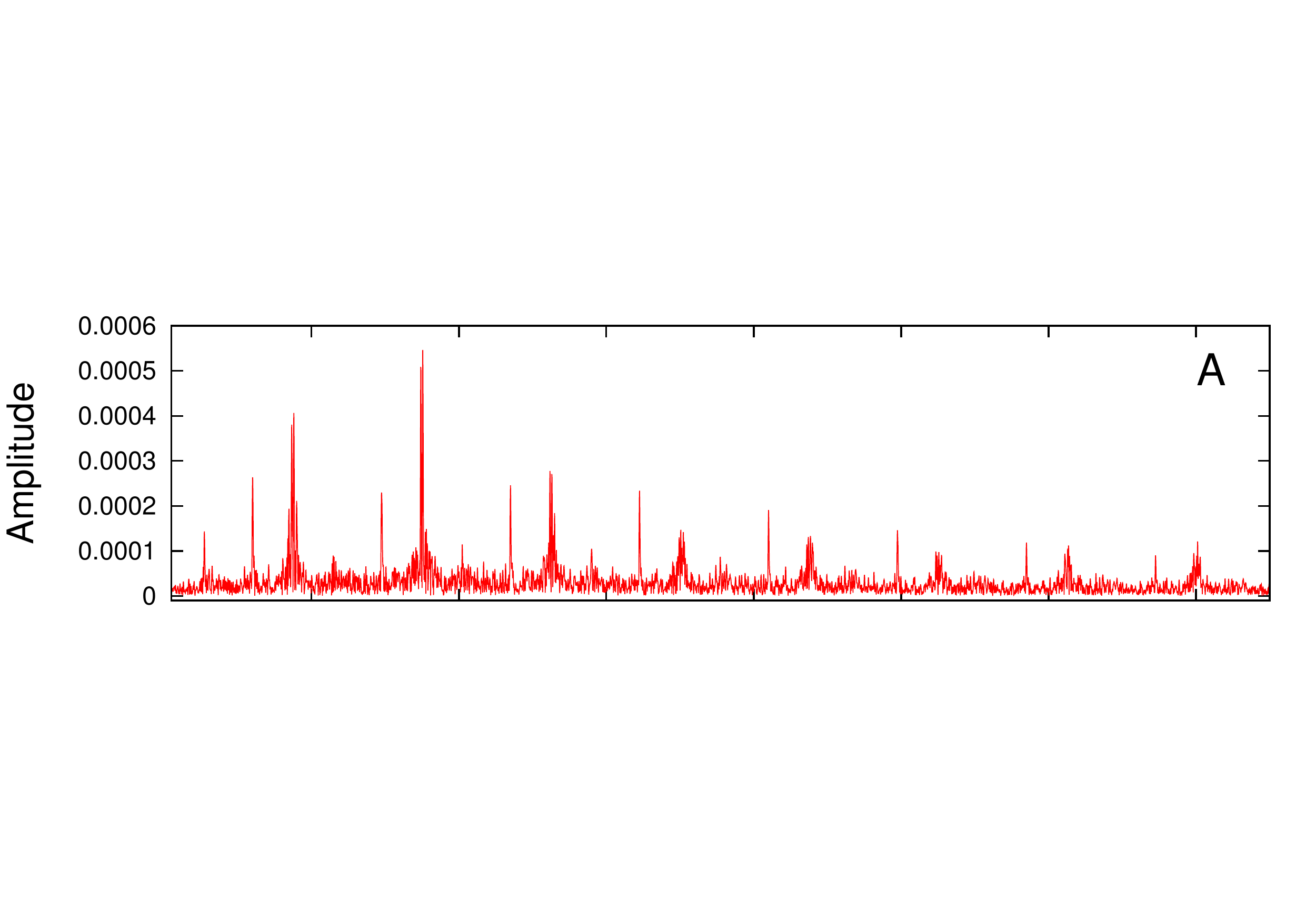}
\includegraphics[scale=.32,angle=270,trim=4.5cm 0 5cm 0]{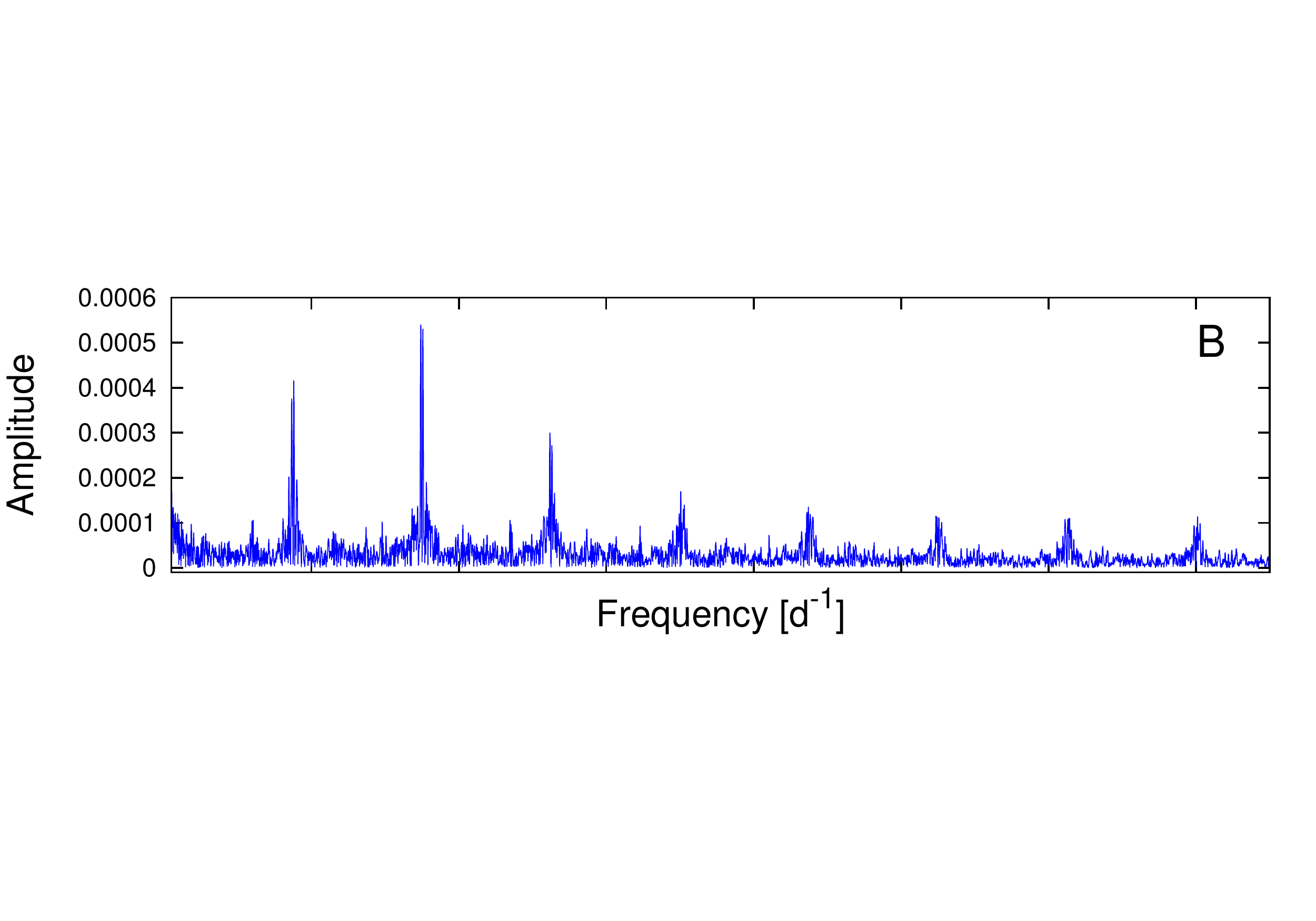}
\includegraphics[scale=.32,angle=270,trim=4.5cm 0 7cm 0]{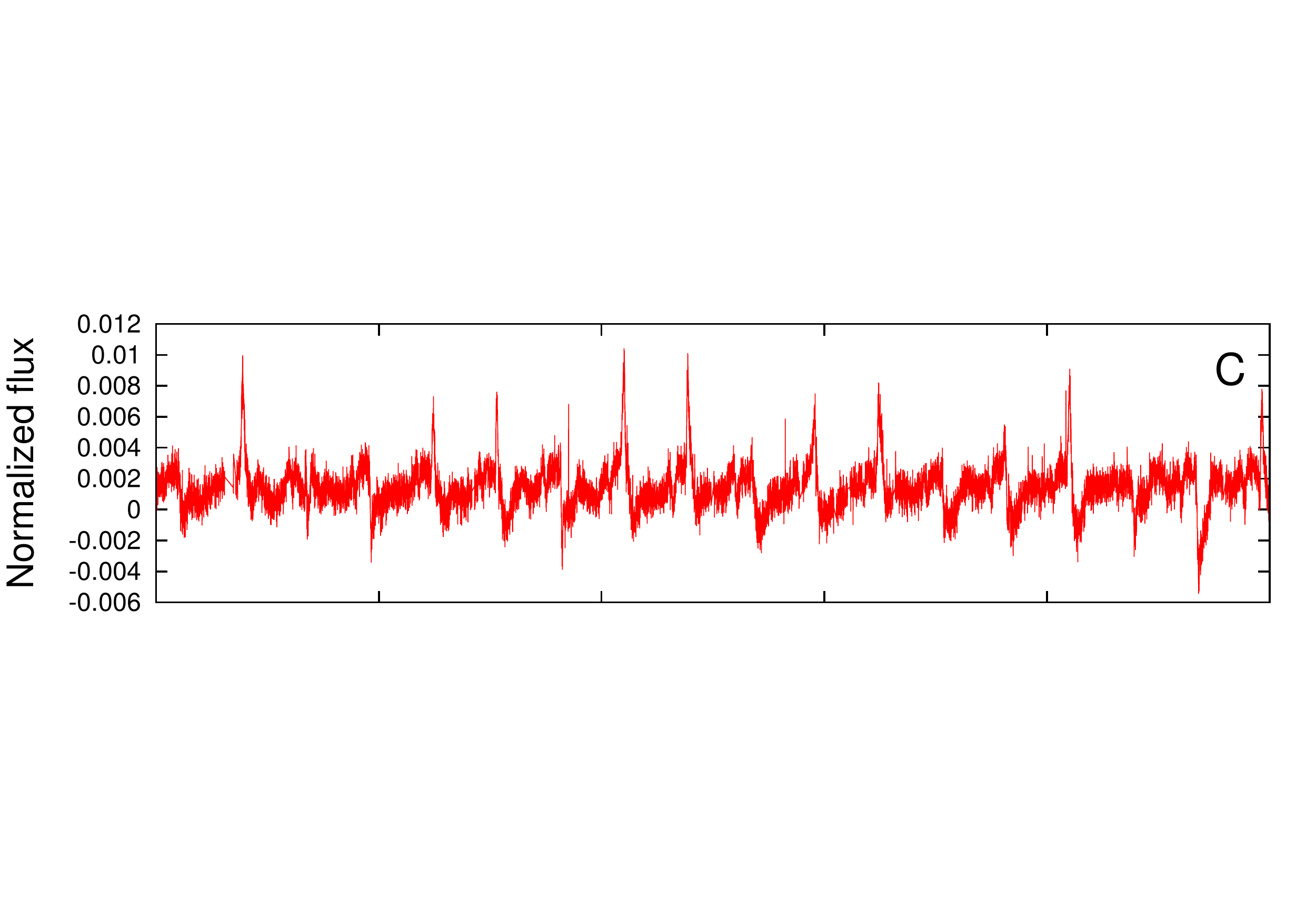}
\includegraphics[scale=.32,angle=270,trim=4.5cm 0 4cm 0]{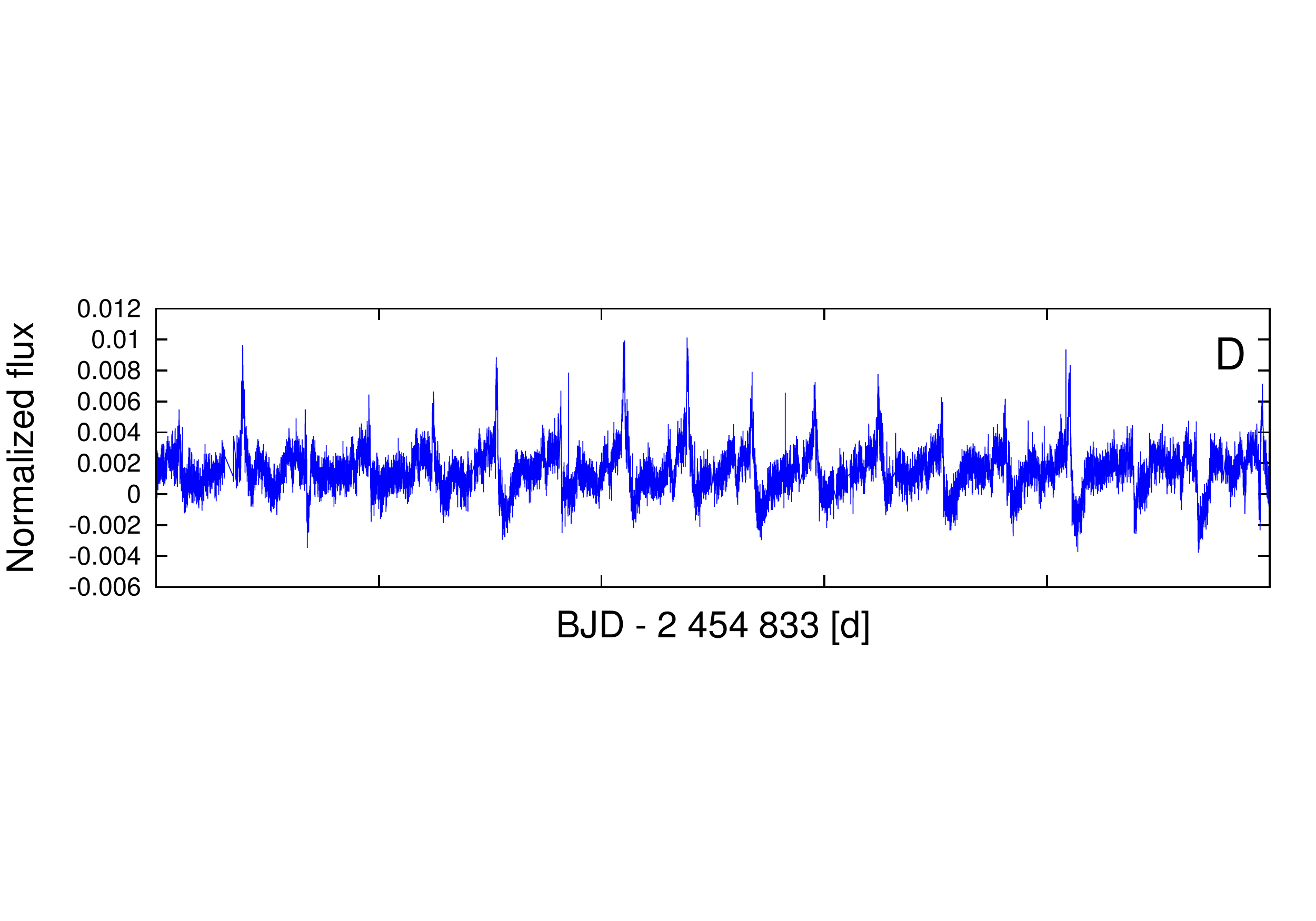}
\caption{The role of the additional frequencies in C2C variations.
Residual spectrum of the nomalized SC flux curve of V894\,Lyr (panel A) and
the same spectrum if we pre-whitened the additional frequencies from the data (panel B).
Part of the residual SC flux curve (panel C) and the same flux curve part after we removed
the additional frequencies (panel D).
}\label{fig:res-add}
\end{figure} 
In the case of previously studied regular C2C light curve variations of the Blazhko stars 
as period doubling \citep{Kolenberg10, Szabo10} or other resonances \citep{Molnar12, Molnar14}
suggest that the extra modes which manifest additional frequencies in the Fourier spectra, can
cause regular C2C variations on the light curves.

Taking these into account, the question arises: what is the relationship between the observed extra 
frequencies and the C2C variation? 
First, there are a number of stars which (e.g. V782\,Cyg, KIC\,6100702, KIC\,7030715, AW\,Dra) show 
significant C2C variations but 
there are no signs of any additional mode frequencies in their SC spectra. In other words, the
excited additional modes can be ruled out as the only reason of the C2C variation.

Second, we tested the role of the additional modes in the C2C variation. 
For this, we used the stars with additional modes (V346\,Lyr, 
V894\,Lyr and  
KIC\,9658012), pre-whitened  all the significant additional frequencies and their linear
combinations from their light curve, then we re-analysed them searching for C2C variations in the same way as 
in Sec.~\ref{sec:C2C} for the original curves. 
We only show the results of V894\,Lyr which is the brightest among these three stars. 
Fig.~\ref{fig:res-add} shows the spectra before (panel A) and after (panel B) pre-whitening 
the additional frequencies from the data. Because of the time dependent amplitudes 
discussed in Sec.~\ref{sec:tfourier}, some frequencies remain after 
the pre-whitening process but with marginal amplitudes.
The normalized flux curves belonging to these spectra are shown in the panels C and D.
The flux curves with and without removing the additional frequencies have very similar shapes 
illustrating that the additional modes only marginally
affect the C2C variations. The dominant random variation seems to be independent
from these modes. 

\section{The presence of the Blazhko effect}

The present hypothesis is that amongst RRab stars only the Blazhko
stars show additional frequencies. This hypothesis was set because 
sooner or later all the non-Blazhko stars showing additional modes 
turned out to display the Blazhko effect \citep{Benko10,Nemec11,Benko15}.
In the previous section, however, we have seen that considerable part 
of the \textit{Kepler} non-Blazhko sample shows additional mode 
pulsation. This is true even if we omit the discovered anomalous RRd stars
from the sample.

The presence or the lack of the Blazhko effect needs a careful investigation.
It is especially relevant now, because a recent result
suggests that the Blazhko incidence ratio among RRab stars could be
as high as 90\% \citep{Kovacs18}.    

\subsection{The O$-$C diagrams}\label{sec:o-c}

The Blazhko effect means simultaneous amplitude and frequency/phase modulation 
with the same frequency or frequencies. If the amplitude of the amplitude
modulation part is high enough this effect can be easily detected.
This is obviously not true for our sample. The amplitude modulations 
if they exist at all must be of very low amplitude. In addition, the amplitudes are 
more sensitive to the instrumental and data handling problems than the
phase, therefore the potential phase variations were carefully tested by 
using a refined version of the classical O$-$C (observed minus calculated) method 
(\citealt{Sterken05} and references therein). 

Traditionally, the O$-$C diagrams are constructed from
definite phase points of a periodic light curve (maxima, minima, etc.).
The exact position of these phase points (the `O' values)
are determined by the e.g. maxima of a least square fitted polynomial, 
or spline function around the predicted (`C') positions.  As 
\cite{Jurcsik01} showed for the sparse data of globular cluster $\omega$ Cen 
RR\,Lyrae the accuracy of O$-$C diagrams can be significantly improved if we 
define a template and the `O' values are determined from the 
least square minimization of the horizontal shifts of
the template at each proper position.
This way we take into account the entire light curve and not just 
parts of it around the critical phases. This method was applied by
\citet{Derekas12} when they detected the random
period jitter of a Cepheid (V1154\,Cyg),  and also by  \citet{Li14} and
\citet{Guggenberger15}
who search for potential light-time effect caused by a companion in 
the {\it Kepler} RR\,Lyrae sample.  
This work used the same implementation of the method what we used in \citet{Benko16},
namely the program of \citet{Derekas12} sligthly adjusted to RR\,Lyrae stars.

\begin{figure}
\includegraphics[width=\linewidth]{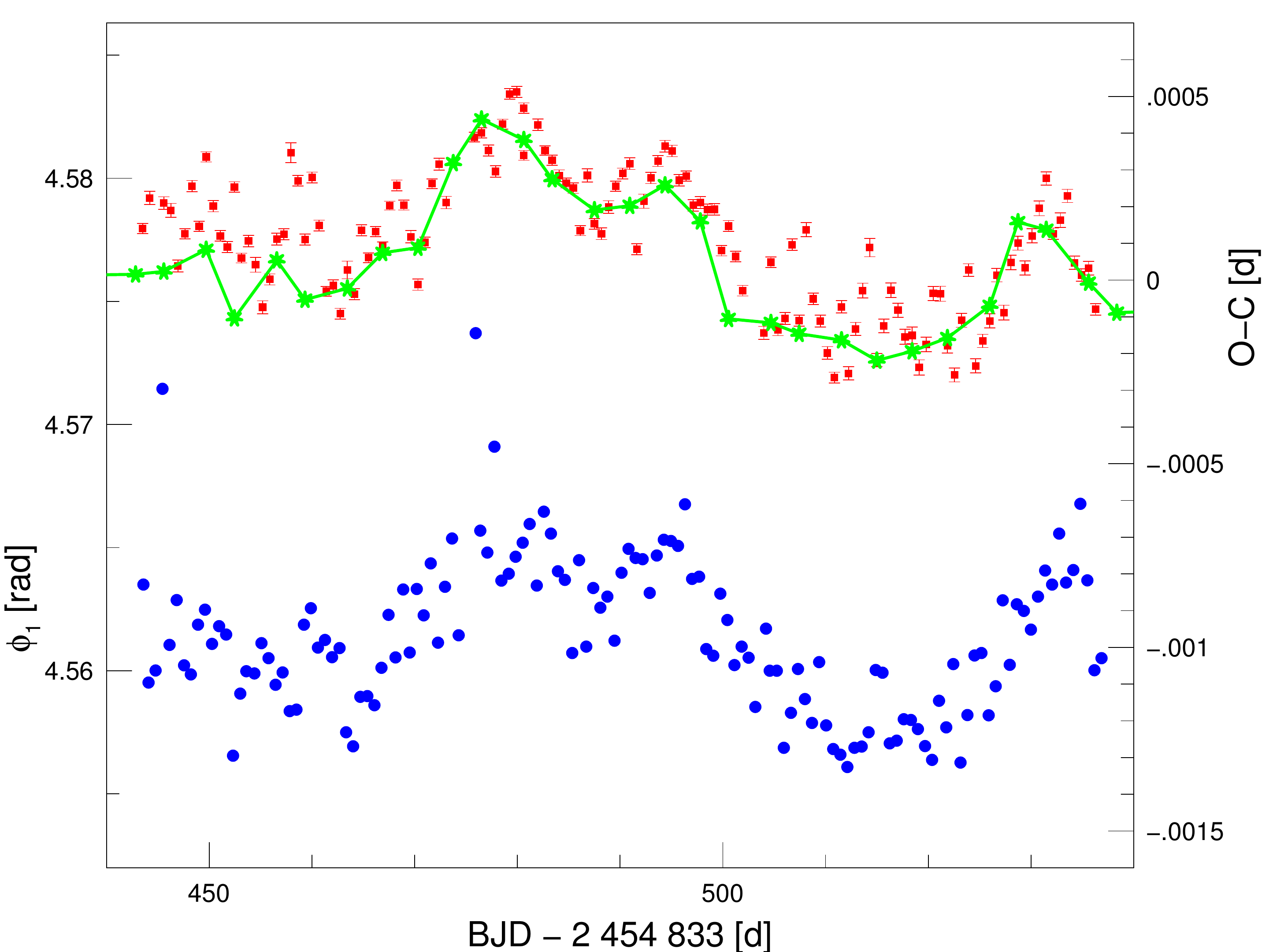}
\caption{O$-$C variation of the SC data of AW\,Dra in the {\it Kepler} Q5 quarter (red
squares with errorbars) with the parallel phase variation 
$\phi_1(E)$ (blue dots) where we transformed the time variation 
into epoch scale. The green astrisks connected with straight lines show the 
O$-$C values calculated from the LC data (see the text for the details).
}\label{fig:O-C_phi}
\end{figure}
O$-$C diagrams were constructed for both the SC and the LC light curves. 
In the case of the SC data each pulsation cycle can be handled separately without any problems,
however,  it does not work for the LC data due to their sparse sampling. For the LC light curves 
five-cycle-long parts were chosen and the template shift values (viz. the `O' values) 
were determined on these intervals. 
This handling means an averaging which smooths the O$-$C curve but proved to be a good compromise.
When we leave the cycle-by-cycle handling we lose only a little information
as it is demonstrated in the upper curve of Fig.~\ref{fig:O-C_phi}
but we win a much longer data set.  
 The green continuous line in Fig.~\ref{fig:O-C_phi} shows the 
O$-$C diagram of the LC data calculated with this manner. 
As we seen the O$-$C diagrams of the LC data
handling by our method are sufficient ever for studying rather short time scale variations as well. 
Accordingly, unless otherwise stated, we describe here the results obtained from the LC data.
   
\begin{table*}
        \centering
        \caption{Some parameters of the sample stars.
ID; period published by \citet{Nemec13} $P_{\mathrm N}$; improved period $P_0$;
period change rate $\dot{P}$, and its accuracy; the highest amplitude additional mode
frequency, its possible identification and signal-to-noise ratio.
The given number of digits of the periods and the frequencies indicate the accuracy.} 
        \label{tab:per}
        \begin{tabular}{r*2{l}*4{r}} 
                \hline
ID & $P_{\mathrm N}$ &   $P_0$ & $\dot{P}$ & $\sigma (\dot{P})$ & Add. fr. & $S/N$ \\
    & (d)             &   (d)   &  $\times 10^{-10}$        &   $\times 10^{-11}$ & (d$^{-1}$) & \\
                \hline
NR\,Lyr&  0.6820264 &        0.6820268   &   12.38   &  6.98 & & \\
V715\,Cyg&  0.47070609&      0.4707059  &   $-$15.13  &  4.29 & & \\
V782\,Cyg&  0.5236377 &      0.5236375  &   4.50   &  3.21 & & \\
V784\,Cyg&  0.5340941 &      0.5340947  &   $-$4.51  &  3.6 & & \\
KIC\,6100702&  0.4881457 &   0.4881452  &   $-$1.24  &  2.56 & & \\
NQ\,Lyr&  0.5877887 &        0.5877889  &   $-$8.43  &  3.94 & $2.323822=f_1$& 4.2\\
FN\,Lyr&  0.52739847&        0.5273986  &   $-$9.31  &  3.78 & & \\
KIC\,7030715&  0.68361247&   0.6836125  &   4.58   &  8.16 & & \\
V349\,Lyr&  0.5070740 &      0.5070742  &   0.41  &  5.90 & & \\
V368\,Lyr&  0.4564851 &      0.4564859  &   $-$11.81  &  2.67 & & \\
V1510\,Cyg&  0.5811436 &     0.5811426  &   27.41   &  5.41 & $1.178286=f_2-f_0$& 7.6\\
V346\,Lyr&  0.5768288 &      0.5768270  &   12.40   &  20.41 & $1.189183=f_2-f_0$& 13.3\\
V894\.Lyr&  0.5713866 &      0.5713865  &   22.66   &  14.24 & $1.198871=f_2-f_0$& 9.5\\
KIC\,9658012&  0.533206  &   0.533195   &   $-$7.87  &  36.08 & $3.164672=f_2$& 11.0\\
KIC\,9717032&  0.5569092 &   0.556908   &   74.14   &  39.43 & & \\
V2470\,Cyg&  0.5485905 &     0.5485897  &   $-$1.21  &  3.11 &  $2.524809=f_1$ & 4.1\\
V1107\,Cyg&  0.5657781 &     0.5657795  &   $-$0.40  &  5.76 & & \\
V839\,Cyg&  0.4337747 &      0.4337742  &   1.39   &  6.45 & & \\ 
AW\,Dra&  0.6872160 &        0.6872186  &   $-$53.39  &  18.26 & & \\
                \hline
        \end{tabular}
\end{table*}

\begin{figure*}
\centering
\begin{minipage}{0.495\textwidth}
\includegraphics[width=\linewidth,{trim=10 0 20 0},clip]{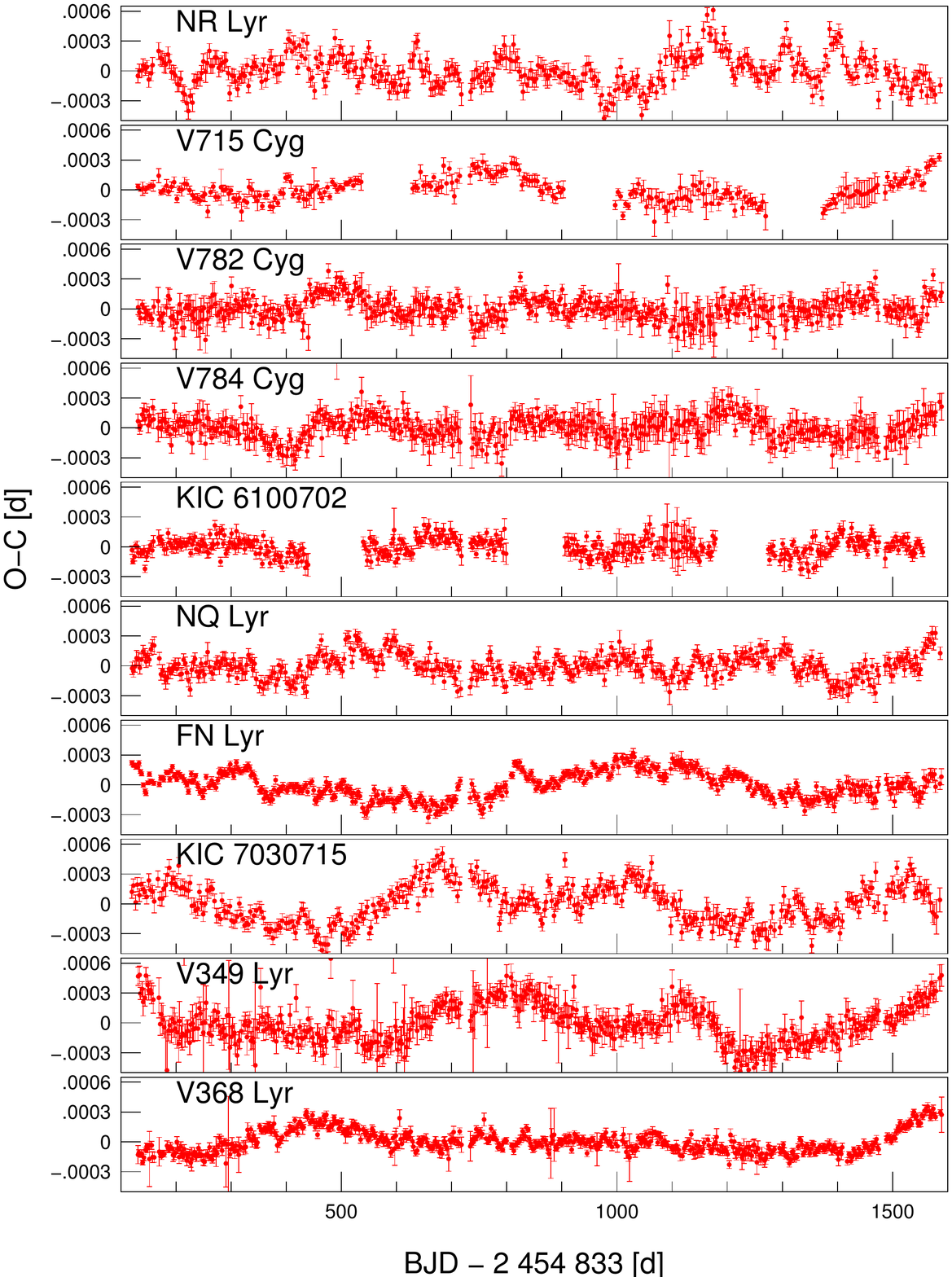}
\end{minipage}\hfill
\begin{minipage}{0.221\textwidth}
\includegraphics[width=\linewidth,{trim=5 0 0 0},clip]{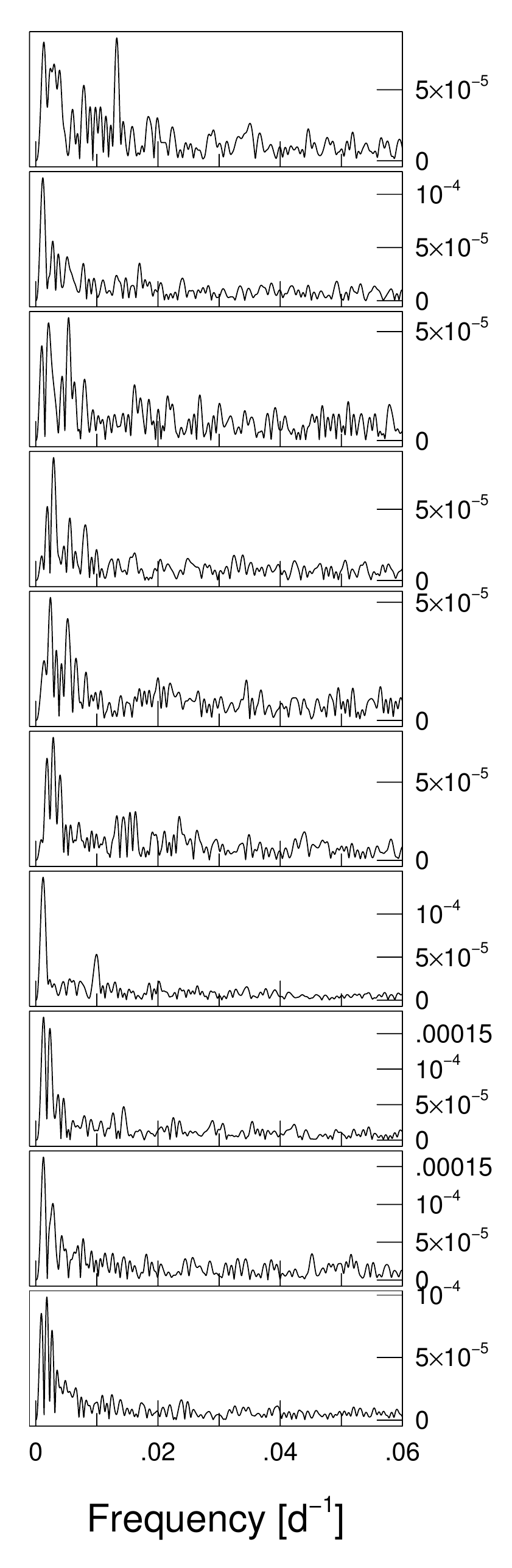}
\end{minipage}
\begin{minipage}{0.249\textwidth}
\includegraphics[width=\linewidth,{trim=5 0 0 0},clip]{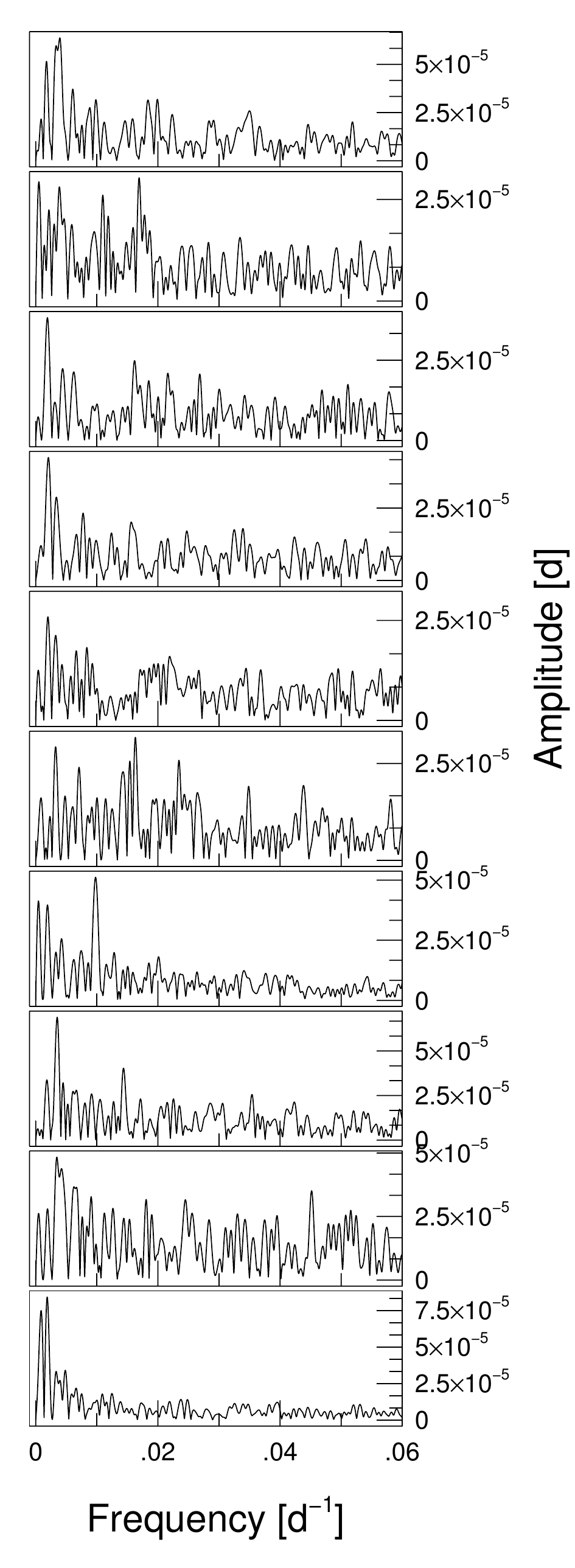}
\end{minipage}
\caption{
O$-$C diagrams of the LC data after removing a 
linear and a quadratic trends (in the left). The Fourier spectra of the
left-hand-side O$-$C diagrams (in the middle) and these spectra
after we removed the detected instrumental frequencies connected to
the {\textit{Kepler}} year (in the right). 
}\label{fig:oc1}
\end{figure*}
\begin{figure*}
\centering
\begin{minipage}{0.495\textwidth}
\includegraphics[width=\linewidth,{trim=10 0 20 0},clip]{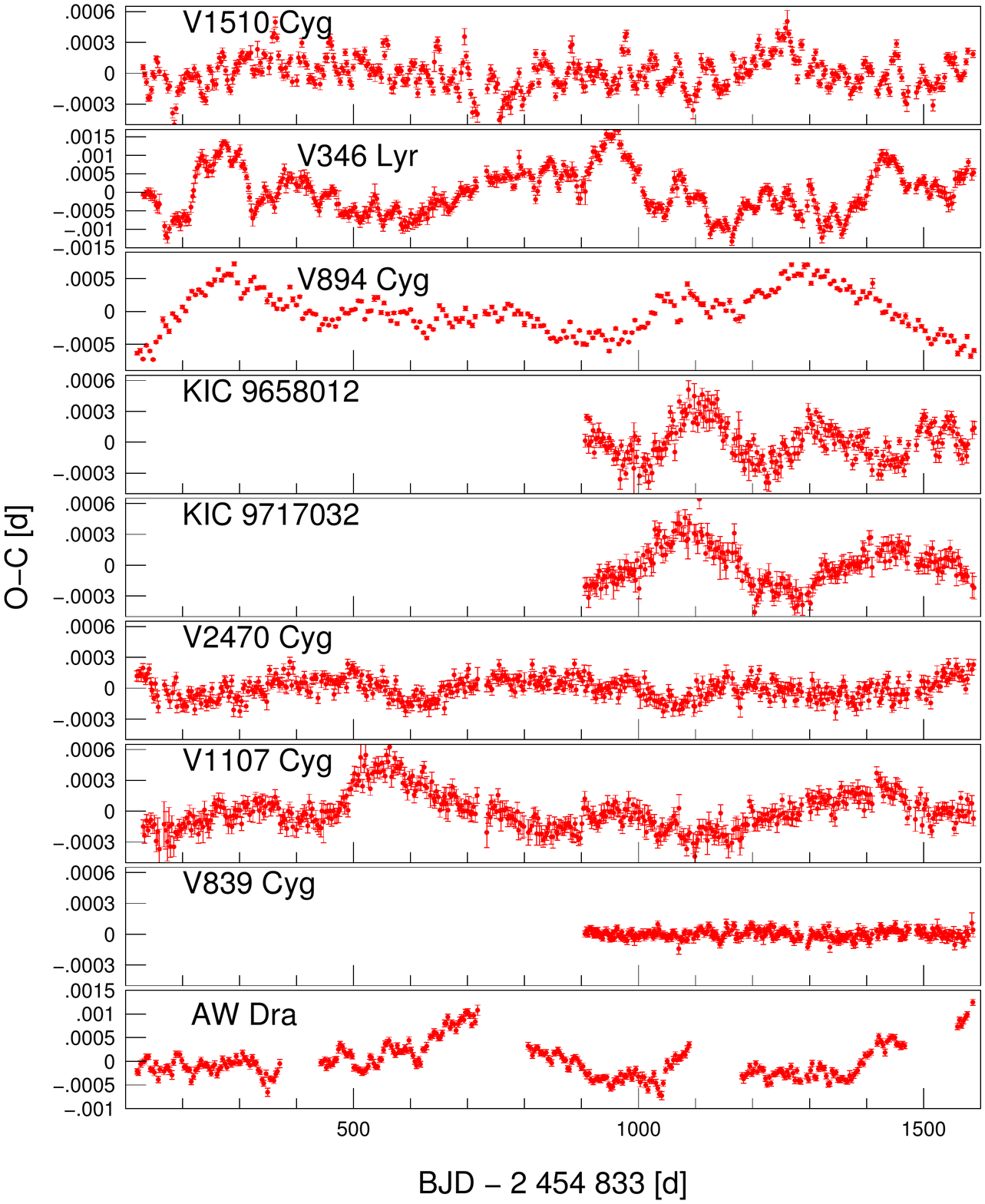}
\end{minipage}\hfill
\begin{minipage}{0.221\textwidth}
\includegraphics[width=\linewidth,{trim=5 0 0 0},clip]{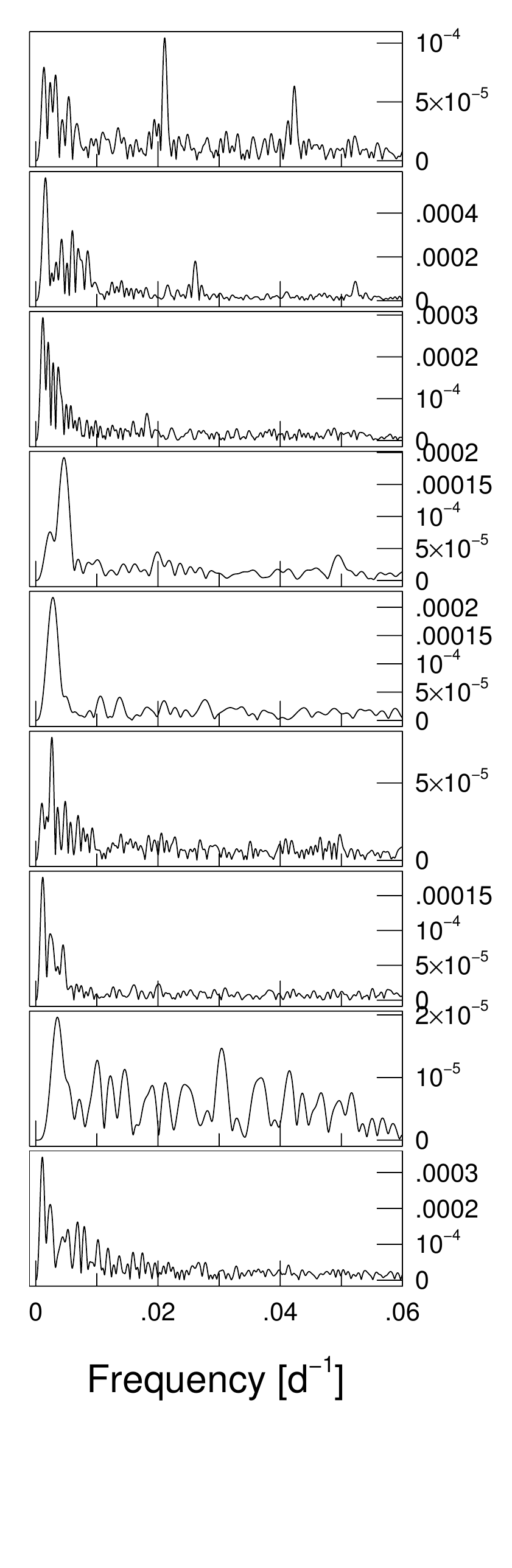}
\end{minipage}
\begin{minipage}{0.249\textwidth}
\includegraphics[width=\linewidth,{trim=5 0 0 0},clip]{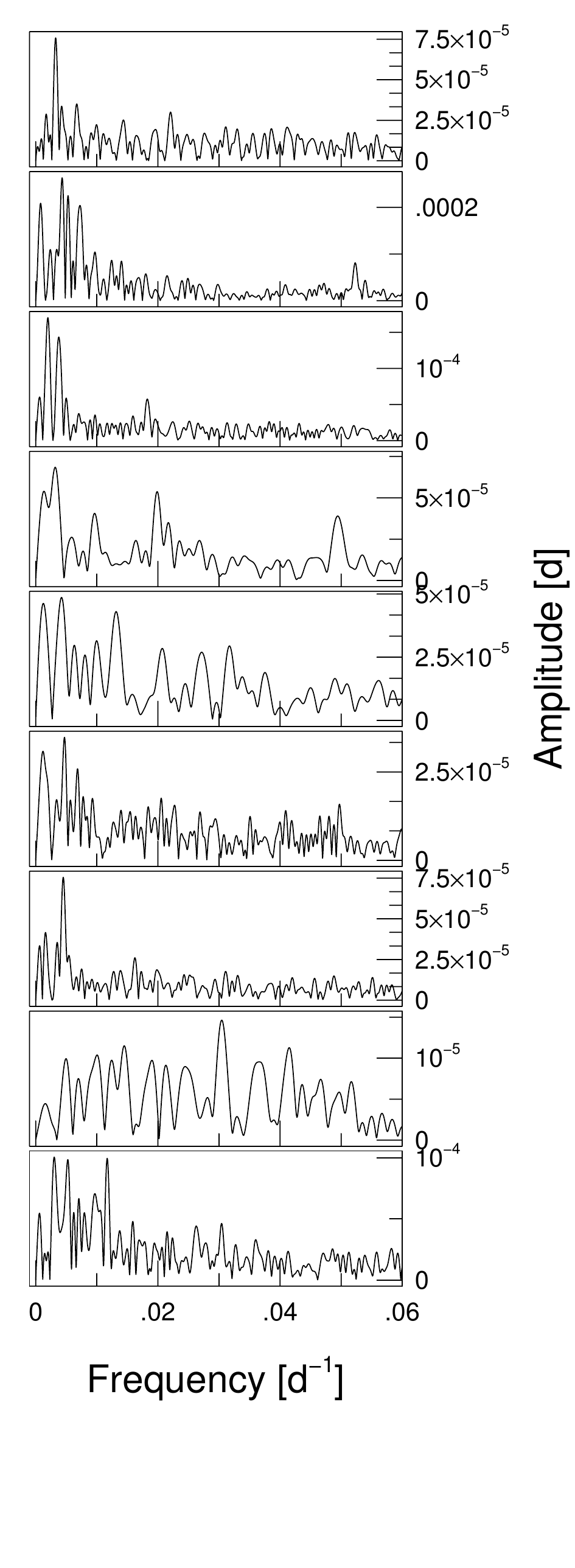}
\end{minipage}
\vspace{-1cm}
\caption{
Continue of Fig.~\ref{fig:oc1}. We call the attention the O$-$C diagrams of three
stars (KIC\, 8344381, KIC\,9591503 and KIC\,11802860) which are plotted
in higher vertical scales.}\label{fig:oc2}
\end{figure*}
The O$-$C diagrams were prepared using the latest and most precise
values of the periods and starting epochs published by \citet{Nemec13}.
The obtained diagrams are dominated many times by a linear trend showing
that the periods need refining. To do this, nonlinear fits containing 35-50
harmonics of the main pulsation frequencies were applied to the complete data sets.
The input frequencies of the fits were calculated 
from \citet{Nemec13} periods $P_{\mathrm N}$ (column 1 in Tab.~\ref{tab:per}). 
The determined accurate average pulsation periods $P_0$ on total observed time 
spans are given in column 2 of Tab.~\ref{tab:per}. The well-known evolutionary period 
change of RR\,Lyrae stars causes the parabolic shape of most O-C diagrams.
By subtracting a quadratic fit as
\[
O-C (t) = \frac{1}{2} P_0 \dot{P} t^2 + const. 
\]
these trends were also eliminated. 
As a by-product, we could determine the  
period change rates $\dot P$ (column 3 in Tab.~\ref{tab:per}).
Their errors  $\sigma(\dot P)$ in column 4 of  Tab.~\ref{tab:per} are 
the RMS error of the fits.  
These $\dot P$ values are between $7\cdot 10^{-9}$ and $4\cdot 10^{-11}$~dd$^{-1}$
which are in good agreement with both the theoretical predictions \citep{Sweigart79, Lee90, 
Dorman92, Pietrinferni04} and the other observed values \citep{Szeidl11, Jurcsik01, Jurcsik12}.

After removing the above mentioned linear and quadratic trends from the data, the O$-$C diagrams are shown in the left panels of Figs.~\ref{fig:oc1} and \ref{fig:oc2}.
\begin{table}
        \centering
        \caption{The detected frequencies in the O$-$C diagrams.
The columns show the star's name; the value of the found frequency $f$; 
its signal-to-noise ratio; the possible identity of $f$; 
and the frequency difference between the detected and the exact instrumental frequencies: 
$\Delta f=\vert f-f_{\mathrm I} \vert$}
        \label{tab:o-c_freq}
        \begin{tabular}{rlrcl} 
                \hline
Name & $f$ & S/N &  $f_{\mathrm I}$ & $\Delta f$\\
     & (d$^{-1}$) &     &                  &  ($\times 10^{-4}$d$^{-1}$) \\
                \hline
NR\,Lyr   &0.0133323   & 6.91  & 5$f_{\mathrm K}$ & 0.88 \\ 
          &0.0013398   & 6.66  & $f_{\mathrm K}/2$ & 0.02 \\
          &0.0078670   & 4.24  & 3$f_{\mathrm K}$ & 1.85 \\
          &0.0039507   & 5.07  & $2f'$  &   \\ 
          &0.0017521    & 4.09  & $f'$ &  \\ 
V715\,Cyg &0.0011696    & 10.06  & $f_{\mathrm K}/2$ & 1.72\\
V782\,Cyg &0.0053867    & 7.21  & 2$f_{\mathrm K}$ &  0.19\\
          &0.0019214    & 4.84  & $f'$  &  \\ 
V784\,Cyg &0.0029159    & 10.15  & $f_{\mathrm K}$ & 2.32 \\
          &0.0055916    & 5.16   & 2$f_{\mathrm K}$ & 2.23 \\
          &0.0081301    & 4.60   & 3$f_{\mathrm K}$ & 0.78 \\
          &0.0020926    & 4.96  & $f'$  &  \\ 
KIC\,6100702 &0.0024056   & 7.40  & $f_{\mathrm K}$ & 2.78 \\
             &0.0052295   & 6.13  & 2$f_{\mathrm K}$ & 1.38 \\
NQ\,Lyr   &0.0028651    & 9.63 &  $f_{\mathrm K}$ & 1.81 \\
          &0.0018274    & 7.99 &  $f'$                 &     \\
          &0.0039181    & 6.68 & $2f'$                &     \\
FN\,Lyr   &0.0012259    & 19.52 &  $f_{\mathrm K}/2$ & 1.16 \\
          &0.0100458    & 7.29  & 4$f_{\mathrm K}$? &   6.90 \\ 
KIC\,7030715&0.0012963    & 13.51 &  $f_{\mathrm K}/2$ & 0.46 \\
            &0.0023198    & 12.27 &  $f_{\mathrm K}$ &   3.64 \\
            &0.0035138    & 5.34  & $2f'$  &  \\ 
V349\,Lyr   &0.0012694    & 11.23 &  $f_{\mathrm K}/2$ & 0.73 \\
            &0.0028134    & 6.99 &  $f_{\mathrm K}$ &    1.29 \\
V368\,Lyr   &0.0018181     & 15.35 & $f'$   &  \\
            &0.0009262    & 13.31 & $f'/2$ &  \\
            &0.0026757    & 11.19 &  $f_{\mathrm K}$ & 0.08 \\
            &0.0036019    & 4.94  & $2f'$ &  \\ 
V1510\,Cyg  &0.0210886     & 9.77  & 8$f_{\mathrm K}$  & 1.03  \\ 
            &0.0013395     & 7.43  & $f_{\mathrm K}/2$ & 0.03 \\ 
            &0.0032629    & 6.81  & $2f'$ & \\ 
            &0.0023698    & 6.22  & $f_{\mathrm K}$ & 3.14 \\ 
            &0.0425550    & 5.93  & 16$f_{\mathrm K}$ & 3.89 \\ 
            &0.0053924    & 5.09  & 2$f_{\mathrm K}$ & 0.24 \\ 
V346\,Lyr &0.0016135  & 15.71  &  $f_{\mathrm K}/2$ &  2.71\\  
	  &0.0059924  &  8.95  & $f''+f_{\mathrm K}/2$ & \\ 
	  &0.0042225  & 7.86 & $f''$ & \\ 
	  &0.0069345  & 6.67 & $2f''-f_{\mathrm K}/2$ & \\ 
	  &0.0085136  & 6.38 & $2f''$ & \\ 
          &0.0052867   & 6.40  & $2f_{\mathrm K}$ & 0.81 \\ 
          &0.0260899   & 5.05  & 10$f_{\mathrm K}$? & 7.10\\ 
          &0.0008239   & 5.93  & $f_{\mathrm K}/4$ & 1.53 \\ 
V894\,Cyg &0.0011922   &  11.08 & $f_{\mathrm K}/2$ & 1.49 \\
          &0.0020097   &   6.15 & $f'$  &  \\ 
          &0.0037809   &   5.18 & $2f'$ &  \\ 
KIC\,9658012&0.0046154   & 12.15  &  2$f_{\mathrm K}$ & 7.52\\
            &0.0032234   &  4.31  &  $2f'$ &  \\ 
KIC\,9717032&0.0027851   & 12.59  &  $f_{\mathrm K}$ & 1.01\\  
V2470\,Cyg  &0.0026576    & 11.05  &  $f_{\mathrm K}$ &  0.26 \\  
            &0.0046337    &  4.84  & $f''$ &  \\ 
            &0.0012380    &  4.29  &  $f_{\mathrm K}/2$ & 1.04 \\ 
V1107\,Cyg  &0.0011673    & 16.51  &  $f_{\mathrm K}/2$ & 1.75 \\
            &0.0023690    & 8.92  &  $f_{\mathrm K}$ &  3.15 \\
            &0.0044633    & 7.43  & $f''$  &  \\
V839\,Cyg   &0.0035263    & 4.51   & $2f'$                 &  \\   
AW\,Dra     &0.0010915    &  11.53  & $f_{\mathrm K}/2$  &  2.51 \\  
            &0.0022854    &  8.62   & $f_{\mathrm K}$    &  3.98 \\  
                \hline
        \end{tabular}
\end{table}
We see two types of variability on the diagrams. On the one hand a global year-scale
$\sim 0.0003$~d amplitude flow can be detected on the other hand a shorter time-scale and 
lower amplitude $\sim 10^{-4}-10^{-5}$~d fluctuations are also presented for several stars.

\subsection{Fourier analysis of the O$-$C diagrams}

For the sake of a more quantitative study, we calculated the Fourier spectra of the O$-$C diagrams using the {\sc MuFrAn} program package \citep{Kollath90}. The obtained
spectra are shown in the middle panels of Figs.~\ref{fig:oc1} and \ref{fig:oc2}.
The Fourier spectra contain well-detectable peak(s) for all stars. 
The significant frequencies are listed in Tab.~\ref{tab:o-c_freq}.
Vast majority of these frequencies are the harmonic or sub-harmonic of 
the \textit{Kepler} frequency $f_{\mathrm K}$
within the Rayleigh frequency resolution limit. (This limit frequency is $6.8\times 10^{-4}$~d$^{-1}$
for the longer time series while for KIC\,9658012, KIC\,9717032 and V839\,Cyg 
it is: $1.47\times 10^{-3}$~d$^{-1}$.)
The appearance of the \textit{Kepler} year in the flux data has already been known \citep{Banyai13}
but here we demonstrated that this instrumental systematics affects the phases. 

\citet{Li14} identified the long periodicities in the O$-$C diagrams 
of FN\,Lyr and V894\,Lyr, using the {\it Kepler} data,
as potential light-time effect caused by companions. 
As we can see in Tab.~ \ref{tab:o-c_freq} the frequency of these variations
agree well with $f_{\mathrm K}/2$ and it can be detected in eight additional spectra.
Therefore, it is probable that all these periodicities has the same
instrumental origin rather than the binarity.

We pre-whitened the data with the significant frequencies, the residual
spectra are shown in the right panels of Figs.~\ref{fig:oc1}-\ref{fig:oc2}.
Harmonics and sub-harmonics of two frequencies: $f' \sim 0.00182$~d$^{-1}$ and
$f'' \sim 0.00422$~d$^{-1}$  appeared either in the raw or the pre-whitened spectra of 
different stars, which shows the instrumental origin of these frequencies.  

There are two stars (FN\,Lyr and V346 Lyr) where the 
identification of their frequency contents
with the different instrumental frequencies is not certain. 
Namely, some of their frequencies differ more than the Rayleigh resolution limit
from the possible instrumental frequencies. Though it was shown by \citet{Kallinger08} that the 
Rayleigh limit is actually an overestimation. In our case, the $\Delta f$ differences between the exact 
frequency values and the measured ones are well below this limit for the certain identifications.
For V346\,Lyr the harmonic of 0.02609~d$^{-1}$
appears in the pre-whitened spectrum at 0.05218~d$^{-1}$. 
This frequency is definitely not identical with the $20f_{\mathrm K}$,
because $\Delta f$ would be 0.0015 with this assumption, which is twice as much 
as the Rayleigh frequency resolution. 
This suggests a non-sinusoidal possible variation of V346\,Lyr. 

The case of V1510\,Cyg seems to be
similar to V346\,Lyr where also unusually high order harmonics of $f_{\mathrm K}$
($8f_{\mathrm K}$ and $16f_{\mathrm K}$) are significant. Similarly to V346\,Lyr,
these frequencies are harmonics, but for V1510\,Cyg these high order harmonics are well within 
the resolution limits, that is, we can not separate such possible 
stellar frequencies from the instrumental effects. 

By definition, the Blazhko effect means simultaneous amplitude and phase 
variations with the same period(s). A good tracer of the amplitude modulation is
the appearing of the modulation frequency in the low frequency region of the 
light curve \citep{Benko10}. From the above three stars only V346\,Lyr shows such a
peak (at 0.02624~d$^{-1}$, S/N=17) and therefore V346\,Lyr is the
only well-settled Blazhko candidate of the sample.

\subsection{The phase variation functions}

In order to check the results of the O$-$C diagrams, we studied
the Fourier phase variation function $\phi_n(t)$ of the LC data.
These functions proved to be useful for seperating the non-Blazhko sample \citep{Nemec11} and
also for discovering the small Blazhko effect of V838\,Cyg and KIC\,11125706 \citep{Nemec13}. 
Practically, the first ten phase variation functions were calculated for each star 
by using the non-linear Fourier fit of {\sc{LCfit}} \citep{lcfit} package
as we did for SC light curves in Sec.~\ref{sec:charC2C}. The only difference was here 
that three pulsation cycles were handled together because of the sparse LC sampling, which provides sufficient number of fitted points (about 60-80). 

As it is known from earlier, the structure of the Fourier phase variation 
function $\phi_1(t)$ is similar to the O$-$C curve (e.g. \citealt{Guggenberger12}). 
In Fig.~\ref{fig:O-C_phi} we show an example for this similarity. 
We plotted both the SC and LC O$-$C variations of AW\,Dra
with the phase variation $\phi_1(t)$. The parallel nature of the three curves are evident.
Since the O$-$C diagrams show the total phase variations of a light curve 
these parallelism means that the first order phase variation $\phi_1(t)$ dominates the
total phase variation. 
Therefore, it is not surprising that
the frequencies identified in the $\phi_1(t)$ Fourier spectra
are equal to one of the frequencies appeared in the 
O$-$C diagram spectra (Table~\ref{tab:o-c_freq}). The frequency content of the
O$-$C spectra and $\phi_1(t)$ functions are not exactly the same, however, if  we 
include the significant frequencies of the second and third order functions 
 $\phi_2(t)$ and  $\phi_3(t)$ as well, we receive all the frequencies of Tab.~\ref{tab:o-c_freq}.

Many higher order phase variation functions ($\phi_n(t), n>5$) show 
small amplitude regular fluctuations. This feature is an artefact viz.
the interaction between the quasi-uniform sampling and the 
periodic pulsation can produce the wagon-wheel or stroposcopic effect 
if the period ratio of the sampling and pulsation signals is about a quotient of
two integer numbers. This dynamical effect
causes the so-called moir\'e pattern on the light curves which
can easily be realised on the sparsely sampled LC data. 
This also implies that the higher order phase variation functions are not suitable for 
detecting any real light curve variations.

\subsection{The Blazhko incidence ratio}

As a summary of this section we can estimate the Blazhko incidence
ratio of the entire \textit{Kepler} RRab sample.

Although many hidden RR\,Lyrae stars were discovered in the original 
\textit{Kepler} field \citep{Hanyecz18} the light curves
of those stars have not published yet, so we can calculate with a 37-element RRab 
sample: 18 known Blazhko stars \citep{Benko14,Benko15} plus the RR\,Lyrae itself
in addition the 19 `non-Blazhko' stars of the present work. 
If we omit the discovered anomalous RRd stars NQ\,Lyr and V2470\,Cyg
it remains 35 stars.

If we take into account the  well-established 
V346\,Lyr as a new Blazhko star we find 
the \textit{Kepler} Blazhko incidence ratio to be 19:35 (55\%).
The hypothesis that additional 
modes could appear only on those RRab stars which show the Blazhko effect are 
disproved since V1510\,Cyg, V894\,Lyr and KIC\,9658012 spectra contain additional mode frequencies.
In the light of the present study the Blazhko nature of KIC\,7021124
is also became dubious since it was based on its long time-scale O$-$C variations alone 
\citep{Benko15}. We demonstrated such variations for all stars in the 
previous Sec.~\ref{sec:o-c}. If we count KIC\,7021124 as a non-Blazhko star we get 
18:35 (51\%) for the incidence ratio.
Though the former ratio is a bit higher than most previous works values but significantly lower
than the ratio of the recent paper of \citet{Kovacs18} who found it >90\%.

\section{Conclusions}

In this study we analysed the non-Blazhko RRab sample of the original
\textit{Kepler} field.

(i) One of the main finding is that up to a certain magnitude limit 
all stars show significant random cycle-to-cycle (C2C) light curve variation.
In other words, the RR\,Lyraes are not perfect clocks.
This phenomenon was suspected long ago but up to now there were only
indirect arguments. Studying the \textit{Kepler} SC data resulted in
direct photometric evidences for the first time.
\begin{itemize}
\item{
The C2C variations concentrate around the light curve maxima but 
other parts of the light curves especially the different phases connecting to the
hydrodynamic shocks in the atmospheres} are also concerned.
The maximal amplitude differences between light curve maxima are $\sim0.005-0.008$~mag
and this value seems to be general for all stars.
\item{
The C2C variations are random.  
The variation proved to be independent both from the Blazhko effect and
the potentially appearing low amplitude additional modes.
}
\end{itemize}

(ii) Low amplitude additional modes were detected for numerous stars.
\begin{itemize}
\item{We classified NQ\,Lyr and V2470\,Cyg as anomalous RRd stars showing their 
fundamental and 
first overtone mode frequencies ($f_0$ and $f_1$) in their spectra with extremely
small amplitude ratios $A(f_1)/A(f_0) = 0.00025$ and 0.00032, respectively. 
}
\item{We identified the second radial overtone frequency and its linear combinations
in the spectra of V1510\,Cyg, V346\,Lyr, V894\,Cyg and KIC\,9658012. For three of them the
highest amplitude additional frequency is the combination frequency 
$f_2-f_0$ which is lower than the fundamental frequency $f_0$.
}
\item{
The time frequency representations illustrate well
the amplitude changes of these additional frequencies. By using these diagrams  
several further stars have been revealed (NR\,Lyr, KIC\,6100702 and FN \,Lyr) in which 
frequencies around the positions of either $f_1$ and/or $f_2$ are temporarily appeared.
}
\end{itemize}

(iii) Analyzing the O$-$C diagrams and their spectra we found 
evident instrumental origin long time-scale phase variations for all stars.
We identified a new Blazhko candidate star (V346\,Lyr) and
the Blazhko incidence rate of the total published \textit{Kepler} RRab sample
found to be between 51 and 55\%.

\section*{Acknowledgements}

This work was supported by the Hungarian National Research, 
Development and Innovation Office by the Grants
NKFIH K-115709, K-119517 and NN-129075. 
AD was supported by the \'UNKP-18-4 New National Excellence Program of 
the Ministry of Human Capacities and the J\'anos Bolyai Research Scholarship 
of the Hungarian Academy of Sciences. AD would like to thank the City
of Szombathely for support under Agreement No. 67.177-21/2016.






\bsp	
\label{lastpage}

\end{document}